\documentclass[fleqn,usenatbib]{mnras}

\usepackage{newtxtext,newtxmath}

\usepackage[T1]{fontenc}

\DeclareRobustCommand{\VAN}[3]{#2}
\let\VANthebibliography\thebibliography
\def\thebibliography{\DeclareRobustCommand{\VAN}[3]{##3}\VANthebibliography}

\usepackage{graphicx}	
\usepackage{amsmath}	

\newcommand{\oiii}{[\ion{O}{iii}]}
\newcommand{\nii}{[\ion{N}{ii}]}
\newcommand{\sii}{[\ion{S}{ii}]}

\defcitealias{kakkad20}{K20}

\title[SUPER VII. H$\alpha$ properties of AGN at Cosmic noon]{SUPER VII. Morphology and Kinematics of H$\alpha$ emission in AGN host galaxies at Cosmic noon using SINFONI}

\author[D. Kakkad et al.]{
D. Kakkad$^{1,2,3}$, \thanks{E-mail: dkakkad@stsci.edu}
V. Mainieri$^{4}$, G. Vietri$^{5,6}$, I. Lamperti$^{7}$, S. Carniani$^{8}$, G. Cresci$^{9}$, C. Harrison$^{10}$, \newauthor A. Marconi$^{9,11}$,  M. Bischetti$^{12,13}$, C. Cicone$^{14}$, C. Circosta$^{15,16}$, B. Husemann$^{17}$, A. Man$^{18,19}$, \newauthor F. Mannucci$^{9}$, H. Netzer$^{20}$, P. Padovani$^{4,21}$, M. Perna$^{7,9}$, A. Puglisi$^{22}$, J. Scholtz$^{23,24}$, G. Tozzi$^{11,9}$, \newauthor C. Vignali$^{25,26}$, L. Zappacosta$^{5}$
\\
$^{1}$Space Telescope Science Institute, 3700 San Martin Drive, Baltimore, MD 21218, USA\\
$^{2}$European Southern Observatory, Alonso de Cordova 3107, Vitacura, Casilla 19001, Santiago de Chile, Chile\\
$^{3}$Department of Physics, University of Oxford, Denys Wilkinson Building, Keble Road, Oxford, OX1 3RH, UK\\
$^{4}$European Southern Observatory, Karl-Schwarzschild-Strasse 2, Garching bei M\"{u}nchen, Germany\\
$^{5}$INAF - Osservatorio Astronomico di Roma, Via Frascati 33, 00040, Monte Porzio Catone, Italy\\
$^{6}$INAF - Istituto di Astrofisica Spaziale e Fisica Cosmica Milano, Via A. Corti 12, 20133, Milano, Italy\\
$^{7}$Centro de Astrobiología (CAB), CSIC–INTA, Ctra. de Ajalvir Km. 4, 28850  Torrej\'{o}n de Ardoz, Madrid, Spain\\
$^{8}$Scuola Normale Superiore, Piazza dei Cavalieri 7, I-56126 Pisa, Italy\\
$^{9}$INAF - Osservatorio Astrofisico di Arcetri, Largo E. Fermi 5, 50125, Firenze, Italy\\
$^{10}$School of Mathematics, Statistics and Physics, Newcastle University, Newcastle upon Tyne, NE1 7RU, UK\\
$^{11}$Dipartimento di Fisica e Astronomia, Universit\'{a} di Firenze, Via G. Sansone 1, Sesto Fiorentino, 50019, Firenze, Italy\\
$^{12}$Dipartimento di Fisica, Universit\'{a} di Trieste, Sezione di Astronomia, Via G.B. Tiepolo 11, I-34131 Trieste, Italy\\
$^{13}$INAF - Osservatorio Astronomico di Trieste, Via G. Tiepolo 11, I-34143 Trieste, Italy\\
$^{14}$Institute of Theoretical Astrophysics, University of Oslo, PO Box 1029, Blindern 0315, Oslo, Norway\\
$^{15}$Department of Physics \& Astronomy, University College London, Gower Street, London, WC1E 6BT, UK\\
$^{16}$European Space Agency (ESA), European Space Astronomy Centre (ESAC), Camino Bajo del Castillo s/n, 28692 Villanueva de la Cañada, Madrid, Spain\\
$^{17}$Max-Planck-Institut f\"{u}r Astronomie, K\"{o}nigstuhl 17, D-69117 Heidelberg, Germany\\
$^{18}$Dunlap Institute for Astronomy and Astrophysics, University of Toronto, 50 St George Street, Toronto ON, M5S 3H4, Canada\\
$^{19}$Department of Physics \& Astronomy, University of British Columbia, 6224 Agricultural Road, Vancouver, BC V6T 1Z1, Canada\\
$^{20}$School of Physics and Astronomy, Tel-Aviv University, Tel Aviv 69978, Israel\\
$^{21}$Affiliated to INAF-Osservatorio di Astrofisica e Scienza dello Spazio, Via Piero Gobetti 93/3, I-40129 Bologna, Italy\\
$^{22}$Centre for Extragalactic Astronomy, Department of Physics, Durham University, South Road, Durham, DH1 3LE, UK\\
$^{23}$Kavli Institute for Cosmology, University of Cambridge, Madingley Road, Cambridge, CB3 0HA, UK\\
$^{24}$Cavendish Laboratory - Astrophysics Group, University of Cambridge, 19 JJ Thompson Avenue, Cambridge, CB3 0HE, UK\\
$^{25}$Dipartimento di Fisica e Astronomia "Augusto Righi", Universit\'{a} degli Studi di Bologna, via P. Gobetti 93/2, 40129 Bologna, Italy\\
$^{26}$INAF-Osservatorio di Astrofisica e Scienza dello Spazio di Bologna, Via Piero Gobetti, 93/3, I-40129, Bologna\\
}

\date{Accepted XXX. Received YYY; in original form ZZZ}

\pubyear{2021}

\begin{document}
\label{firstpage}
\pagerange{\pageref{firstpage}--\pageref{lastpage}}
\maketitle

\begin{abstract}
We present spatially resolved H$\alpha$ properties of 21 type 1 AGN host galaxies at z$\sim$2 derived from the SUPER survey. These targets were observed with the adaptive optics capabilities of the SINFONI spectrograph, a near-infrared integral field spectrograph, that provided a median spatial resolution of 0.3\arcsec~ ($\sim$2 kpc). We model the H$\alpha$ emission line profile in each pixel to investigate whether it traces gas in the narrow line region or if it is associated with star formation. To do this, we first investigate the presence of resolved H$\alpha$ emission by removing the contribution of the AGN PSF. We find extended H$\alpha$ emission in sixteen out of the 21 type 1 AGN host galaxies (76\%). Based on the BPT diagnostics, optical line flux ratios and the line widths (FWHM), we show that the H$\alpha$ emission in five galaxies is ionised by the AGN (30\%), in four galaxies by star formation (25\%) and for the rest (45\%), the ionisation source is unconstrained. Two galaxies show extended H$\alpha$ FWHM $>$600 km/s, which is interpreted as a part of an AGN-driven outflow. Morphological and kinematic maps of H$\alpha$ emission in targets with sufficient signal-to-noise ratio suggest the presence of rotationally supported disks in six galaxies and possible presence of companions in four galaxies. In two galaxies, we find an anti-correlation between the locations of extended H$\alpha$ emission and \oiii-based ionised outflows, indicating possible negative feedback at play. However, in the majority of galaxies, we do not find evidence of outflows impacting H$\alpha$ based star formation.
\end{abstract}




\section{Introduction} \label{sect1}

The conventional $\Lambda$ cold dark matter ($\Lambda$CDM) paradigm supports a hierarchical growth of objects i.e., smaller objects are formed first, then they merge successively into larger bodies. Within this picture, stars and galaxies are formed when baryons fall into dark matter potential wells, resulting in shocks followed by radiative cooling of gas \citep[e.g.,][]{white78, hopkins06}. Although this paradigm had initial success to describe a cosmological model for galaxy evolution, feedback processes needed to be invoked to regulate the formation of stars in the interstellar medium (ISM). One of the prominent sources of feedback in massive galaxies comes from the centrally located supermassive black holes \citep[e.g.,][]{kormendy95, richstone98, silk98}. 

Models predict that the feedback from these black holes is a result of radio jets and/or radiation pressure driven winds from the accretion of gas and dust from the ISM \citep[e.g.,][]{soltan82, yu02, fabian12}. During the accretion process, the luminosity of these black holes, also called active galactic nuclei (AGN) at this stage, can outshine the overall luminosity of the host galaxy itself. The net effect of these winds could be to relocate or eject cold molecular gas from the galaxy, dissociate or heat this gas and/or possibly prevent the inflow of gas from the halo into the ISM. Feedback processes from both star formation and AGN are largely used in state-of-the-art cosmological simulations \citep[e.g.,][]{springel05, hirschmann14, vogelsberger14, schaye15, steinborn15, dubois16, pillepich18}. An indirect manifestation of the influence that the AGN has on the growth history of its host galaxy is represented by the observed AGN-galaxy correlations, such as the black hole mass versus the stellar velocity dispersion ($M_{\rm BH}-\sigma$) and the black hole mass and bulge mass relation ($M_{\rm BH}-M_{\rm bulge}$) \citep[e.g.,][]{gebhardt00, gultekin09, caglar20}. 

Spectroscopic observations of AGN host galaxies over the past decades have revealed the ubiquitous presence of fast outflows with speeds $>$1000 km s$^{-1}$ in multiple gas phases across a wide range of redshifts \citep[e.g.,][]{greene11, venturi18, rojas20, perna21, puglisi21, ramos-almeida21, tozzi21, gatkine22}. Several studies found that these outflows are either driven by the AGN or star formation, based on the presence of correlations between the outflow and AGN or host galaxy properties \citep[e.g.,][]{carniani15, fiore17, fluetsch19, kakkad22}. However, there are also studies with different samples that report the absence of such scaling relations \citep[e.g.,][]{davies20, baron20}. 

To investigate the feedback effects generated by these outflows, we can study direct or indirect tracers of the host galaxy's molecular gas supply and/or star formation. Several studies in the literature have investigated the overall content and distribution of cold molecular gas with some results suggesting a systematic difference in the molecular gas fractions between AGN and non-AGN galaxies \citep[e.g.,][]{brusa18, kakkad17, circosta21, ellison21}, while several others (mostly at low redshift) do not find the same result \citep[e.g.,][]{husemann17, rosario18, jarvis20, koss21}. The molecular gas observations described above have the limitation that the results are mostly based on individual tracers (such as CO(2-1)), which can give an incomplete view into the ISM because the molecular gas itself can exist in multiple phases or can be traced via different excitation levels of CO. Furthermore, most of these studies do not spatially resolve the molecular gas as the CO measurements are obtained from integrated apertures. Lastly, the apertures of extraction are not always uniform which leads to conflicting results in the literature. 

Whether the outflows from the AGN have an impact on the host galaxy or not can also be revealed by investigating their impact on star formation (see review by \citet{harrison18} and references therein). Cosmological simulations predict a diverse set of scenarios where the star formation is regulated by the AGN, which may not result in an obvious observational signature of this feedback effect on the host galaxies of the AGN population \citep[e.g.,][]{ward22}. We briefly mention some examples here and we refer the reader to the references for further details. Using RAMSES-RT code, \citet{costa18} show that the outflows have the possibility to quench overall star formation \citep[see also ][]{dubois13, beckmann17}. Similar results are obtained in other suite of codes where the outflows have been shown to prevent the formation of new stars rather than shutting down ongoing star formation as a mechanism of quenching \citep[e.g.,][]{pillepich18}. On the other hand, \citet{zubovas13a} shows that it is possible even for stars to be ejected from the host galaxies, when dense shells formed due to the shocks and compression from impact of AGN outflows form stars along a radial trajectory. Such pressure-regulated star formation has been reported in several theoretical studies and high resolution hydrodynamical simulations \citep[e.g.,][]{ishibashi12, silk13, zubovas13b, dugan14, bieri16}. The burst of star formation is then followed by a period of quenching. Therefore, the AGN outflows could have the simultaneous capability of both suppressing and enhancing star formation \citep[e.g.,][]{zubovas17}. We also note that some of the results from the literature also report conditions where these outflows show limited impact of AGN feedback on the gas and star formation in the disk \citep[e.g.,][]{gabor14, roos15}. In summary, simulations predict several possible scenarios on the impact of outflows on star formation and that there is no universal answer to how exactly AGN feedback regulates star formation for each galaxy. 

The results from theory and simulations reported above can perhaps explain the diversity in the results, and often conflicting ones, from an observational perspective \citep[see review on positive and negative feedback in AGN host galaxies][]{cresci18}. Gas ejection by AGN winds or star formation quenching in AGN host galaxies has been reported in several low redshift and high redshift galaxies \citep[e.g.,][]{cano-diaz12, alatalo15, guillard15, carniani16, baron18, george19}. Similar to the predictions made in \citet{pillepich18}, using VLT/XSHOOTER and ALMA observations of a z$\sim$2.5 massive radio galaxy, \citet{man19} reported that most of the molecular gas is consumed by stars, followed by the removal of residual gas by the AGN. On the other hand, positive feedback has also been observed in the extragalactic sources, including dwarf galaxies \citep[e.g.,][]{gaibler12, rauch13, salome15, maiolino17, gallagher19, nesvadba20, perna20, bessiere22, schutte22}, with some targets that show a simultaneous presence of positive and negative feedback via spatially resolved integral field spectroscopic observations at rest-frame optical wavelengths \citep[e.g.,][]{cresci15, shin19}. Not all AGN host galaxies with outflows show the impact of star formation suppression or enhancement, especially when accounting for studies that make use of integrated spectra \citep[i.e., a global outlook into the outflows versus star formation paradigm][]{balmaverde16} or when considering dust obscured star formation \citep[e.g.,][]{scholtz20, scholtz21, lamperti21}. Indeed, the presence of negative or positive feedback in AGN depends upon the conditions of the ISM, outflows and/or jet power \citep[e.g.,][]{kalfountzou17}, which could also explain the diverse sets of results reported in the literature on the observational front. 

Lastly, star formation itself can be calculated via multiple methods, each tracing different time scales of star formation \citep[e.g.,][]{kennicutt98, battisti15, boquien15, catalan-torrecilla15, xie19, calzetti20, michiyama20, kim22, vietri22}. Furthermore, the results may further change depending on whether spatially resolved or integrated spectral measurements were made. For instance, star formation rate (SFR) calculated from the rest-frame ultraviolet wavelengths probe direct stellar light, while SFR calculated from the far-infrared wavelengths probes stellar light reprocessed by the dust \citep[e.g.,][]{shivaei16}. In the examples mentioned above, H$\alpha$ is often used as a star formation tracer for many extra-galactic studies. Although the H$\alpha$ line might be a reliable tracer in normal star forming galaxies across a wide range of redshifts, in the AGN host galaxies, there is a possibility of contamination by the AGN ionisation. Furthermore, we note that emission in FUV or optical wavelengths trace unobscured star formation and therefore, dust obscuration effects need to be accounted for to get a complete picture of the star formation in host galaxies \citep[see also][]{alaghband-zadeh16}.

In this paper, we try to overcome some of the limitations described above. We will characterise spatially-resolved properties of the H$\alpha$ emission in a large sample of moderate to high luminosity AGN at z$\sim$2 (Log $L_{\rm bol} =$ 45.4--47.9 erg s$^{-1}$). These AGN show the presence of high velocity ionised gas outflows in their host galaxies, which was inferred from the \oiii$\lambda$5007 emission \citep{kakkad20}. We investigate whether the H$\alpha$ emission is dominated by star formation or AGN ionisation. In the case where the H$\alpha$ emission is dominated by AGN ionisation, we will determine whether this emission is also a part of ionised gas outflows in the narrow line region (NLR). Finally, using the spatially resolved H$\alpha$ and \oiii$\lambda$5007 emission (\oiii ~hereafter), we investigate whether the outflows and star formation are co-located and whether this can reveal the presence of positive or negative feedback in these galaxies. 

We adopt the following $\Lambda$CDM parameters throughout this paper: $H_{\rm o}$ = 70 km s$^{-1}$, $\Omega_{\rm M}$ = 0.3 and $\Omega_{\Lambda}$ = 0.7. In all the maps in this paper, North is up and East is to left. 

\section{Sample, Observations and data reduction} \label{sect2}

The sample in this study is part of the SINFONI survey for Unveiling the Physics and Effect of Radiative feedback \citep[see][]{circosta18, kakkad20, mainieri21}. SUPER is an ESO large programme with the SINFONI instrument on board the VLT \citep{eisenhauer03} designed to characterise the properties of outflows in the ionised gas phase and their impact on host galaxies in a sample of 39 X-ray selected AGN at z$\sim$2 in AGN host galaxies at z$\sim$2. The survey is designed to trace ionised gas kinematics in the least biased manner by probing a wide range in AGN bolometric luminosity that span up to four orders of magnitude. The survey takes advantage of the Adaptive Optics (AO) module and reaches an angular resolution of $\sim$0.2\arcsec, resolving ionised gas kinematics and potentially star formation down to $\sim$2 kpc scales. Here we provide a brief description of the sample chosen for this paper and defer the reader to \citet{circosta18} and \citet{vietri20} for more details about the parent sample.

The 39 X-ray selected AGN ($L_{\rm 2-10 keV}$ > 10$^{42}$ erg s$^{-1}$) were obtained from shallow and deep fields: { \it Chandra} Deep Field South \citep[e.g.,][]{luo17}, COSMOS-Legacy \citep[e.g.,][]{civano16}, XMM-XXL \citep[e..g,][]{georgakakis11, liu16, menzel16}, Stripe 82 X-ray survey \citep[e.g.,][]{lamassa16} and WISE/SDSS selected Hyper-luminous quasar sample \citep[e.g.,][]{bischetti17, vietri18}. Out of these 39 AGN, 22 are classified as type 1 (56\%) based on the presence of broad H$\beta$, H$\alpha$ and/or MgII lines, characteristic of emission from the Broad Line Region (BLR). A wide range of ancillary data has allowed us to accurately compute various AGN properties such as the AGN bolometric luminosity ($L_{\rm bol}$), the black hole mass ($M_{\rm BH}$), X-ray column density ($N_{\rm H}$) in most targets and host properties such as SFR and stellar mass ($M_{\ast}$) in type 2 AGN. Estimation of host galaxy properties in type 1 AGN has been challenging as the AGN emission dominates the Spectral Energy Distribution (SED) for a wide range of wavelengths, preventing a robust estimation of SFR and stellar mass. Furthermore, CO(3-2) and dust continuum follow-up using sub-mm data from ALMA has also provided constraints on the molecular gas and dust properties for a sub-sample of SUPER targets \citep{circosta21, lamperti21}. 

The SINFONI observations for the SUPER survey were carried out in the H-band and K-band (or the H+K band in the case of bad weather) to trace the rest-frame optical emission lines: H$\beta$, \oiii$\lambda$4959, 5007, \nii$\lambda$6549, 6585, H$\alpha$ and \sii$\lambda\lambda$6716, 6731. The spatial resolution of the Adaptive Optics-assisted observations in the K-band grating reaches down to 0.2\arcsec ~(median 0.3\arcsec), similar to the H-band data, which corresponds to a physical scale of $\sim$2 kpc at z$\sim$2. The H+K grating observations have a spatial resolution of $\sim$0.9\arcsec ($\sim$7.5 kpc) and did not have sufficient offset along the X-direction while nodding during the observations. This implied that a spatially resolved analysis was only performed along the Y-direction. Further details about the observations, data reduction, and results from the \oiii ~analysis of the H-band data of type 1 SUPER targets are available in the first SINFONI data release paper \citep{kakkad20}. This paper focuses on the analysis of the H$\alpha$ emission line in the 21 type 1 SUPER targets, observed with the K-band or H+K band grating of the SINFONI spectrograph. We focus on the type 1 AGN to have a better estimate of PSF smearing effects by modelling the emission from the BLR. We have excluded the type 1 AGN S82X2106 since no signal was detected in our K-band observations. The targets in this paper have the following ranges in their black hole and host galaxy properties: log $M_{\ast}/M_{\odot} \sim$  10.38--11.20, SFR $<$94--686 $M_{\odot}$ yr$^{-1}$, log $L_{\rm bol}$/[erg s$^{-1}$] $\sim$ 45.4--47.9, log $M_{\rm BH}/M_{\odot} \sim$ 8.3--10.7, and log $N_{\rm H}$/cm$^{2} \sim <$21.25--24.1. Tables with full list of these values for each target are available in \citet{circosta18}, \citet{kakkad20}, and \citet{vietri20}. 
 
\section{Analysis} \label{sect3}

One of the primary goals of this paper is to investigate whether the observed H$\alpha$ emission is ionised by star formation or the AGN. We also compare the H$\alpha$ morphology and kinematics from the K-band data with that of the \oiii ~line presented in \citet{kakkad20} to infer if the ionised gas outflows have an impact on the unobscured star formation. 

In \citet{kakkad20}, we characterised the properties of the ionised gas kinematics in the NLR of the type 1 AGN from the SUPER sample using the \oiii$\lambda$5007 line in the H-band data. We describe the main results from the H-band data briefly here before presenting the analysis methods used in this paper. The non-parametric \oiii ~velocity dispersion, $w_{80}$, of the SUPER sample was always above 600 km s$^{-1}$, a velocity-cut usually indicative of the presence of ionised gas outflows driven by the AGN. The AGN as a source of outflow was also confirmed by the presence of strong correlations between the different ionised gas outflow properies ($w_{80}$, maximum velocity, $v_{\rm max}$ and mass outflow rates) and the AGN properties ($L_{\rm X}$(2-10 keV), $L_{\rm bol}$, $M_{\rm BH}$ and $\lambda_{\rm Edd}$). The ionised gas emission is extended in seven out of eleven targets for which the signal-to-noise in the \oiii ~emission was sufficient enough to perform a PSF de-convolution. The outflows associated with the ionised gas is extended up to $\sim$6 kpc from the AGN location, although $<$10\% of these outflows have the potential to escape the gravitational pull of the host galaxy.

\subsection{Modelling of the integrated spectrum} \label{sect3.1}

\begin{figure*}
\centering
\includegraphics[scale=0.28]{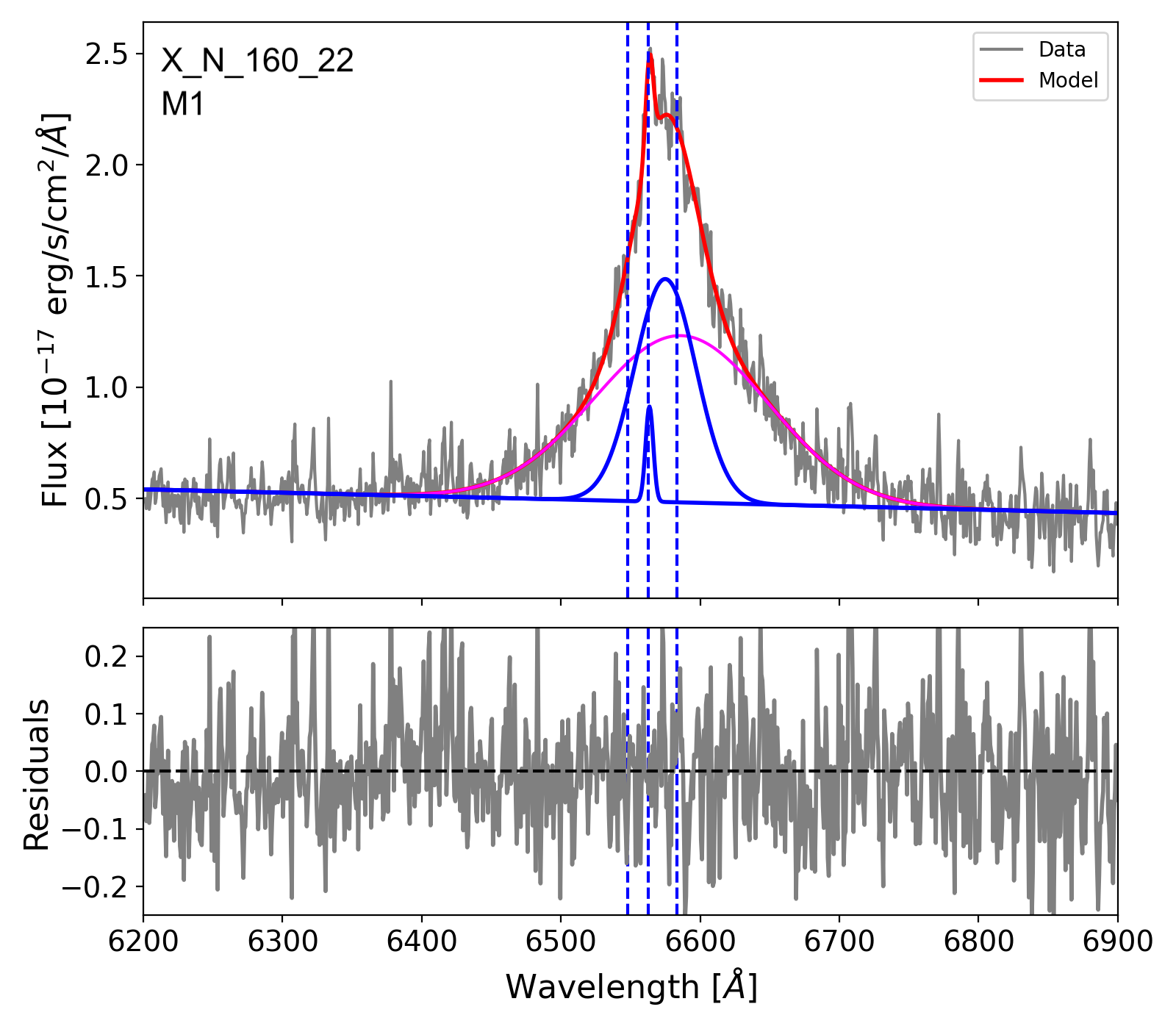}
\includegraphics[scale=0.28]{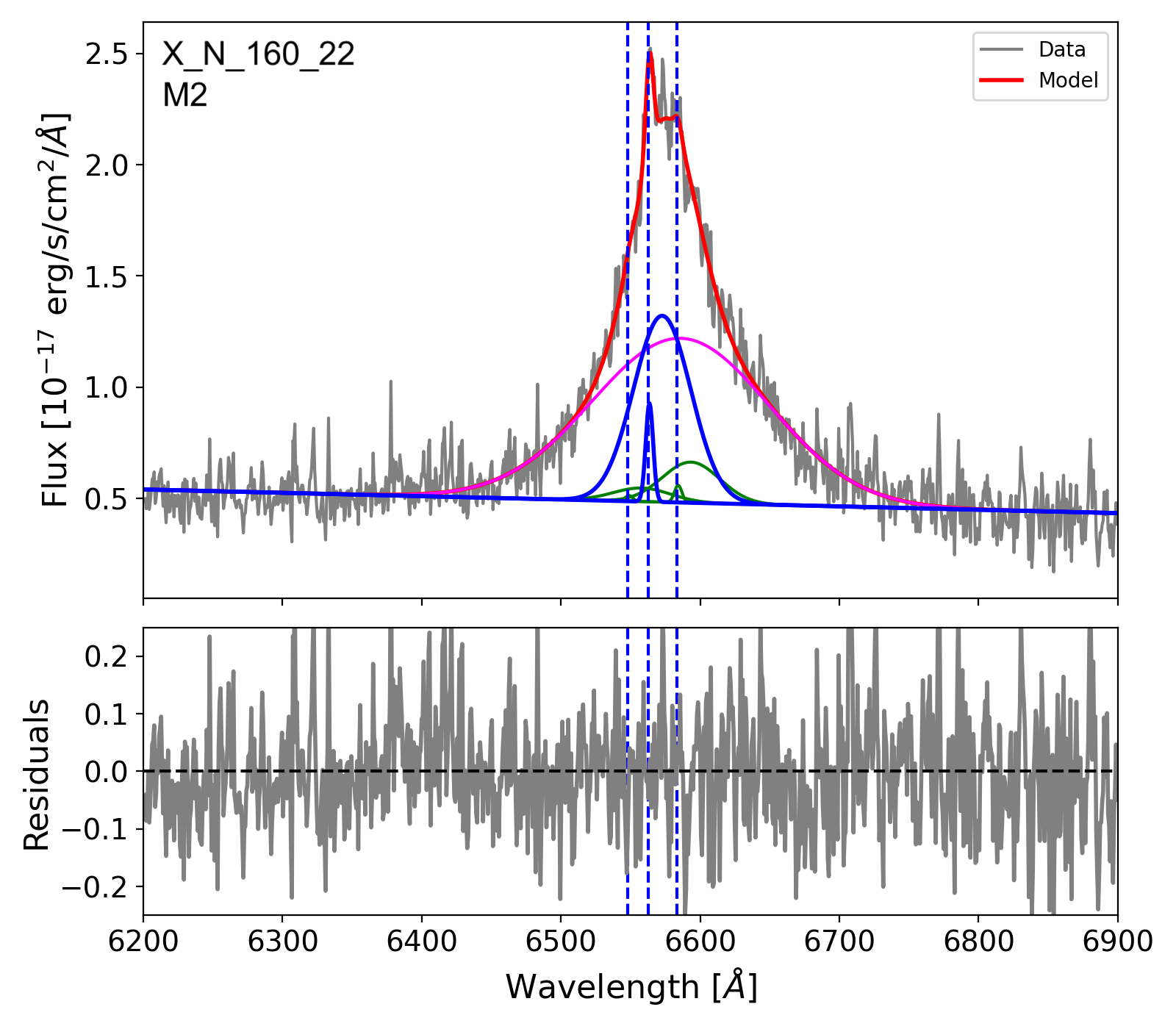}
\includegraphics[scale=0.28]{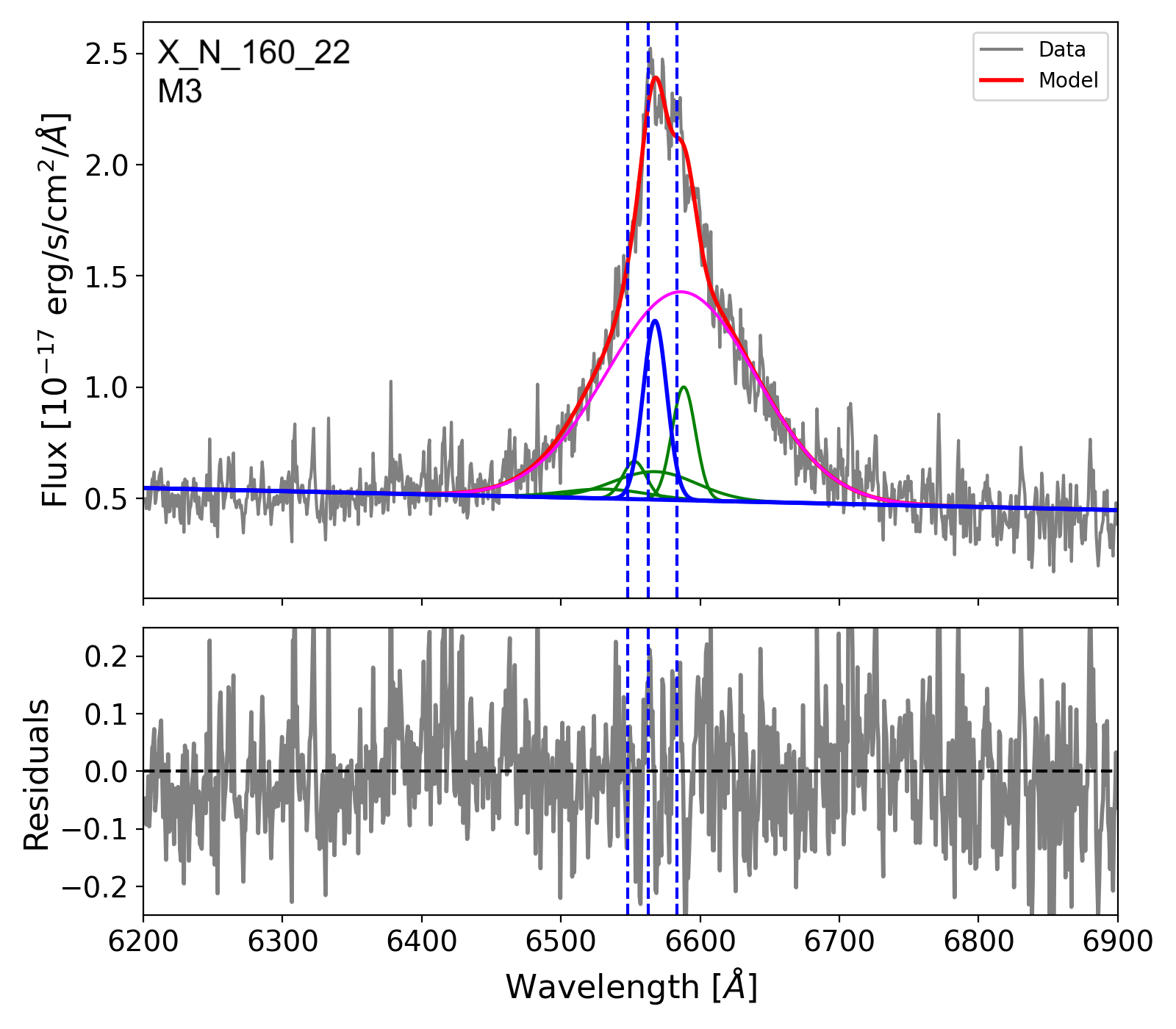}
\caption{The top panel shows the integrated spectrum of X\_N\_160\_22 extracted from a circular aperture of 0.9\arcsec ~diameter, showing the H$\alpha$ complex i.e., \nii$\lambda\lambda$6549,6585 and H$\alpha$ lines blended together. The left panel shows the line fitting model M1, the middle panel M2 and the right panel M3. The bottom panels show the residuals from the emission line modelling. Further details on the three models are described in Sect. \ref{sect3.1}. The figure highlights the differences in the Gaussian models when imposing different constraints on the fitting routines and that the H$\alpha$ complex modelling is often degenerate. In all the panels, the grey curve shows the raw spectrum extracted from the cube (or the residuals in the bottom panel), the red curve shows the overall model, the blue curves show the H$\alpha$ NLR models, the magenta curve represents the H$\alpha$ BLR emission and the green curve shows the \nii$\lambda\lambda$6549,6585 emission. The X-axis shows the rest-frame wavelength and the Y-axis shows the observed flux. The vertical dashed blue lines show the expected location of \nii$\lambda\lambda$6549,6585 and H$\alpha$ lines based on the redshift of this target. The \sii$\lambda\lambda$6716, 6731 lines remain undetected in most galaxies. }
\label{fig:intspec_X_N_160_22}
\end{figure*}

We first construct a model for the integrated spectrum before performing a pixel-by-pixel analysis of the emission lines in the K-band data. We extract the integrated spectrum for each object using a circular aperture centred on the AGN. The peak of the K-band continuum emission is used as an estimate for the AGN location and the aperture size was chosen to include at least $\sim$95\% of the total emission. We used the spectrum extracted from an object-free region to estimate the error on the spectrum. We focused the modelling of the extracted spectrum around the H$\alpha$ and \sii ~emission line regions.

Modelling the H$\alpha$ emission line profiles of high-redshift type 1 AGN host galaxies is not straightforward as the \nii ~and H$\alpha$ emission lines are blended into one broad feature. Each emission line is modelled using multiple Gaussian functions. The H$\alpha$ emission consists of contributions from clouds from the BLR, NLR, and emission from the host galaxy. The NLR component(s) of \nii$\lambda$6549, 6585 emission (as well as \sii$\lambda$6716, 6731) can be blended with the different components of the H$\alpha$ emission. As a result of such complex nature of the emission around the H$\alpha$ line (referred to as the H$\alpha$ complex hereafter), there is no unique solution to the emission line model. Therefore, constraints need to be imposed on different Gaussian parameters in order to obtain a unique solution. In this section, we explore the different ways of fitting the H$\alpha$ complex in the type 1 AGN host galaxies, based on methods previously adopted in the literature. We will then present the arguments that justify the fitting methodology adopted in this paper. 

We model the H$\alpha$ complex using three fitting methods, which we will call M1, M2 and M3. The three methods differ on some of the constraints imposed on the emission lines within the H$\alpha$ complex. These constraints are based on the results of the H$\beta$ and \oiii ~emission line models presented in \citet{kakkad20}. We start describing the assumptions common to the three methods used to model the H$\alpha$ complex. The integrated spectrum is modelled between the rest-frame wavelengths 6200--6900 \r{A} which includes the AGN continuum emission,  \nii$\lambda\lambda$6549,6585, H$\alpha$ and \sii$\lambda\lambda$6716,6731 emission lines. The BLR component of H$\alpha$ is modelled using a broad Gaussian or a broken power law, the choice depending on the model used to reproduce the H$\beta$ line in the H-band data \citep[see also][]{vietri20}. The NLR components of H$\alpha$ and \nii$\lambda6549, 6585$ are modelled using multiple Gaussian functions. A maximum of two Gaussians were used to reproduce the non-BLR components of each emission line (\nii$\lambda$6549, 6585 ~and H$\alpha$), equal to the number used for the \oiii ~line. In the case when two Gaussians are used for the NLR component, the Gaussian with the lower width is labelled as the narrow component, while the one with larger width is labelled as the broad component, without setting strict lower or upper limits for each component. To distinguish between the broad Gaussian component from the NLR and the one from the BLR, we will use the term broad Gaussian for the former and BLR-Gaussian for the latter. The widths of the narrow (broad) Gaussian components of \nii ~and H$\alpha$ are coupled to each other and the centroid difference between each component is fixed based on the expected positions in the rest-frame spectra. Furthermore, the relative ratio between the fluxes of \nii$\lambda$6549 and \nii$\lambda$6585 is fixed to the expected theoretical value of 1:3 \citep[e.g.,][]{osterbrock06}. We do not include fitting models for an iron component or \sii$\lambda\lambda$6716,6731, as they remain undetected or their contribution is negligible in all the galaxies.

\begin{table*}
\centering
\caption{This table reports the H$\alpha$ emission line fitting parameters using the M3 model described in Sect. \ref{sect3.1} \citep[see also][]{vietri20} and whether the emission is extended. (1) Target name, (2) The diameter of the circular aperture used for spectral extraction, Columns (3) \& (4) report the width (FWHM) of the individual NLR Gaussian components of the H$\alpha$ line (narrow NLR = $v_{1}$ and broad NLR = $v_{2}$), (5) width (FWHM) of the BLR Gaussian component of H$\alpha$. Columns (6), (7) \& (8) report the luminosity of the two NLR components (narrow NLR = $L_{1}$ and broad NLR = $L_{2}$) and the BLR component ($L_{\rm BLR}$), respectively, (9) reports the presence or absence of extended H$\alpha$ emission. The errors are obtained using a Monte Carlo approach and the values reported are the 1$\sigma$ error limits. $^{*}$Targets with the same line fitting results as the M2 model (Table \ref{table:intspec_results_kakkad_model}). Line modelling for all the targets has been done between rest-frame wavelengths 6200--6900 \AA, except J1441+0454 for which the fitting is performed between rest-frame wavelengths 6380--6950 \AA.}
\label{table:intspec_results_vietri_model}
\begin{tabular}{cccccccccc}
\hline
Target & Aperture & \multicolumn{3}{c}{FWHM} & \multicolumn{3}{c}{$L_{\rm H\alpha}$} & Extended H$\alpha$\\
& & $v_{1}$ & $v_{2}$ & $v_{\rm BLR}$ & $L_{1}$ & $L_{2}$ & $L_{\rm BLR}$\\
 & arcsec & km/s & km/s & km/s & erg/s & erg/s & erg/s & \\
 (1) & (2) & (3) & (4) & (5) & (6) & (7) & (8) & \\
\hline\hline
X\_N\_160\_22 & 0.9 & 915$\pm$195 & -- & 5730$\pm$156 & 43.44$\pm$0.78 & -- & 44.30$\pm$0.03 & Yes\\
X\_N\_81\_44 & 0.9 & 530$\pm$162 & 2000$\pm$213 & 6621$\pm$324 & 42.82$\pm$0.19 & 43.72$\pm$0.12 & 44.26$\pm$0.02 & Yes\\
X\_N\_53\_3$^{*}$ & 0.8 & 345$\pm$213 & -- & 4577$\pm$187 & 42.04$\pm$0.22 & -- & 43.70$\pm$0.01 & Yes\\
X\_N\_66\_23$^{*}$ & 0.9 & 150$\pm$150 & -- & 5640$\pm$190 & 41.45$\pm$7.8 & -- & 43.72$\pm$0.01 & May be\\
X\_N\_35\_20$^{*}$ & 0.5 & 532$\pm$56 & -- & 6549$\pm$1188 & 41.97$\pm$0.08 & -- & 42.68$\pm$0.08 & No\\
X\_N\_12\_26$^{*}$ & 1.0 & 544$\pm$42 & -- & 4615$\pm$99 & 42.80$\pm$0.04 & -- & 43.95$\pm$0.01 & Yes\\
X\_N\_44\_64$^{*}$ & 0.5 & 411$\pm$17 & -- & 7688$\pm$560 & 42.37$\pm$0.02 & -- & 43.03$\pm$0.02 & No\\
X\_N\_4\_48$^{*}$ & 0.7 & 473$\pm$53 & -- & 7596$\pm$303 & 42.79$\pm$0.06 & -- & 44.18$\pm$0.01 & Yes\\
X\_N\_102\_35 & 0.3 & -- & 1735$\pm$480 & 5418$\pm$246 & -- & 42.93$\pm$0.24 & 43.94$\pm$0.03 & No\\
X\_N\_115\_23 & 0.9 & 413$\pm$55 & 1400$\pm$267 & 7031$\pm$148 & 42.93$\pm$0.08 & 43.25$\pm$0.09 & 44.15$\pm$0.01 & Yes\\
cid\_166 & 0.9 & 395$\pm$126 & 1875$\pm$226 & 6881$\pm$104 & 42.48$\pm$0.27 & 43.76$\pm$0.11 & 44.60$\pm$0.01 & May be\\
cid\_1605$^{*}$ & 0.3 & 516$\pm$110 & -- & 3802$\pm$80 & 41.96$\pm$0.15 & -- & 43.27$\pm$0.01 & Yes\\
cid\_346$^{*}$ & 0.9 & 301$\pm$43 & 2916$\pm$156 & 7556$\pm$592 & 42.60$\pm$0.08 & 43.59$\pm$0.16 & 43.86$\pm$0.05 & Yes\\
cid\_1205$^{*}$ & 0.8 & 446$\pm$89 & -- & 5023$\pm$183 & 42.04$\pm$0.12 & -- & 43.46$\pm$0.01 & No\\
cid\_467$^{*}$ & 0.3 & 575$\pm$61 & -- & 8750$\pm$285 & 42.37$\pm$0.05 & -- & 43.86$\pm$0.01 & Yes\\
J1333+1649 & 1.0 & 615$\pm$572 & 2760$\pm$771 & 6217$\pm$467 & 42.77$\pm$0.67 & 44.74$\pm$1.28 & 45.53$\pm$0.02 & Yes\\
J1441+0454 & 1.0 & -- & 1000$\pm$85 & 5262$\pm$90 & -- & 44.18$\pm$0.03 & 44.92$\pm$0.02 & No\\
J1549+1245 & 1.1 & 376$\pm$221 & 1385$\pm$298 & 7914$\pm$97 & 43.40$\pm$0.30 & 44.36$\pm$0.20 & 45.63$\pm$0.01 & May be\\
S82X1905 & 1.0 & 660$\pm$58 & -- & 4935$\pm$99 & 43.01$\pm$0.04 & -- & 44.12$\pm$0.01 & Yes\\
S82X1940 & 0.8 & 380$\pm$124 & 1428$\pm$710 & 4145$\pm$434 & 42.76$\pm$0.25 & 42.80$\pm$4.53 & 44.25$\pm$0.11 & Yes\\
S82X2058& 0.9 & 363$\pm$128 & 1675$\pm$228 & 6616$\pm$184 & 42.13$\pm$0.30 & 42.56$\pm$3.06 & 44.09$\pm$0.02 & Yes\\
\hline
\end{tabular}
\end{table*}

We now describe the differences in the fitting constraints between the three methods. In the M1 method, we first fit the H$\alpha$ complex using only the H$\alpha$ components. The \nii ~line components are added only when the models from the H$\alpha$ only fit result in significant residuals on visual inspection. This approach was previously adopted in the literature for the analysis of the rest-frame optical spectra of one of the high-z quasars \citep[e.g.,][]{carniani16}, where the \nii ~emission is assumed to be undetected if no significant residuals are present after modelling with only the H$\alpha$ components. In the M2 method, \nii ~emission line model is included, irrespective of the presence or absence of residuals from the H$\alpha$-only fit. The M2 method assumes that the \nii ~emission is always present within the H$\alpha$ complex. Similar to the M1 method, no other constraints from the H-band spectra are imposed apart from the common constraints described in the previous paragraph. The M2 method is explored to gauge the variation in the fluxes and widths of various H$\alpha$ components upon inclusion of the \nii ~lines. Lastly in the M3 method, we fit the \nii ~line in the H$\alpha$ complex, similar to the M2 method. However, three additional constraints are imposed during the line fitting procedure: (1) The maximum allowed width of the narrow H$\alpha$ component is the width of the narrow \oiii ~line from the H-band line fitting results. (2) The difference in the centroid of narrow and broad Gaussian components of H$\alpha$ is kept the same as the difference in the centroid of narrow and broad Gaussian components of the \oiii ~line. (3) The width of the broad H$\alpha$ component is kept the same as that of the broad \oiii ~components. The method M3 is similar to the fitting model presented in \citet{vietri20}, with the exception that \sii$\lambda\lambda$6716,6731 emission line components were not included. Figure \ref{fig:intspec_X_N_160_22} shows the integrated spectrum and the emission line modelling of X\_N\_160\_22 as an example, using the three methods (left panel: M1, middle panel: M2 and right panel: M3). The integrated spectra of the rest of the targets are shown in the Appendix \ref{sect:appendix} (Figs. \ref{fig:intspec_alltargets1} and \ref{fig:intspec_alltargets2}). Below, we present arguments to support the selection of the M3 method as our baseline methodology in this paper.

\begin{figure*}
\centering
\includegraphics[scale=0.37]{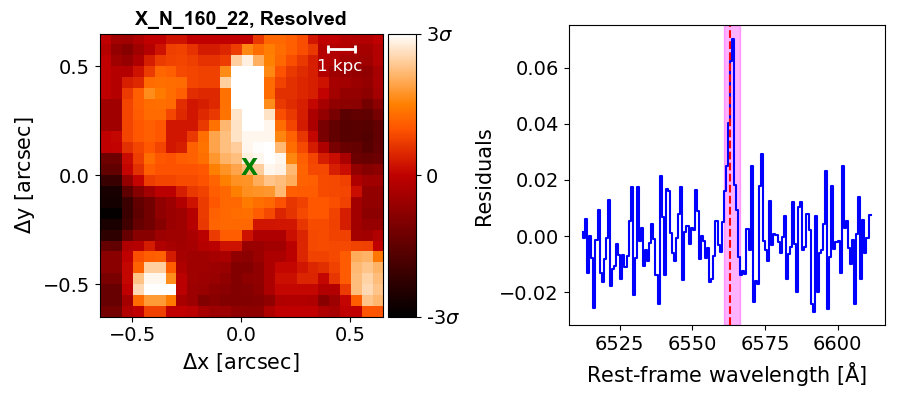}
\includegraphics[scale=0.37]{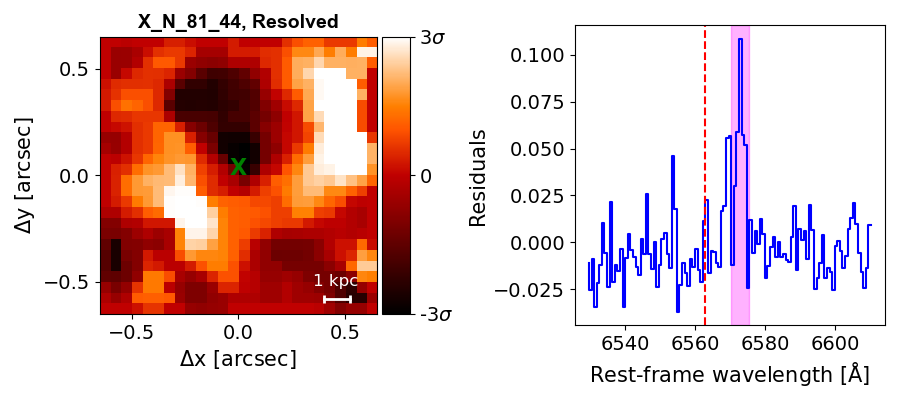}\\
\includegraphics[scale=0.37]{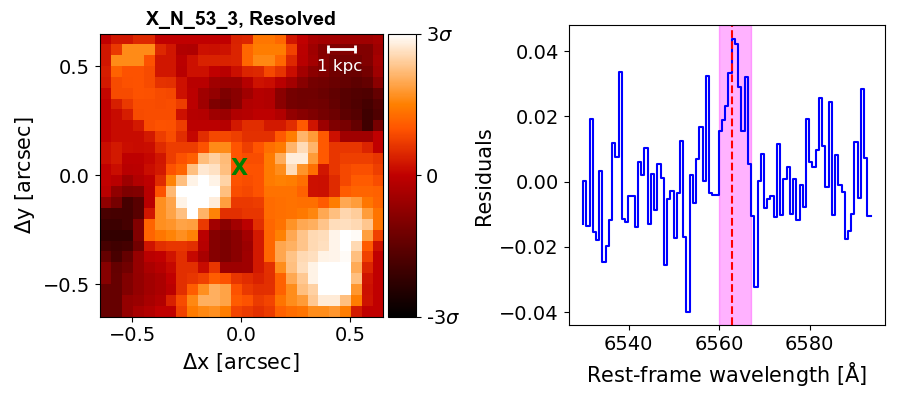}
\includegraphics[scale=0.37]{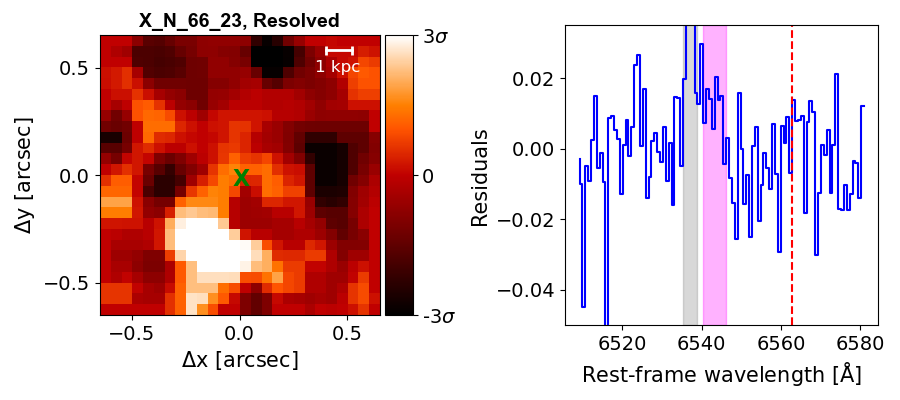}\\
\includegraphics[scale=0.37]{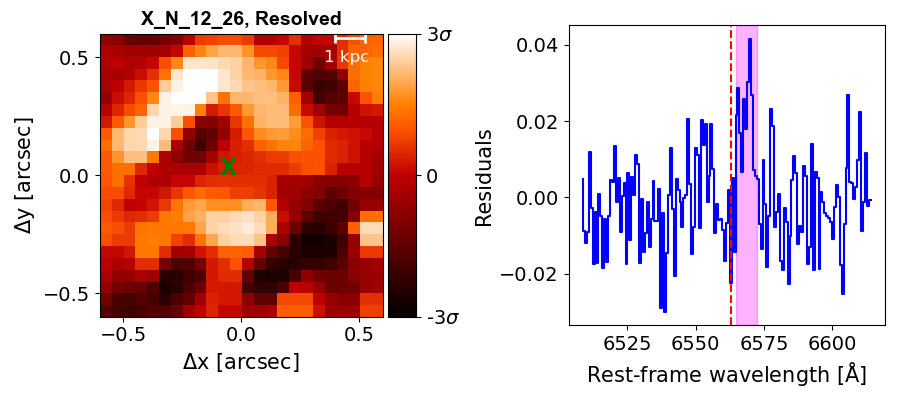}
\includegraphics[scale=0.37]{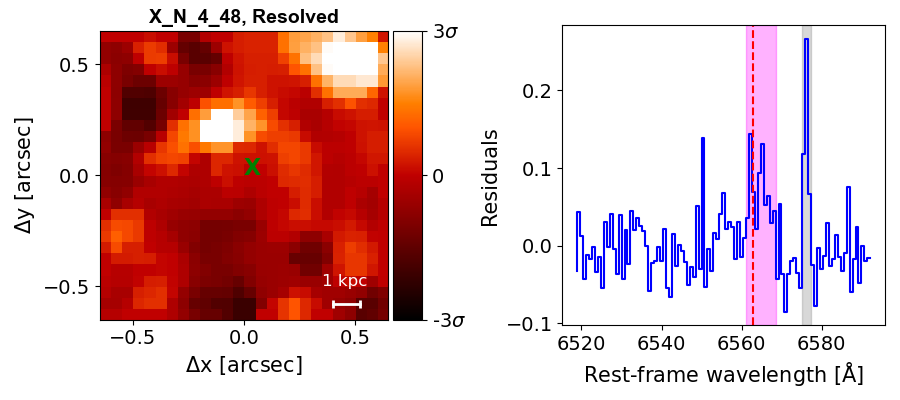}\\
\includegraphics[scale=0.37]{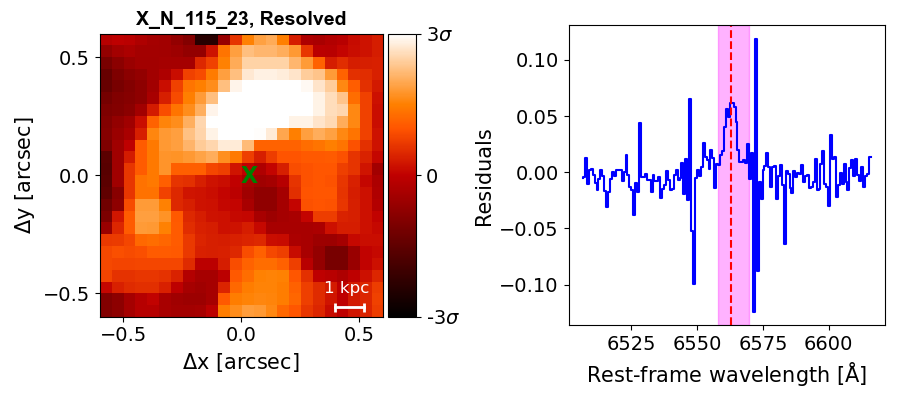}
\includegraphics[scale=0.37]{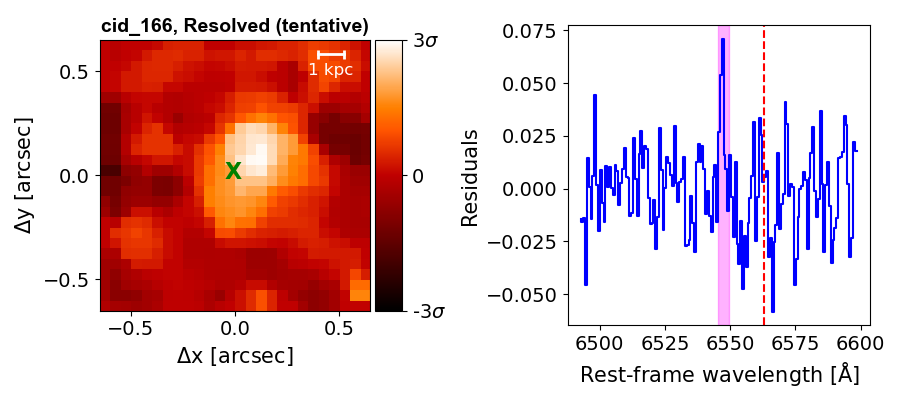}\\
\includegraphics[scale=0.37]{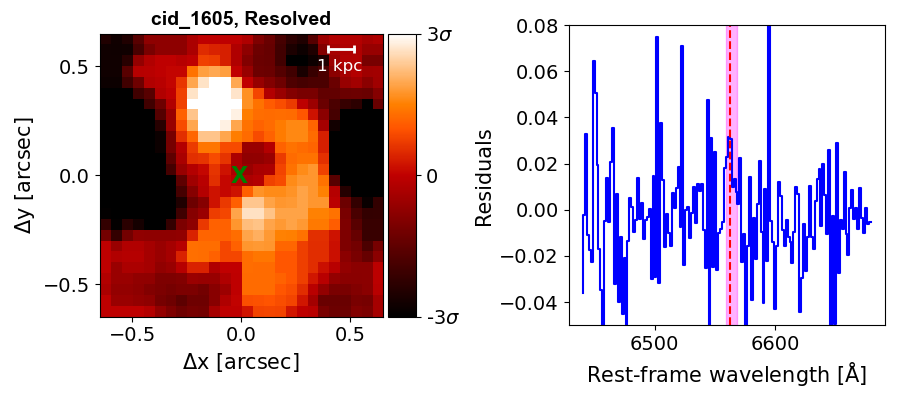}
\includegraphics[scale=0.37]{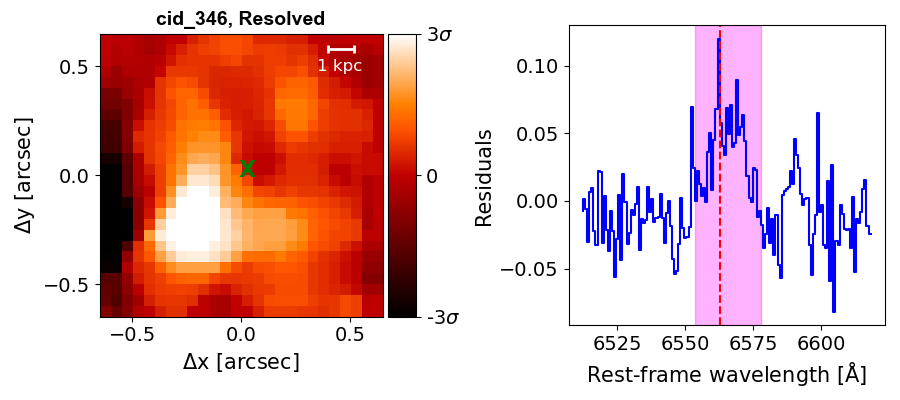}\\
\caption{The plots in this figure show the results of the PSF-subtraction procedure described in Sect. \ref{sect3.2}. The maps show the PSF-subtracted H$\alpha$ channel map and the corresponding right panels show the spectrum extracted from regions above 3$\sigma$ noise levels (white regions in the channel maps). The noise levels in each map is obtained from object-free regions. The "X" in the maps mark the peak of the K-band continuum emission which is used as a proxy for the AGN location. The magenta shaded region in the spectra show the channels that were collapsed to obtain the PSF-subtracted images. The vertical red dashed line shows the expected location of the H$\alpha$ line based on the redshift of the respective targets. The presence of structure in the PSF-subtracted map and a visible detection of the emission line in the spectra would suggest an extended H$\alpha$ feature. Each target in this figure show signatures of extended H$\alpha$ emission (labelled as "Resolved"), except X\_N\_66\_23 and cid\_166 (labelled as "Resolved?") where the extension is a possibility but unconfirmed with the current data. Further details are given in Sect. \ref{sect3.2}.}
\label{fig:BPT_PSF}
\end{figure*}

\begin{figure*}
\centering
\includegraphics[scale=0.37]{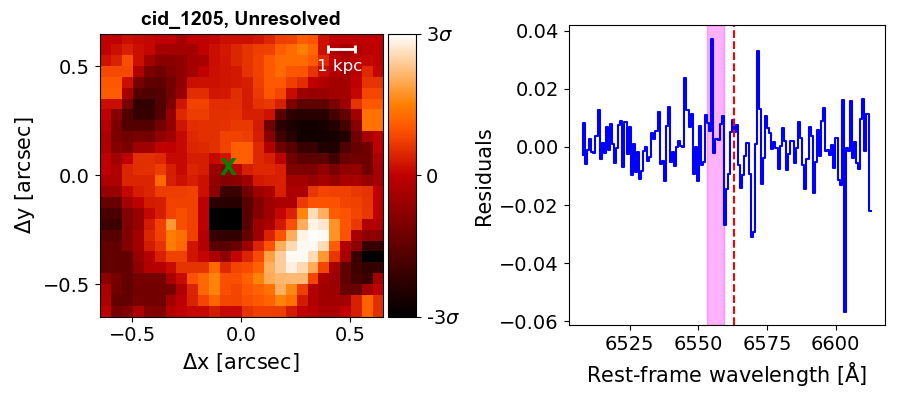}
\includegraphics[scale=0.37]{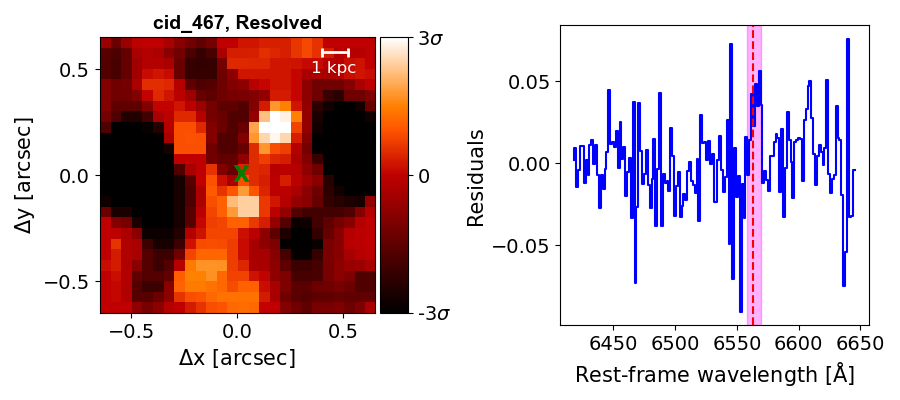}\\
\includegraphics[scale=0.37]{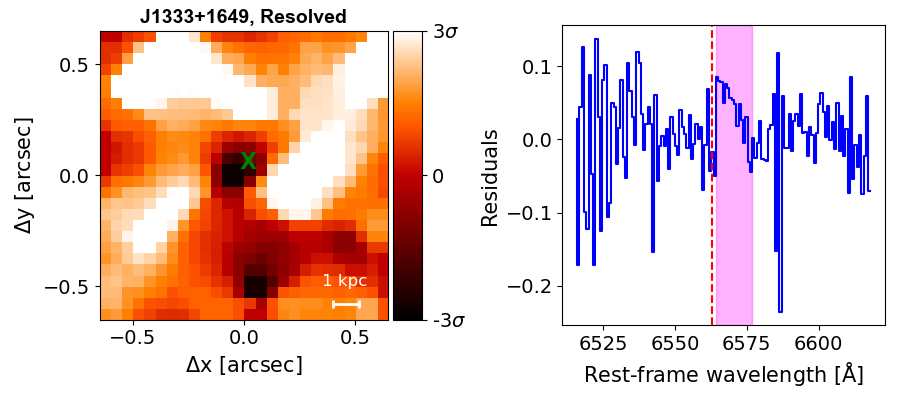}
\includegraphics[scale=0.37]{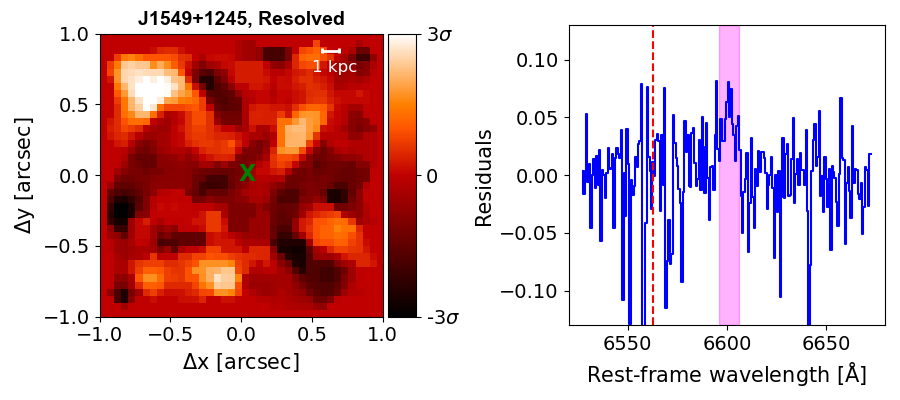}\\
\includegraphics[scale=0.37]{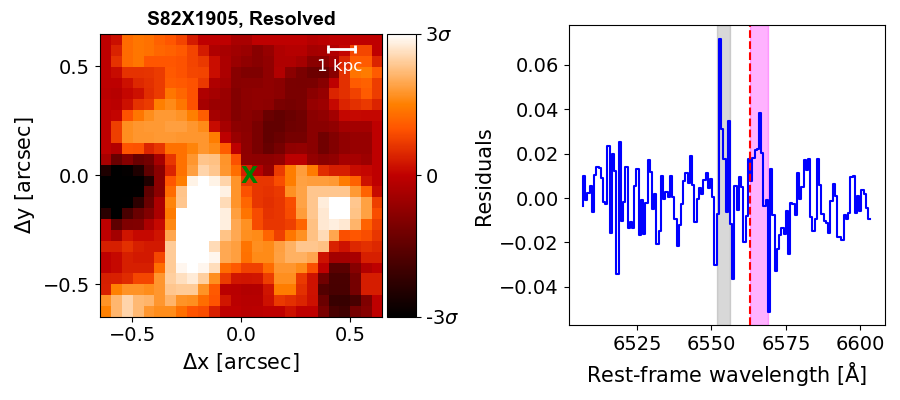}
\includegraphics[scale=0.37]{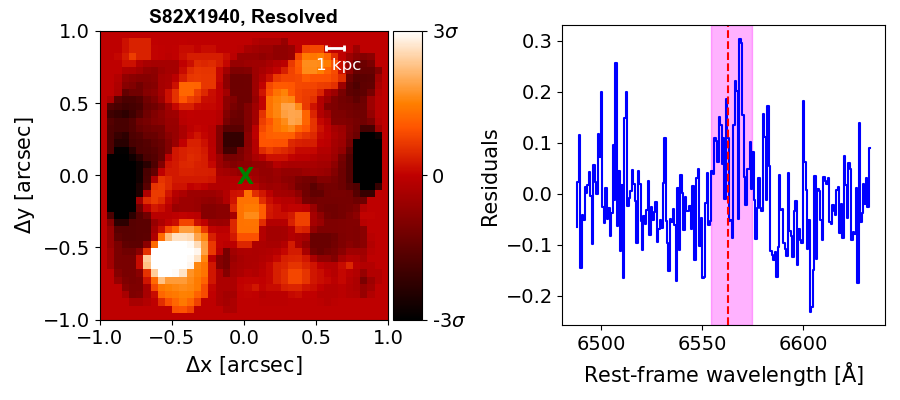}\\
\includegraphics[scale=0.37]{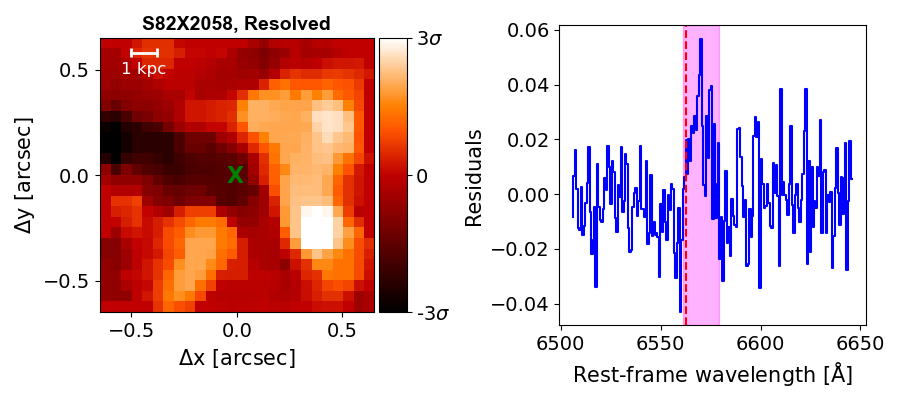}
\caption{Same as Fig. \ref{fig:BPT_PSF}, for the rest of the targets presented in this paper. All the targets, except cid\_1205 and J1549+1245, show confirmed evidence of extended H$\alpha$ emission. We detected no extended emission in cid\_1205 and the detection in J1549+1245 is tentative.}
\label{fig:BPT_PSF2}
\end{figure*}

\begin{figure}
\centering
\includegraphics[width=0.47\textwidth]{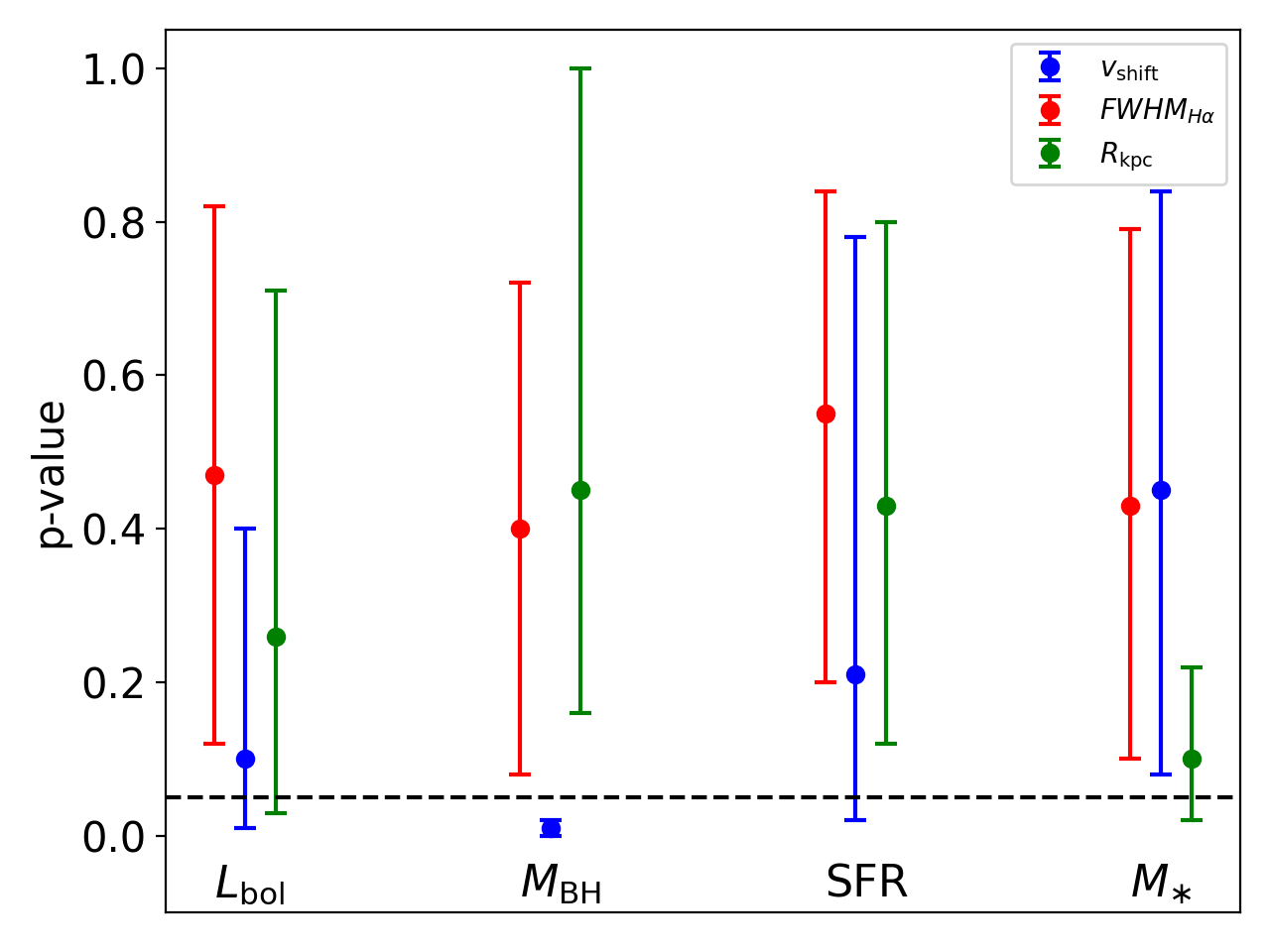}
\caption{The plot shows the null-hypothesis probability for non-correlation between different extended H$\alpha$ properties and the properties of the black hole and the host galaxy. The dashed black line shows the 5\% line for p-value, used as a demarcation between the presence or absence of correlations.}
\label{fig:scaling}
\end{figure}

Tables \ref{table:intspec_results_vietri_model}, \ref{table:intspec_results_kakkad_model} and \ref{table:intspec_results_carniani_model} report the fitting parameter values obtained using the three methods described above (M3, M2 and M1, respectively). In most of the type 1 AGN, the \nii ~emission is required as significant residuals remain from the H$\alpha$-only fitting procedure. Therefore, in the cases where such residuals are not observed from the H$\alpha$-only fit, we assume that the \nii ~lines are actually blended with the overall emission line structure. The presence of \nii ~lines in most high redshift galaxies is also supported by the results of the SINS/zC-SINF survey \citep[e.g.,][]{forster-schreiber18} where $\sim$65\% of the star forming galaxies had \nii ~detections. However, the M2 method return very high values for the width of the BLR component of the H$\alpha$ line ($>$9000 km s$^{-1}$ in some cases), which would result in unrealistically high black hole masses. This is a direct consequence of leaving all the emission line component centroids and widths free to vary. The presence of unrealistic widths and degeneracy in the results are solved using the M3 method. Therefore, hereafter we will use the M3 method in this paper to model the H$\alpha$ complex, which is also similar to the methodology adopted in previous papers from the SUPER survey \citep[e.g.,][]{vietri20, lamperti21}.

We employ a Monte Carlo (MC) approach to determine the errors in the fitting model, where we add noise to the integrated model and repeat the fitting procedure 100 times to determine the 1$\sigma$ errors on different parameters \citep[see e.g.,][]{perna15, brusa16, kakkad16}. The input noise for the MC is determined from the rms noise in an emission line-free region in the raw spectrum.

\subsection{Extended H\texorpdfstring{$\alpha$}{} emission} \label{sect3.2}

\begin{table*}
\centering
\caption{Properties of the extended H$\alpha$ emission presented in this paper. (1) Target name, (2) The maximum distance from the AGN location where the extended H$\alpha$ is detected at $\geq3\sigma$ after the PSF subtraction (see Figs. \ref{fig:BPT_PSF} and \ref{fig:BPT_PSF2}, (3) The width of the PSF-subtracted H$\alpha$ emission line from the extended regions, (4) The extended H$\alpha$ centroid in the spectral space with respect to the expected location of H$\alpha$ based on the \oiii ~redshift from \citet{kakkad20}, (5) The flux ratio, \nii/H$\alpha$ from the extended H$\alpha$ spectra from the K-band data, (6) The flux ratio \oiii/H$\beta$ from the extended \oiii ~spectra from the H-band data \citep{kakkad20}, (7) The ionisation source of the extended H$\alpha$ emission line regions based on its location in the BPT diagram or the \nii/H$\alpha$ flux ratio. Further details in Section \ref{sect3.3}. This table does not show the data for X\_N\_35\_20, X\_N\_44\_64, X\_N\_102\_35, cid\_1205, cid\_467 and J1441+0454, as these targets are unresolved.}
\begin{tabular}{ccccccc}
\hline
Target & R$_{\rm H\alpha}$ & v$_{\rm H\alpha}$ & $v_{\rm shift}$ & log \nii/H$\alpha$ & log \oiii/H$\beta$ & Ionisation source\\
 & kpc & km s$^{-1}$ & km s$^{-1}$\\
(1) & (2) & (3) & (4) & (5) & (6)& (7)\\
\hline\hline
X\_N\_160\_22  & 4.5$\pm$0.4 & 112$\pm$30 & 42$\pm$10 & $<$-0.7 & -- & SF\\
X\_N\_81\_44 & 3.5$\pm$0.4 & 113$\pm$15 & 485$\pm$5 & $<$-0.8 & -- & SF\\
X\_N\_53\_3 & 5.2$\pm$0.4 & 199$\pm$44 & 33$\pm$17 & -0.6$^{+0.2}_{-0.4}$ & -- &  unconstrained\\
X\_N\_66\_23 & 4.0$\pm$0.4 & 370$\pm$109 & -1200$\pm$300 & $<$-0.9 & -- & SF\\
X\_N\_12\_26 & 4.5$\pm$0.4 & 241$\pm$45 & 286$\pm$21 & $<$-0.6 & -- & unconstrained\\
X\_N\_4\_48 & 6.4$\pm$0.4 & 256$\pm$58 & 78$\pm$29 & $<$-0.4 & -- & unconstrained\\
X\_N\_115\_23 & 4.6$\pm$0.4 & 270$\pm$35 & 19$\pm$15 & $<$-0.8 & 0.9$\pm$0.3 & AGN\\
cid\_166 & 2.6$\pm$0.4 & 73$\pm$16 & -713$\pm$9 & $<$-0.5 & -- & unconstrained\\
cid\_1605 & 3.9$\pm$0.4 & 351$\pm$140 & 10$\pm$1 & $<$-0.3 & -- & unconstrained\\
cid\_346 & 4.7$\pm$0.4 & 628$\pm$45 & 134$\pm$20 & $<$-0.4 & $>$0.5 & AGN\\
cid\_467 & 2.6$\pm$0.4 & 350$\pm$65 & 180$\pm$25 & $<$-0.7 & -- & SF \\
J1333+1649 & 5.7$\pm$0.4 & 302$\pm$67 & 260$\pm$22 & $<$-0.5 & 0.7$\pm$0.2 & AGN\\
J1549+1245 & 9.0$\pm$0.4 & 408$\pm$92 & 1720$\pm$200 & $<$-0.4 & -- & unconstrained\\
S82X1905 & 4.4$\pm$0.4 & 142$\pm$28 & 143$\pm$12 & $<$-0.6 & $>$1.0 & AGN\\
S82X1940 & 7.8$\pm$0.4 & 784$\pm$98 & 202$\pm$120 & $<$-0.2 & -- & AGN\\
S82X2058 & 4.4$\pm$0.4 & 436$\pm$68 & 330$\pm$30 & $<$-0.4 & -- & unconstrained\\
\hline
\end{tabular}
\label{table:halpha_results}
\end{table*}

Before performing a pixel-by-pixel analysis of the H$\alpha$ spectrum of the type 1 targets, it is important to quantify the contribution of beam-smearing from the AGN PSF to the observed H$\alpha$ emission. This procedure, also called PSF-subtraction, isolates the underlying contributions from the host galaxy. Beam smearing results in emission line regions to mimic radial profiles consistent with PSF profiles, and consequently appear artificially extended \citep[see e.g.,][]{carniani15, husemann16, luo19}. We subtract this contribution from the AGN-PSF using methods similar to those employed in the analysis of the extended \oiii ~emission in the H-band data in \citet{kakkad20}. 

If the H$\alpha$ emission is unresolved, then the spectrum at any distance from the AGN will be the same as the spectrum at the AGN location, except for an overall scaling factor across the spectrum \citep[e.g.,][]{jahnke04}. Therefore, we first model the spectrum extracted at the AGN location (circular aperture of diameter 0.1\arcsec ~centred on the K-band continuum peak). We will refer to this spectrum as the "nuclear model". The nuclear model is subtracted from every pixel across the SINFONI field-of-view, only allowing a variation in the overall normalisation factor of the spectrum. The kinematic parameters of different Gaussian components (i.e., the line centroid and the line width) are kept fixed with respect to the nuclear model. After the subtraction of the PSF model, we collapsed the channels in the residual data cube at the expected location of narrow H$\alpha$ emission. The width of the channel window is optimised for each target separately, based on where we obtain the maximum residual H$\alpha$ emission. A noisy map (an image with net zero residuals) indicates that the underlying H$\alpha$ emission is unresolved, while systematic patterns in these residual maps would suggest that the H$\alpha$ emission is extended. This extended H$\alpha$ emission can be due to star-formation in the host galaxy, AGN emission or a combination of the two. The nature and origin of this extended emission will be discussed further in Sect. \ref{sect3.3}. 

Figures \ref{fig:BPT_PSF} and \ref{fig:BPT_PSF2} shows the results of the PSF-subtraction method for each target presented in this paper. We do not show the PSF-subtraction results for the following four targets: X\_N\_35\_20, X\_N\_44\_64, X\_N\_102\_35 and J1441+0454. Targets X\_N\_35\_20 and X\_N\_44\_64 had low S/N in their H$\alpha$ line in the integrated spectrum to obtain a reliable estimate on the H$\alpha$ BLR flux and consequently the PSF. X\_N\_102\_35 was observed in the H+K band and no residuals were observed in the Y-direction. Lastly, the location of the H$\alpha$ line in J1441+0454 is contaminated by telluric emission and the PSF-subtraction did not yield a reliable detection. 

All the maps in Figs. \ref{fig:BPT_PSF} and \ref{fig:BPT_PSF2} show the PSF-subtracted channel maps at the expected location of H$\alpha$ emission. The colour map in the H$\alpha$ images is set between $\pm$3$\sigma$, where $\sigma$ is the noise level in the map determined from object-free locations. The panels on the right side of the maps show the spectrum extracted from regions at or above 3$\sigma$ in the H$\alpha$ channel maps. The magenta shaded regions in the spectra show the channels that were collapsed to obtain the PSF-subtracted H$\alpha$ map. The vertical dotted red line in the spectra show the expected location of the H$\alpha$ line based on the \oiii ~redshift values reported in \citet{kakkad20}.

We confirm the presence of extended H$\alpha$ emission in 13 out of the 21 galaxies and 3 galaxies (X\_N\_66\_23, cid\_166 and J1549+1245) show tentative evidence of extended emission. The detection in X\_N\_66\_23 is classified as a tentative one, as the extended emission is observed at a blue-shift of $\sim$-1200 km s$^{-1}$ from the expected location of the H$\alpha$ line in the spectra. Similarly, the extended emission in J1549+1245 is detected at a distance of 9 kpc from the AGN location and the spectral position of the detection is redshifted by $\sim$+1800 km s$^{-1}$ with respect to the expected location of the H$\alpha$ line. Although currently available archival optical observations do not suggest the presence of companions, this may be due to their limited spatial resolution. In fact, ALMA CO observations of J1549+1245 have shown the presence of a possible companion south of the AGN location \citep[see][]{bischetti21}. Such extreme velocity shifts have been observed in previous CO observations of high redshift X-ray AGN sources \citep[e.g.,][]{carniani17}. The extended H$\alpha$ detections in X\_N\_66\_23 and J1549+1245 may suggest emission from tidal tails. In the case of cid\_166, the detection is almost at the limit of the spectral resolution of SINFONI. Furthermore, the observed residual is blue-shifted by $\sim$700 km s$^{-1}$ from the expected location of the H$\alpha$ emission and therefore, the extended emission in cid\_166 has been classified as a tentative detection. cid\_1205 is the only target among the ones observed with AO and with sufficient S/N for the PSF-subtraction analysis that shows no extended H$\alpha$ emission. 

Overall, we find a higher fraction of targets with extended H$\alpha$ emission ($\sim$76\% including the tentative detections) than extended \oiii ~emission ($\sim$35\%). The spatial extent of the H$\alpha$ emission (maximum distance between the AGN location and the H$\alpha$ emission in the PSF-subtracted maps), $R_{\rm H\alpha}$ is in the range 3--9 kpc, with a mean value of $\sim$5 kpc. Due to the low signal-to-noise after the PSF-subtraction, we model the extended H$\alpha$ residual spectrum using single Gaussian functions and define two parameters: the width (FWHM) of the extended H$\alpha$ emission, $FWHM_{\rm H\alpha}$ and the velocity shift, $v_{\rm shift}$, between the expected location of the H$\alpha$ line and actual location of the extended H$\alpha$ line in the spectra. We find $FWHM_{\rm H\alpha}$ in the range 73--784 km s$^{-1}$ with a mean width of 312 km s$^{-1}$. The velocity shift, $v_{\rm shift}$ is in the range -1200 -- 1720 km s$^{-1}$, with the majority of the targets showing redshifted H$\alpha$ emission. Only two targets, X\_N\_66\_23 and cid\_166, display blue-shifted extended H$\alpha$ emission and both of these targets are classified as having a tentative extended emission. The properties of the extended H$\alpha$ emission are summarized in Table \ref{table:halpha_results}.

We investigated whether the properties of the extended  H$\alpha$ emission, namely $R_{\rm H\alpha}$, $FWHM_{\rm H\alpha}$ and $v_{\rm shift}$ show any correlations with the AGN or host galaxy properties such as $L_{\rm bol}$, $M_{\rm BH}$, SFR and $M_{\ast}$. We derived the Pearson coefficient and the p-value (null hypothesis probability for non-correlation) for the individual relations. The p-values for different correlations are summarised in Fig. \ref{fig:scaling}. We define a correlation to exist based on p$<$0.05. Accounting for the errors,  $v_{\rm shift}$ correlates with $M_{\rm BH}$ and possibly also with $L_{\rm bol}$,  and SFR. $R_{\rm kpc}$ also shows correlation with $M_{\ast}$ and possibly with $L_{\rm bol}$. No correlations are found between $FWHM_{\rm H\alpha}$ and the AGN or host galaxy properties. However, we note that since these are type 1 targets, the host galaxy properties, namely SFR and $M_{\ast}$, are highly unconstrained and are only available for 5 targets \citep[see][]{circosta18}. Overall, the $v_{\rm shift}$ parameter seems to correlate strongly with the AGN properties suggesting that the observed velocity shift is possibly due to outflowing ionised gas traced with H$\alpha$. On the other hand, the size of the extended region correlates with the galaxy mass, consistent with the correlations reported between the size and stellar mass of galaxies \citep[e.g.,][]{van-der-wel14}. Furthermore, the lack of correlations or weaker correlations in Fig. \ref{fig:scaling} may also suggest that the observed H$\alpha$ emission in some galaxies might also be tracing a companion galaxy i.e., not associated with the X-ray source or the host galaxy itself. The presence of companions will be further discussed in Sect. \ref{sect4}.

\subsection{Ionisation source of the extended H\texorpdfstring{$\alpha$}{} emission} \label{sect3.3}

\begin{figure}
\centering
\includegraphics[scale=0.33]{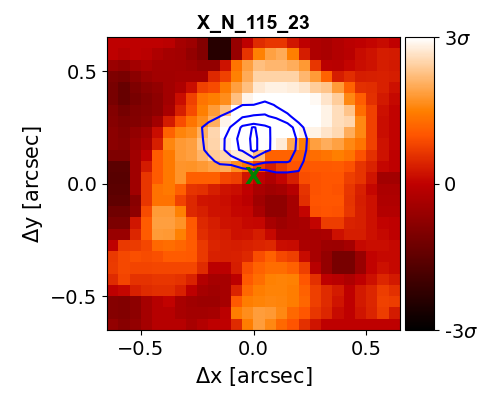}
\includegraphics[scale=0.33]{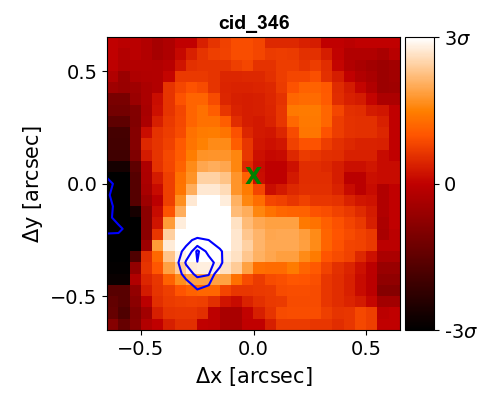}\\
\includegraphics[scale=0.33]{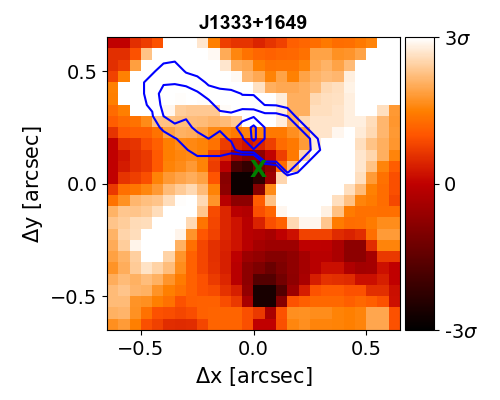}
\includegraphics[scale=0.33]{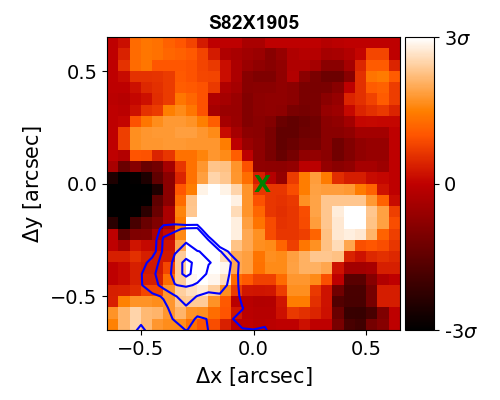}\\
\includegraphics[scale=0.4]{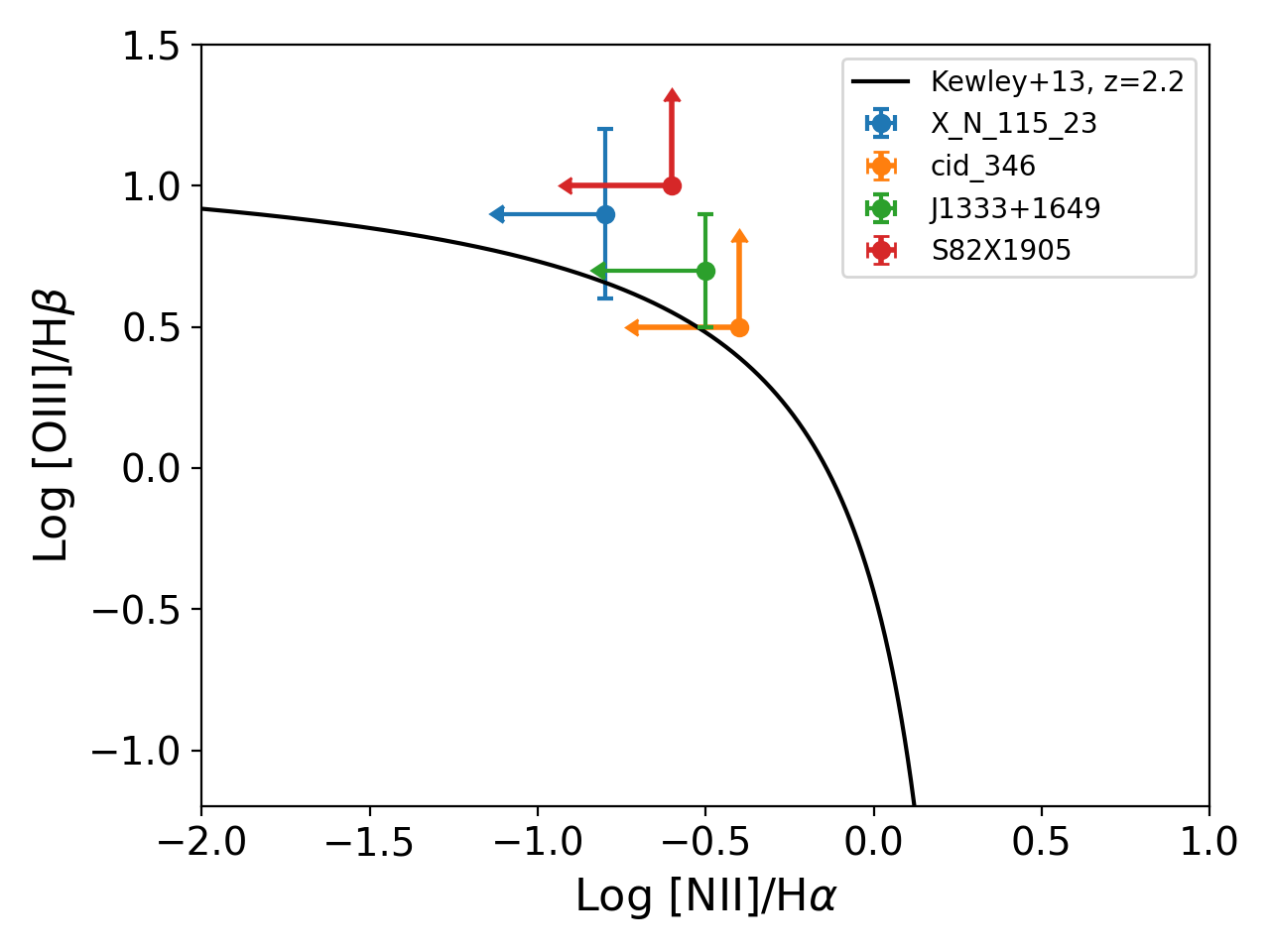}
\caption{The background images in the maps in the top two rows shows the PSF-subtracted H$\alpha$ images and the overlaid blue contours show the locations of extended \oiii ~emission. The contour levels are at 98\%, 90\%, 70\% and 50\% of the peak \oiii ~flux. In four targets, X\_N\_115\_23, cid\_346, J1333+1649 and S82X1905, the spatial location of the extended H$\alpha$ emission coincides with that of the \oiii ~emission. This made it possible to place the extended H$\alpha$ emission of these four galaxies in the BPT diagram shown in the bottom panel. In all these four galaxies, the extended emission is ionised by the AGN. The black line shows the division between star forming and AGN ionisation from \citet{kewley13}. Targets, X\_N\_81\_44 and J1549+1245 do not show the same location of the \oiii ~and H$\alpha$ emission.}
\label{fig:BPT}
\end{figure}

\begin{figure}
\centering
\includegraphics[scale=0.33]{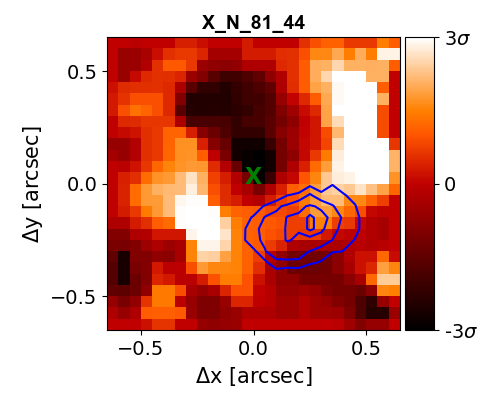}
\includegraphics[scale=0.33]{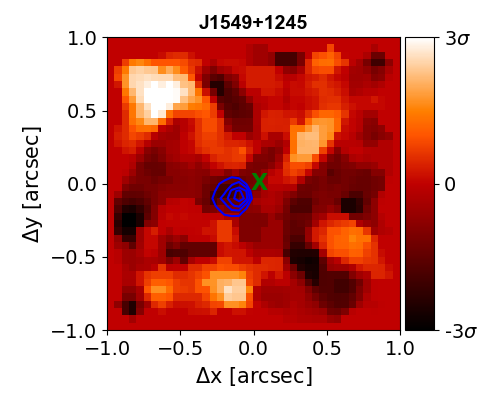}
\caption{Same as Fig. \ref{fig:BPT} but for targets, X\_N\_81\_44 and J1549+1245, that do not show the same location of the \oiii ~and H$\alpha$ emission.}
\label{fig:BPT2}
\end{figure}

We now discuss the ionisation source of the extended H$\alpha$ emission, i.e., whether the emission is ionised by the AGN, star formation or a combination of the two, based on the \oiii ~results reported in \citet{kakkad20} and PSF-subtracted images and spectra shown in Figs. \ref{fig:BPT_PSF} and \ref{fig:BPT_PSF2}. We use three different techniques to infer the ionisation source: (1) We compare the extended H$\alpha$ morphology with the \oiii ~morphology. A similarity between the two emission regions could suggest that the extended H$\alpha$ emission traces gas in the NLR. (2) As described in Sect. \ref{sect3.2}, we fit the single Gaussian functions to the extended H$\alpha$ emission to derive the line width. A large width (e.g., $>$600 km s$^{-1}$) would mean H$\alpha$ emission traces outflowing gas. This is supported by the maximum gas velocity of $\sim$500 km s$^{-1}$ observed in the case of star formation driven outflows in high redshift galaxies \citep[see review by][]{forster-schreiber20}, and therefore using 600 km s$^{-1}$ as a cut off for AGN-driven outflows is a conservative assumption. However, we note that galaxy interactions can also increase the line widths \citep[e.g.,][]{puglisi21}. Due to the relatively low S/N of the H$\alpha$ lines compared to integrated spectra, we do not attempt to fit multiple Gaussian functions to these spectra. (3) Lastly, the single Gaussian fits to the extended H$\alpha$ emission (Sect. \ref{sect3.2}, Fig. \ref{fig:BPT_PSF}) were used to estimate the flux of the \nii ~and H$\alpha$ lines. The flux ratios were then placed on the classical Baldwin, Phillips \& Telervich \citep[BPT, see][]{baldwin81, veilleux87} diagnostic diagrams to estimate the probable source of ionisation from expectations based on the literature \citep[e.g.,][]{kauffmann03}. For the majority of the galaxies, extended \oiii ~and H$\beta$ emission lines are not detected at the location of extended H$\alpha$ emission, therefore placing them on the BPT diagrams  was not possible.

We were able to plot resolved BPT maps for four galaxies, namely X\_N\_115\_23, cid\_346, J1333+1649 and S82X1905, as extended \oiii ~emission is also detected at the location of extended H$\alpha$ as shown in the top two rows of Fig. \ref{fig:BPT}. The bottom panel in Fig. \ref{fig:BPT} shows the location of extended emission in these four galaxies in the \nii-BPT diagram. In these four galaxies the spatial coincidence of the extended H$\alpha$ and \oiii ~emission suggests that we are tracing the NLR rather than the host galaxy disk. That the ionization source for the extended H$\alpha$ is the AGN rather than star-formation is further supported by the location of these galaxies in the BPT diagram (bottom panel in Fig. \ref{fig:BPT}). The \nii ~line remains undetected in the extended H$\alpha$ region for all the galaxies and therefore, we estimate an upper limit for the \nii/H$\alpha$ ratio. In the case of cid\_346 and S82X1905, we also do not detect H$\beta$ emission in the extended regions and therefore, a lower limit to the \oiii/H$\beta$ line ratio is estimated. All four galaxies are above the region of the BPT where we expect to have star-formation as the main ionization source. In addition, for  cid\_346, the width of the extended H$\alpha$ emission is 628$\pm$45 km s$^{-1}$, which is above the limit usually used to distinguish between an outflowing and non-outflowing gas. This suggests that the observed H$\alpha$ emission is a part of the NLR outflow. This is also supported by the fact that the ionised outflow is also detected in the \oiii ~line at the same location as the extended H$\alpha$ \citep[see][]{kakkad20}. We also note here that the BPT results shown here for cid\_346 may be different compared to the ones presented in \citet{lamperti21}, as the extraction apertures for the PSF and the extended H$\alpha$ emission may be different. 

For all the other galaxies, as the \oiii ~and H$\beta$ lines remain undetected, they could not be placed in the classical BPT diagram. Therefore, we estimate the ionisation source based on the limit derived for the \nii/H$\alpha$ line. Four targets, X\_N\_160\_22, X\_N\_81\_44, X\_N\_66\_23 and cid\_467, have log(\nii/H$\alpha$) values $<$-0.7 and with such low \nii/H$\alpha$ line ratios, it is less likely for these sources to be ionised by the AGN  \citep[see the results from low redshift galaxy sample in ][]{kauffmann03}. For one of them, X\_N\_81\_44, we also have evidence that the \oiii ~and H$\alpha$ emission are extended, but are not spatially co-located (Fig. \ref{fig:BPT2}, left panel). We would therefore classify the extended emission in these four galaxies as most probably ionized by star-formation.

The galaxies where the log(\nii/H$\alpha$) line ratio $>$-0.7, the source of ionisation could not be constrained for the extended regions with the current data. The target, J1549+1245 falls under this category, where the extended \oiii ~and H$\alpha$ emission are not spatially co-located (Fig. \ref{fig:BPT2}, right panel). Therefore, the ionisation source of these targets are labelled as "unconstrained" in Table \ref{table:halpha_results}. An exception is S82X1940 shows an extended H$\alpha$ emission towards the SE direction from the AGN location (Fig. \ref{fig:BPT_PSF2}). This target displays the largest H$\alpha$ line width in the extended region of $\sim$784$\pm$98 km s$^{-1}$. The \oiii ~emission line analysis of this target already shows the presence of an ionised outflow \citep{kakkad20} and therefore, the H$\alpha$ emission most likely also traces the outflowing gas driven by the AGN. Based on this evidence,  we conclude that the observed extended H$\alpha$ emission in S82X1940 is part of the NLR ionised by the AGN. In all the cases presented above, we do not rule out the possibility of a companion galaxy in a merger with the AGN host galaxy.

In summary, out of the 16 type 1 AGN in the SUPER survey that show extended H$\alpha$ emission, four galaxies ($\sim$25\%, $\sim$18\% if we include all type 1 AGN, including unresolved galaxies) show that the extended H$\alpha$ emission is most likely ionised by star formation and in five galaxies ($\sim$30\%, $\sim$23\% if all the type 1 AGN are included) the ionisation is dominated by the AGN. In the remaining 7 galaxies, the ionisation source of the extended H$\alpha$ emission remains unconstrained. In two galaxies (cid\_346 and S82X1940), the width of the H$\alpha$ line $>$600 km s$^{-1}$, suggesting that the H$\alpha$ emission is tracing ionised outflow driven by the AGN. However, we cannot exclude the possibility that the turbulence in the ISM is related to an on-going merger event, for which higher resolution and deeper data is required. We also note that the width alone should not be used as an indicator for star forming or AGN origin to the emission, as the spectra extracted from extended emission line regions are highly limited by noise.

From the analysis presented in this section, we conclude that H$\alpha$ emission in high redshift galaxies does not necessarily trace star formation, but can also be associated with AGN ionisation or outflows or possible companions. 

\begin{figure*}
\centering
\includegraphics[scale=0.45]{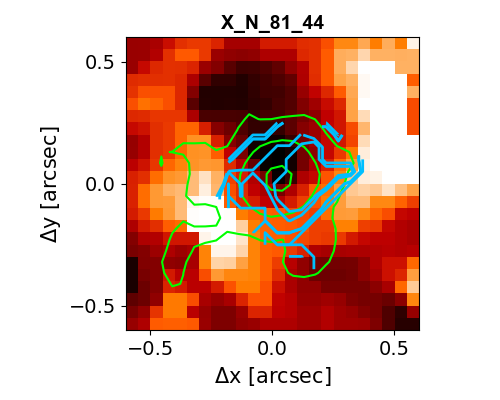}
\includegraphics[scale=0.45]{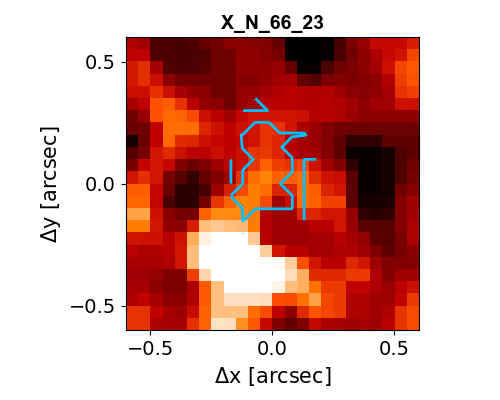}\\
\caption{The figure shows the spatial locations of the H$\alpha$ emission and \oiii ~based ionised outflows in X\_N\_81\_44 (left panel) and X\_N\_66\_23 (right panel). The background image shows the PSF-subtracted H$\alpha$ images from the maps in Fig. \ref{fig:BPT_PSF}. The blue contours trace the \oiii ~outflows ($w_{80}>$600 km s$^{-1}$) at levels 600, 700 and 750 km s$^{-1}$ in X\_N\_81\_44; and 850 and 1050 km s$^{-1}$ in X\_N\_66\_23. The extended H$\alpha$ emission in these galaxies is absent in the direction with highest $w_{80}$ values. The green contours in X\_N\_81\_44 show the archival ALMA band 7 dust continuum emission (from \citet{lamperti21}), that traces dust reheated star formation. The implications of these observations are furth er discussed in Sect. \ref{sect3.3}.}
\label{fig:HA_OIII_w80}
\end{figure*}

\begin{figure}
\centering
\includegraphics[scale=0.38]{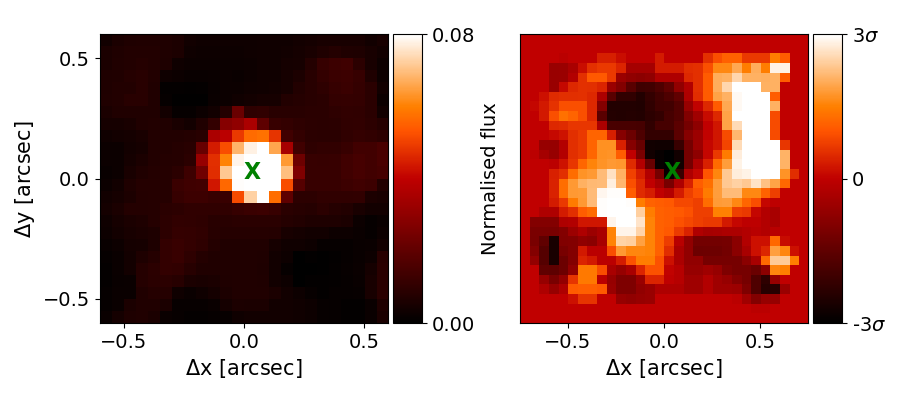}
\caption{The left panel shows the narrow H$\alpha$ flux map of X\_N\_81\_44 obtained from the pixel-by-pixel Gaussian fit of the H$\alpha$ complex. The right panel, on the other hand, shows the PSF-subtracted H$\alpha$ channel map of the same target. The figure highlights that the PSF dominates the bulk of the emission close to the AGN location (marked by the black star).} 
\label{fig:HA_narrow_PSF_comparison}
\end{figure}

\subsection{Impact of ionised outflows on unobscured star formation} \label{sect3.4}

To quantify the impact that ionised outflows may have on star formation, we compare the spatial locations of ionised outflows, traced using the \oiii ~emission presented in \citet{kakkad20}, with the extended H$\alpha$ emission presented in this paper. We perform this analysis on targets where both the \oiii ~and the H$\alpha$ emission is extended. Furthermore, we limit this analysis to those targets where the extended H$\alpha$ emission is consistent with ionisation by star formation, in order to use it as a tracer of short time-scale ($\sim$10 Myr) unobscured star formation. Based on the analysis presented in Sects. \ref{sect3.2} and \ref{sect3.3}, two targets fulfil these criteria: X\_N\_81\_44 and X\_N\_66\_23. Figure \ref{fig:HA_OIII_w80} shows the extended H$\alpha$ map tracing star formation in these two targets, overlaid by \oiii-based outflow velocity contours in blue. Only the contours with \oiii ~$w_{80}$ values above 600 km s$^{-1}$ are shown in Fig. \ref{fig:HA_OIII_w80}. The astrometry of the H-band and K-band images was registered using the AGN continuum peak emission in the respective data cubes. 

We observe a spatial anti-correlation between the locations of high velocity \oiii ~outflows ($w_{80}>$600 km s$^{-1}$) and unobscured star formation in X\_N\_81\_44 and X\_N\_66\_23. These results might suggest that star formation is being actively shut down in regions with high velocity outflows, a result also previously reported in the literature \citep[e.g.,][]{cano-diaz12, carniani16}. However, we cannot rule out a scenario where the high velocity winds compress gas ahead of it, resulting in triggering of star formation in the edges of the outflow \citep[e.g.,][]{cresci15}. We note here that the spatial anti-correlation between high velocity outflows and H$\alpha$ emission is observed only in the PSF-subtracted H$\alpha$ images and not in the narrow H$\alpha$ maps obtained using the pixel-by-pixel fit, which will be described in Sect. \ref{sect4}. The left panel in Fig. \ref{fig:HA_narrow_PSF_comparison} shows the H$\alpha$ flux map of X\_N\_81\_44, obtained from the narrow Gaussian component, which shows that most of the emission is concentrated close to the AGN location. However, the emission close to the AGN is dominated by the PSF smearing effect. The right panel, on the other hand, shows the PSF-subtracted H$\alpha$ channel map, which removes any emission that might be affected by beam smearing, as described earlier in Sect. \ref{sect3.2}. The results in Fig. \ref{fig:HA_narrow_PSF_comparison} may explain some of the observed differences, such as the presence or absence of such spatial anti-correlations found in other high-z AGN host galaxies \citep[e.g.,][]{carniani16, scholtz21}. 

We note that the spatial resolution of the SINFONI observations prevents us from investigating the morphology of the H$\alpha$ emission within the 2 kpc PSF element which was subtracted in the maps shown in Fig. \ref{fig:HA_OIII_w80}. In other words, the SINFONI data is unable to trace star formation on scales smaller than 2 kpc at these redshifts and therefore, we cannot exclude the presence of on-going star formation in the central cavity in the maps in Fig. \ref{fig:HA_OIII_w80}. Observations with upcoming facilities such as ELT/HARMONI will provide the necessary spatial resolution to resolve regions in sub-kiloparsec scales in these galaxies, where there could possibly be underlying emission from the host galaxy.

In X\_N\_81\_44, we also compare the outflow locations with archival high resolution ($\sim$0.2\arcsec) ALMA Band 7 (870 $\mu$m observed-frame, $\sim$260 $\mu$m rest-frame) dust continuum maps from \citet{lamperti21}. The dust continuum emission is overlaid as red contours in Fig. \ref{fig:HA_OIII_w80}. The Spectral Energy Distribution (SED) analysis of X\_N\_81\_44 suggests that $>$99\% of the rest-frame 260$\mu$m emission is from dust heated by star formation and $<$1\% contribution from the AGN-heated dust and synchrotron emission \citep[see][]{lamperti21}. We can, therefore, use the dust continuum maps to trace longer time scale star formation from dust reheated by the UV radiation from stars over the last $\sim$100 Myrs.

High redshift galaxies are known to host copious amount of dust, which can have a significant impact on how we interpret the results on the spatial distribution of outflow versus star formation \citep[e.g.,][]{fujimoto18, lamperti21, scholtz21}. Star formation tracers using rest-frame optical emission lines, such as H$\alpha$, are susceptible to dust obscuration and previous work has shown that the dust distribution (obtained from far-infrared ALMA observations, for instance) and H$\alpha$ distributions (or rest-frame optical continuum distribution in case of narrow band HST observations) are different in high redshift galaxies \citep{hodge16, lang19, chen20}. In general, H$\alpha$ distributions tend to be more extended compared to the dust emission. There is a large concentration of star formation dominated infrared emission at the centre of galaxies that one needs to account for \citep[e.g.,][]{hao11}. This is the case also for the ALMA dust continuum emission (red contours) for X\_N\_81\_44 (left panel in Fig. \ref{fig:HA_OIII_w80}) that peaks in the central region and extends towards the H$\alpha$ emission in the SE direction from the AGN. Therefore, based on the ALMA map, there is no clear anti-correlation between the location of the ionised outflow and regions with active star formation. However, we note that the dust obscured star formation traced by sub-mm or far-infrared observations represents star formation over long timescales of $\sim$100 Myrs. On the contrary, the outflow timescale is much shorter \citep[10$^5$ to a few Myrs, e.g.,][]{schawinski15}. Therefore, we cannot exclude that the possible impact of the outflow on star formation may be diluted by the large time range to which the star formation is sensitive to. Furthermore, the ALMA dust-continuum observations \citep{lamperti21} were not able to resolve spatial scales below $\sim$2 kpc and therefore, we cannot exclude the presence of cavities below this physical scale. This limitation could be mitigated with spatially resolved mid-infrared observations with JWST/MIRI. 

\begin{figure*}
\centering
\includegraphics[scale=0.3]{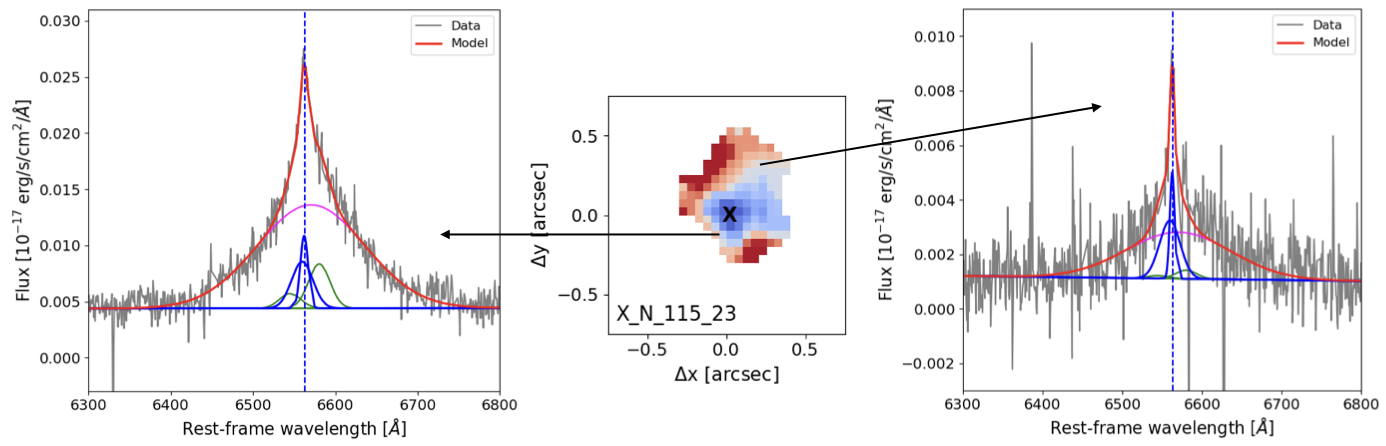}
\caption{An example of line modelling in individual pixels in X\_N\_115\_23. The middle panel shows the H$\alpha$ centroid map and the left and right panel show the extracted spectrum and the emission line model in two pixels located by the arrows.} 
\label{fig:pixel_fit_example}
\end{figure*}

\begin{figure*}
\centering
\includegraphics[scale=0.38]{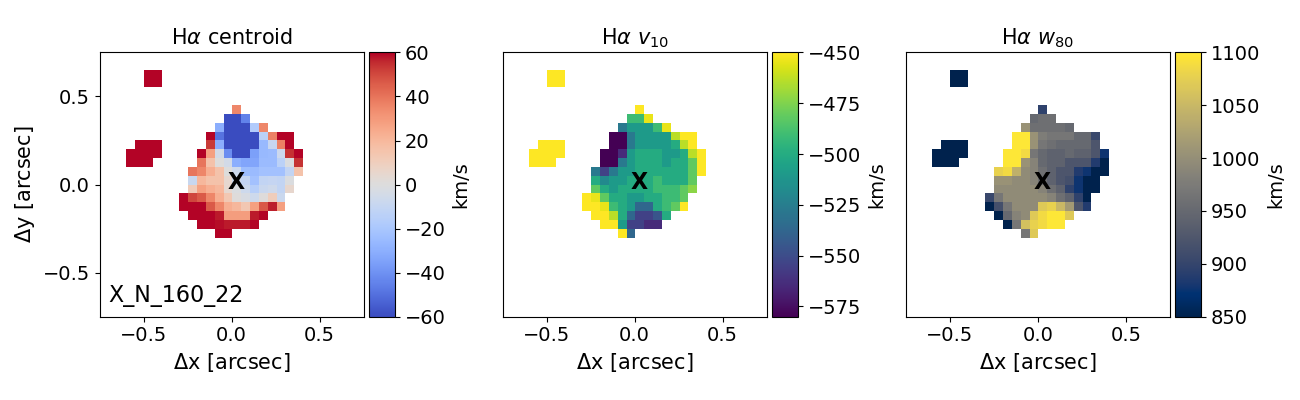}
\includegraphics[scale=0.38]{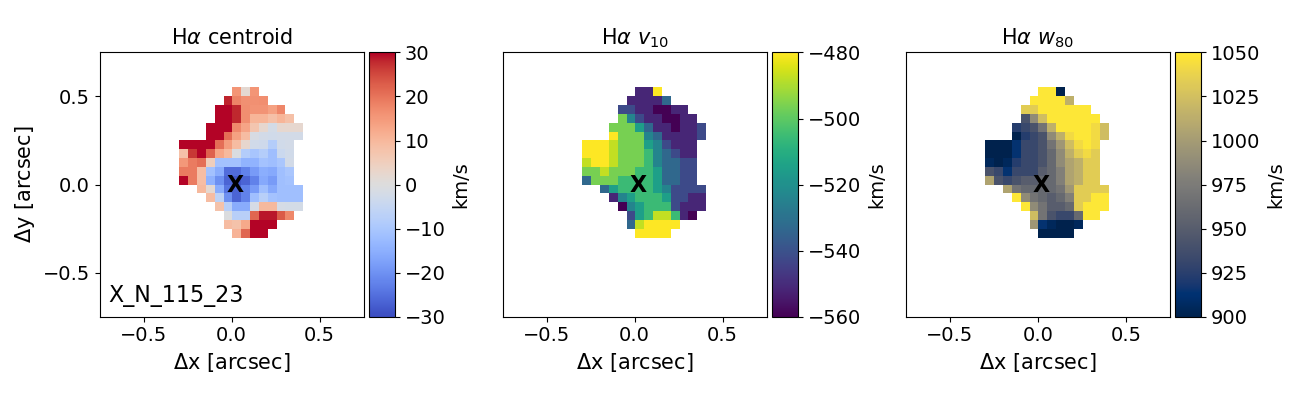}
\includegraphics[scale=0.38]{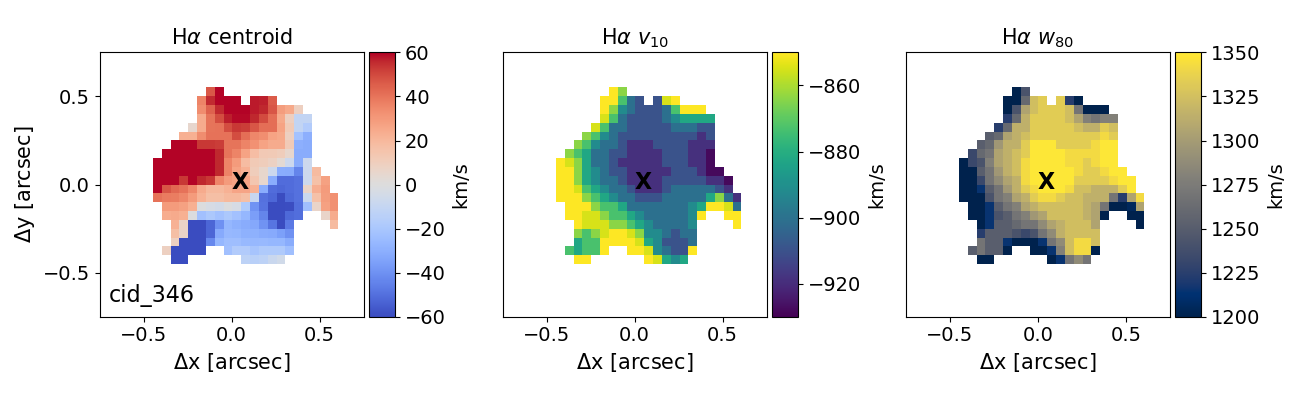}
\includegraphics[scale=0.38]{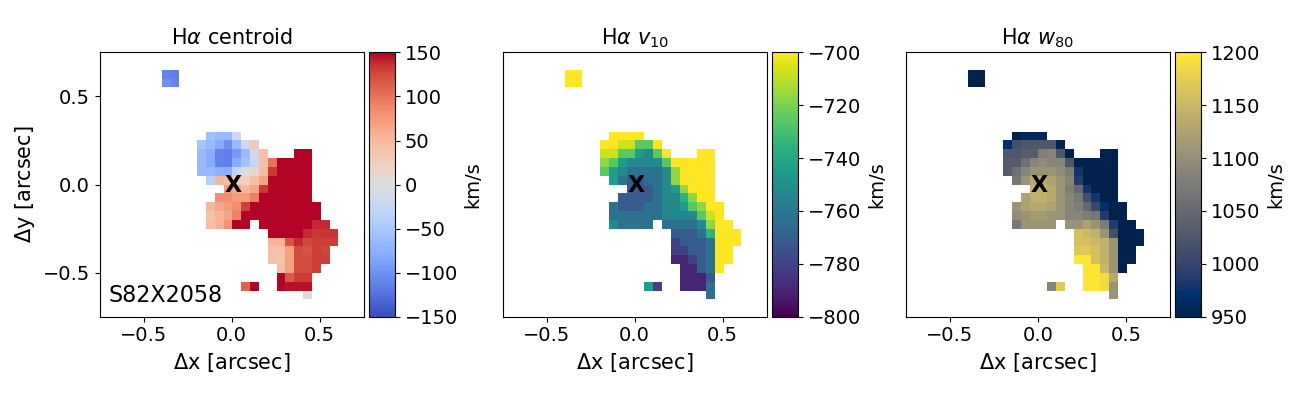}
\caption{H$\alpha$ NLR component velocity maps of (from top to bottom) X\_N\_160\_22, X\_N\_115\_23, cid\_346 and S82X2058 as examples. The plots for the rest of the targets are moved to the appendix \ref{sect:appendix} (Fig. \ref{fig:velocity_maps_2}). The left panels show the H$\alpha$ centroid map, the middle panels the $v_{10}$ map and the right panels the $w_{80}$ maps. The targets show a diverse set of H$\alpha$ velocity distributions. X\_N\_160\_22, cid\_346 and S82X2058 show smooth velocity gradients in their H$\alpha$ centroid profiles. S82X2058 shows an extended structure towards the SW of the nucleus, suggesting a possible faint companion. The H$\alpha$ centroid profile of X\_N\_115\_23 does not show a smooth rotation-like velocity gradient as in other galaxies, which is consistent with the finding that the extended H$\alpha$ emission is a part of the NLR (Fig. \ref{fig:BPT_PSF}). These maps are discussed further in Sect. \ref{sect4}.} 
\label{fig:velocity_maps}
\end{figure*}

\begin{figure}
\centering
\includegraphics[scale=0.62]{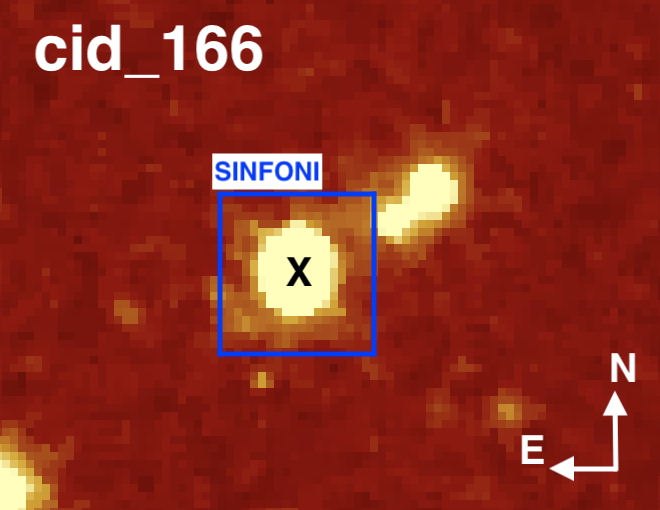}
\includegraphics[scale=0.3]{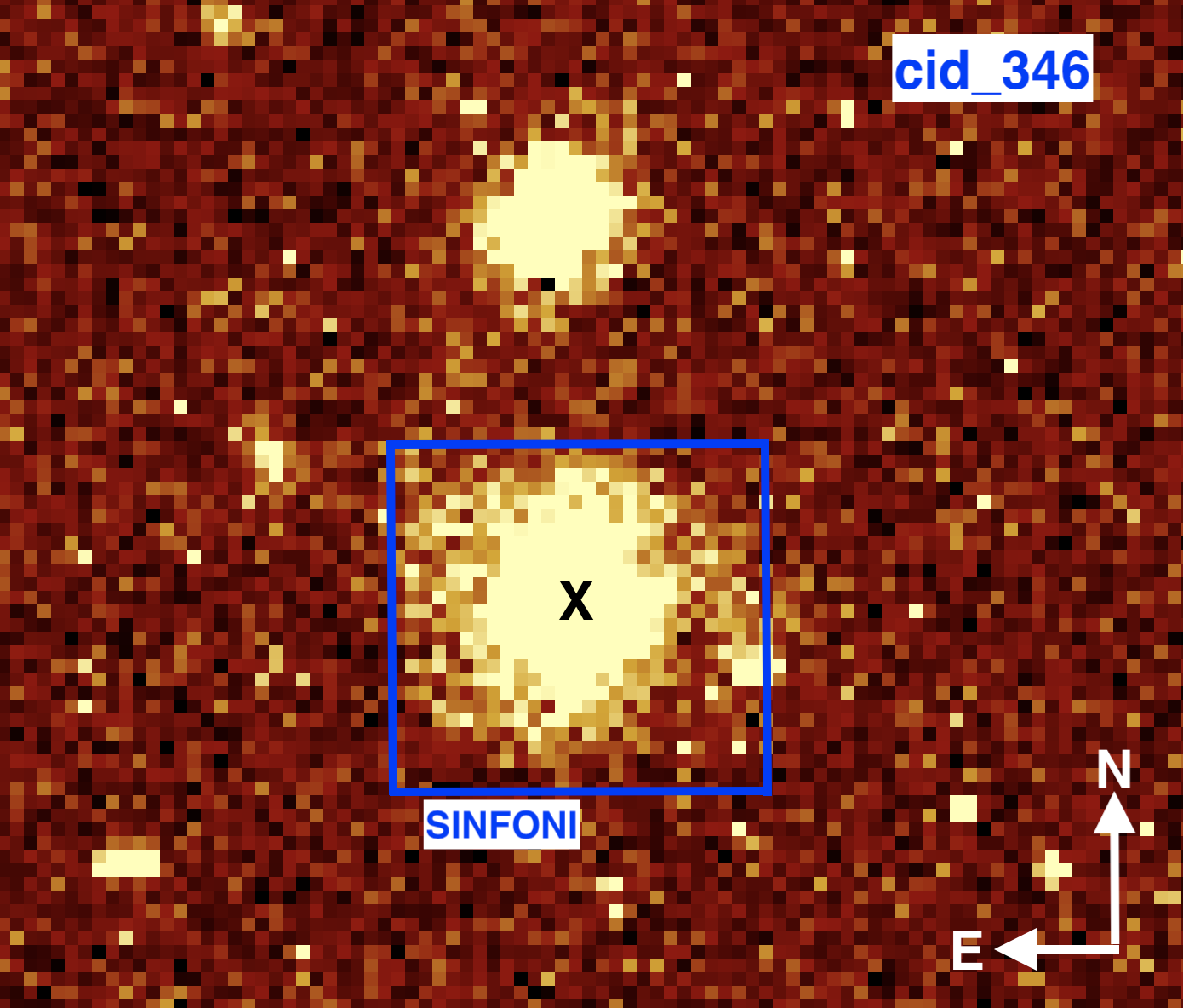}
\caption{Archival HST/WFC3 images of cid\_166 (top panel) and cid\_346 (bottom panel), both type 1 AGN in the SUPER survey. The HST images are obtained with the F160W filter (near-infrared) and both the images suggest the presence of a companion. The cid\_166 image clearly shows a stream linking the two galaxies. The blue square in each of the panels show the SINFONI 3\arcsec$\times$3\arcsec ~FoV. Therefore, current observations presented in this paper do not capture the presence of these companions.}
\label{fig:mergers}
\end{figure}

\section{Kinematic properties of narrow H\texorpdfstring{$\alpha$}{} emission} \label{sect4}

In this section, we describe the pixel-by-pixel H$\alpha$ emission line modelling for the targets that show extended H$\alpha$ emission after subtracting the PSF, as described earlier in Sect. \ref{sect3.2}. We primarily derive the kinematic maps from the BLR component-subtracted emission line models, namely the  H$\alpha$ centroid map (also called moment 1 map) and non-parametric velocity and velocity dispersion maps i.e. $v_{10}$ and $w_{80}$ maps \citep[see e.g.,][]{harrison14, cresci15}. We compare these H$\alpha$ maps with that of the \oiii ~emission in the H-band data to determine whether the H$\alpha$ emission mimics the kinematics of the \oiii ~emitting gas. This comparison can be used to further distinguish between a star formation or NLR origin to the H$\alpha$ emission.

As described earlier in Sect. \ref{sect3.1}, the H$\alpha$ complex is composed of several Gaussian functions, each representing the individual components of the H$\alpha$ and \nii$\lambda\lambda$6549, 6585 emission lines. To avoid degeneracy during the pixel-by-pixel fit, we employ the following constraints on the individual Gaussian components to model the spatial distribution of the narrow component of the H$\alpha$ emission, conventionally treated as tracing the host galaxy star formation. All the constraints described here are fixed or varied relative to the integrated spectrum fitting results using the M3 model, as described in Sect. \ref{sect3.1}. As the BLR emission is unresolved, we fix the centroid and the width of the BLR components and only allow a variation in its peak value. In all but two targets (cid\_346 and S82X1940), the spectrum extracted from the extended regions in the PSF-subtracted cube shows a H$\alpha$ width (FWHM) of $<$600 km s$^{-1}$. In these targets where the H$\alpha$ FWHM$<$600 km s$^{-1}$, we fix the centroid and width of the broad NLR component and only allow a variation in its peak (similar to the constraint imposed on the BLR component). The centroid, width and the peak of the narrow NLR component of the H$\alpha$ line is allowed to vary across the field of view. In the case of cid\_346 and S82X1940, we also allow small variations in the broad NLR component. 

We attempt the pixel-by-pixel analysis of the H$\alpha$ line only for the 15 targets that show extended H$\alpha$ emission from the analysis presented in Sect. \ref{sect3.2}. Out of these 15 targets, 4 targets (X\_N\_53\_3, X\_N\_66\_23, cid\_166 and cid\_1605) did not have sufficient S/N per pixel to constrain the parameters of individual Gaussian components. We note that deriving the PSF-subtracted maps was possible despite the low S/N in these 4 targets as we relied on the channel maps and not on Gaussian fits to verify the presence or absence of residuals. The H$\alpha$ complex in the case of J1333+1649 is contaminated by telluric features in the wings of the profile and in the case of J1549+1245, the individual Gaussian components remained unconstrained, which made the pixel-by-pixel analysis for these two targets unreliable. Fig. \ref{fig:pixel_fit_example} shows an example of the line fitting in individual pixels for X\_N\_115\_23 as an example.

In figure \ref{fig:velocity_maps}, we show the BLR-subtracted H$\alpha$ kinematic maps of X\_N\_160\_22, X\_N\_115\_23, cid\_346 and S82X2058 as examples, while similar plots for other targets are moved to the Appendix \ref{sect:appendix}. The left panels in Fig. \ref{fig:velocity_maps} shows the H$\alpha$ centroid map, the middle panel shows the 10th percentile velocity, $v_{10}$ and the right panel shows the H$\alpha$ width containing 80\% of the line flux. The kinematic centre for the centroid map is obtained from the location of the narrow H$\alpha$ emission in the integrated spectrum (Sect. \ref{sect3.1}). Folding the information derived in Sect. \ref{sect3.3}, we now discuss the possible origin of the observed kinematic properties of the H$\alpha$ line. We attempt to distinguish between rotating disk signatures, mergers and outflowing components. For a rotationally supported disk, we expect a smooth gradient in the line centroid maps and a centrally peaked velocity dispersion (in this case, $w_{80}$). On the other hand, galaxies undergoing mergers or outflows will not necessarily show a centrally peaked velocity dispersion map and/or an ordered velocity gradient in the centroid map. We discuss a few examples of targets shown in Fig. \ref{fig:velocity_maps}.

X\_N\_160\_22 does not show extended \oiii ~emission, but shows extended H$\alpha$ emission towards the north of the AGN location, evident in the PSF-subtracted H$\alpha$ channel map in Fig. \ref{fig:BPT_PSF} and the centroid map in the left panel of Fig. \ref{fig:velocity_maps}. The centroid map shows a smooth velocity gradient between $\pm$60 km s$^{-1}$. The $v_{10}$ and $w_{80}$ maps show maximum velocities of $<$-600 km s$^{-1}$ and $>$1000 km s$^{-1}$, suggesting the presence of AGN outflows, especially close to the AGN location. Considering that the extended H$\alpha$ emission is consistent with star formation, the H$\alpha$ velocity maps of X\_N\_160\_22 in Fig. \ref{fig:velocity_maps} could be interpreted as a superposition of a rotating medium (centroid map that shows a smooth gradient) and a turbulent medium ($v_{10}$ and $w_{80}$ maps that show high velocities near the AGN location).

In the case of X\_N\_115\_23, the H$\alpha$ centroid value at the AGN location appears blue-shifted and the $w_{80}$ map does not show a centrally peaked profile, but shows a smooth gradient. The velocities in the $w_{80}$ map reaches values between $\sim$850 -- $>$1000 km s$^{-1}$ suggesting that H$\alpha$ emission is tracing gas which is part of an outflow in the NLR. This is also  supported by the fact that the H$\alpha$ and \oiii ~emission are spatially extended in the same direction and that the extended emission is ionised by the AGN (see BPT map in Fig. \ref{fig:BPT}). 

cid\_346 shows a smooth velocity gradient in the centroid map with the major axis orientated roughly along the NE direction and the $w_{80}$ map peaks at $\sim$1300 km s$^{-1}$ at the AGN location. The large dispersion value cannot be explained purely as ordered rotation and therefore, we infer that the observed H$\alpha$ emission is a combination of rotation from the host galaxy and AGN outflows. Although the extended H$\alpha$ emission in this galaxy falls under the AGN ionised region in Fig. \ref{fig:BPT}, the presence of upper limits in the \nii/H$\alpha$ ratio could suggest that star formation may play a role in the ionisation and kinematics of the observed extended emission. However, the current data does not allow us to definitely confirm this. 

Lastly, S82X2058 shows a smooth velocity gradient in its H$\alpha$ centroid map, but the centroid velocity profile is extended more towards the SE direction from the AGN location. Accounting for the localised peak in the $w_{80}$ map at the AGN location, this target shows a rotating system probably in an interaction with another galaxy. 

To summarize the results on the spatially resolved kinematic analysis of the H$\alpha$ line: Six galaxies (X\_N\_160\_22, X\_N\_81\_44, X\_N\_4\_48, cid\_346, S82X1940 and S82X2058) show smooth velocity gradients in their H$\alpha$ centroid maps. The velocity dispersion maps in these galaxies ($w_{80}$) show localized peaks\footnote{The $w_{80}$ maps are sensitive to the line models derived from the multiple Gaussian fits. Also the IFU data is noise limited, especially in the outskirts of the galaxies and therefore the spatial profiles in these velocity might not always look smooth, but can suggest underlying patterns.}. This suggests that in these six galaxies, part or most of the H$\alpha$ emission traces the rotating gas. On the contrary, X\_N\_12\_26, X\_N\_115\_23 and S82X1905 do  not have smooth gradients in their centroid maps. The extended H$\alpha$ emission in X\_N\_115\_23 and S82X1905 is consistent with ionisation by the AGN (Sect. \ref{sect3.3}) and therefore the absence of smooth gradient in the H$\alpha$ centroid profile confirms that the H$\alpha$ emission is dominated by AGN emission. The extended H$\alpha$ emission in X\_N\_12\_26 is highly asymmetric around the AGN location, which may suggests the presence of companion galaxies. This is not a unique case: from the H$\alpha$ kinematic maps, there are indications of companions also in the case of X\_N\_4\_48, S82X1940 and S82X2058. 

Several galaxies presented in this paper show extended H$\alpha$ emission blue-shifted or redshifted by several hundred km s$^{-1}$ compared to the systemic values. For instance, J1549+1245 shows an extended H$\alpha$ component at $\sim$10 kpc from the AGN location, but this component is red-shifted by $\sim$1800 km s$^{-1}$ on extracting the spectra (Figure \ref{fig:BPT_PSF2}). Although the presence of large scale outflows cannot be ruled out in this case, the presence of such extended components might also suggest the presence of companions or the AGN host galaxies in the process of a merger. Observations of some of the type 1 SUPER AGN in other wavelengths already suggest the presence of companions. For instance, recent ALMA CO(4-3) observations of J1549+1245 confirmed the presence of a companion at $>3\sigma$ significance \citep[see][]{bischetti21} and therefore, the observed extended H$\alpha$ could be a part of a tidal tail. Furthermore, archival HST observations of cid\_166 and cid\_346 already indicate possible presence of galaxies nearby (Fig. \ref{fig:mergers}), which suggests that galaxy interactions may be a common phenomenon at these redshifts. We note here that cid\_346 is surrounded by an extended CO halo, which could also point to an over-density of galaxies around this source \citep[see][]{cicone21}. Although AGN hosted in interacting systems have been reported to be rare at 1$<$z$<$2 \citep[e.g.,][]{cisternas11, mainieri11}, this may also be due to observational biases or the limitation to obtain better spatial resolution and high sensitivity at high redshift. In fact, recent ALMA observations have revealed evidence of quasars residing in interacting galaxies \citep[e.g.,][]{trakhtenbrot17, banerji21}. 

\section{Summary and conclusions} \label{sect5}

We presented the spatially resolved properties of the H$\alpha$ emission from the K-band ($\sim$2--2.4$\mu$m) VLT/SINFONI observations of 21 type 1 AGN at z$\sim$2.2 derived from the SUPER survey. The adaptive optics assisted observations provided a spatial resolution of $\sim$2 kpc that allowed us to infer the source of ionisation in the extended H$\alpha$ regions, after accounting for beam-smearing effects due to the PSF. We used resolved BPT analysis, \nii/H$\alpha$ ratios of the extended emission and a comparison between the H$\alpha$ kinematics with that of the \oiii ~emission, obtained from the H-band observations presented in \citet{kakkad20}, to infer if the H$\alpha$ emission traces star formation. Lastly, we investigate if star-forming regions and high velocity ionised outflows are spatially anti-correlated. The main results of this paper are summarised below:

\begin{itemize}
\item We tested three methods of H$\alpha$ emission line modelling in this paper because the blending of \nii$\lambda\lambda$6549, 6585 and H$\alpha$ emission line components in the integrated spectra of type 1 AGN host galaxies at z$\sim$2 can lead to degenerate results. We conclude that in order to obtain unique solutions, the kinematic components of the H$\alpha$ emission need to be coupled with the \oiii ~emission. 
\item The vast majority of the quasars show extended H$\alpha$ emission ($\sim$76\%). The extended emission is observed out to $\sim$3--9 kpc with a mean value of $\sim$5 kpc. The width (FWHM) of the H$\alpha$ spectrum, extracted from the extended regions, is in the range 73--784 km s$^{-1}$ with a mean width of 312 km s$^{-1}$.
\item We find a correlation between the velocity shift, defined by the difference between the extended H$\alpha$ location in the spectra with the expected location based on its redshift, with the AGN luminosity and black hole mass. This could indicate that the extended H$\alpha$ emission is dominated by outflowing gas. The size of the extended H$\alpha$ emission, on the other hand, shows possible correlations with the stellar mass of the host galaxy, consistent with size-mass relations reported in the literature.
\item In four galaxies, we were able to constrain the flux ratios \oiii/H$\beta$ and \nii/H$\alpha$ at the location of extended H$\alpha$ emission and place them on the classical BPT maps. The H$\alpha$ emission in these four galaxies is consistent with ionisation by star formation. For the rest of the targets, we use \nii/H$\alpha$ ratio and the H$\alpha$ FWHM to identify the source of ionisation. Overall, the extended H$\alpha$ emission in 4 galaxies is consistent with ionisation by star formation ($\sim$25\% of the targets that show extended emission), in 5 galaxies by the AGN ($\sim$30\%) and in 7 galaxies, the ionisation source could not be constrained ($\sim$45\%). 
\item We find a variety of dynamical properties in the host galaxies as traced by the H$\alpha$ emission. Six out of nine galaxies for which pixel-by-pixel emission line fitting was possible show smooth velocity gradients in the H$\alpha$ centroid maps and their BPT diagrams are also consistent with ionisation by star formation, suggesting that the H$\alpha$ emission traces the host rotation and not the NLR. In two galaxies, the H$\alpha$ morphology and kinematic maps are similar to that of \oiii ~and along with the fact that their BPT location is consistent with AGN ionisation, the H$\alpha$ emission in these two galaxies traces the NLR. In four galaxies, the morphological and kinematic characteristics of the H$\alpha$ emission may indicate the presence of companions or AGN hosts being in mergers.
\item In two galaxies, we find evidence of negative AGN feedback on scales $>$2 kpc as the extended H$\alpha$ emission in these galaxies avoid regions with high velocity \oiii-based ionised outflows. For the rest of the targets, there is no strong evidence of AGN outflows having an impact on the host galaxy star formation using the current data. ALMA Band 7 observations of a fraction of galaxies show the prevalence of dust heated star formation at the centre of the galaxy close to the AGN and the ionised outflow. The current SINFONI observations are not sensitive to resolutions $<$2 kpc, therefore future high spatial resolution observations with ELT-class IFS instruments such as HARMONI will be required to constrain the real impact of the outflows by sampling recent star formation in the proximity of the AGN location.
\end{itemize}

The current SINFONI K-band observations of most of the galaxies in the SUPER survey have an on-source exposure time of $\sim$1--2 hours. Deeper observations are, therefore, required with current or future facilities such as rest-frame HST/FUV imaging, VLT/ERIS, VLT/HAWKI-GRAAL and ELT/HARMONI to distinguish between the presence of outflows/inflows and possible presence of companions. Finally, upcoming JWST observations in Cycle-1 (ID 2177) will use the mid-infrared IFS capabilities of JWST/MIRI \citep[e.g.,][]{rieke15}, that will allow us to detect dust-obscured star formation using the PAH 6.2 $\mu$m emission, at a spatial resolution similar to those of SINFONI observations presented in this paper.

\section*{Acknowledgements}
The authors would like to thank the anonymous referee for comments that improved the paper. CMH acknowledges funding from an United Kingdom Research and Innovation grant (code: MR/V022830/1). A.P. gratefully acknowledges financial support from STFC through grants ST/T000244/1 and ST/P000541/1. Based on observations collected at the European organisation for Astronomical Research in the Southern Hemisphere under ESO programme 196.A-0377. For the purpose of open access, the authors have applied a Creative Commons Attribution (CC-BY) license to any author accepted version arising.

\section*{Data Availability}
The IFS data products will be made publicly available via ESO Phase 3 data release. 



\bibliographystyle{mnras}
\bibliography{reference} 



\appendix

\section{Line fitting results using different methods} \label{sect:appendix}

\begin{table*}
\centering
\caption{H$\alpha$ line fitting parameters for the M2 model. The line fitting for J1441+0454 remained unconstrained and hence, not reported in this table.}
\label{table:intspec_results_kakkad_model}
\begin{tabular}{ccccccccc}
\hline
Target & $\lambda_{\rm range}$ & Aperture & \multicolumn{3}{c}{FWHM} & \multicolumn{3}{c}{$L_{\rm H\alpha}$}\\
& & & $v_{1}$ & $v_{2}$ & $v_{\rm BLR}$ & $L_{1}$ & $L_{2}$ & $L_{\rm BLR}$ \\
 & \AA & arcsec & km/s & km/s & km/s & erg/s & erg/s & erg/s\\
\hline
X\_N\_160\_22 & 6200--6900 & 0.9 & 287$\pm$53 & 2211$\pm$278 & 6807$\pm$260 & 42.68$\pm$0.13 & 43.84$\pm$2.90 & 44.28$\pm$0.01\\
X\_N\_81\_44 & 6200--6900 & 0.9 & 990$\pm$82 & 2929$\pm$95 & 8250$\pm$313 & 43.27$\pm$0.06 & 44.02$\pm$0.14 & 44.16$\pm$0.02\\
X\_N\_53\_3 & 6200--6900 & 0.8 & 345$\pm$213 & -- & 4577$\pm$187 & 42.04$\pm$0.22 & -- & 43.70$\pm$0.01\\
X\_N\_66\_23 & 6200--6900 & 0.9 & -- & -- & 5640$\pm$182 & -- & -- & 43.72$\pm$0.01\\
X\_N\_35\_20 & 6200--6900 & 0.5 & 532$\pm$63 & -- & 6549$\pm$972 & 41.97$\pm$0.06 & -- & 42.68$\pm$0.05\\
X\_N\_12\_26 & 6200-6900 & 1.0 & 544$\pm$32 & -- & 4615$\pm$100 & 42.80$\pm$0.04 & -- & 43.95$\pm$0.01\\
X\_N\_44\_64 & 6200--6900 & 0.5 & 411$\pm$17 & -- & 7688$\pm$560 & 42.37$\pm$0.02 & -- & 43.03$\pm$0.02\\
X\_N\_4\_48 & 6200--6900 & 0.7 & 473$\pm$53 & -- & 7596$\pm$303 & 42.79$\pm$0.06 & -- & 44.18$\pm$0.01\\
X\_N\_102\_35 & 6200--6900 & 0.3 & 535$\pm$151 & 3443$\pm$92 & 9070$\pm$588 & -- & 43.71$\pm$0.17 & 43.67$\pm$0.02\\
X\_N\_115\_23 & 6200--6900 & 0.9 & 376$\pm$52 & 1184$\pm$139 & 6862$\pm$162 & 42.85$\pm$0.11 & 43.25$\pm$0.07 & 44.16$\pm$0.01\\
cid\_166 & 6200--6900 & 0.9 & 458$\pm$49 & 1968$\pm$260 & 6943$\pm$95 & 42.66$\pm$0.16 & 43.77$\pm$0.13 & 44.60$\pm$0.01\\
cid\_1605 & 6200--6900 & 0.3 & 516$\pm$110 & -- & 3802$\pm$80 & 41.96$\pm$0.15 & -- & 43.27$\pm$0.01\\
cid\_346 & 6200--6900 & 0.9 & 301$\pm$43 & 2884$\pm$156 & 7298$\pm$592 & 42.60$\pm$0.08 & 43.60$\pm$0.16 & 43.89$\pm$0.05\\
cid\_1205 & 6200--6900 & 0.8 & 446$\pm$89 & -- & 5023$\pm$183 & 42.04$\pm$0.12 & -- & 43.46$\pm$0.01\\
cid\_467 & 6200--6900 & 0.3 & 575$\pm$61 & -- & 8750$\pm$285 & 42.37$\pm$0.05 & -- & 43.86$\pm$0.01\\
J1333+1649 & 6350--7000 & 0.9 & 1059$\pm$415 & 4246$\pm$320 & 8839$\pm$304 & 43.87$\pm$0.16 & 45.28$\pm$3.27 & 45.30$\pm$0.03\\
J1549+1245 & 6200--6900 & 1.1 & 1045$\pm$35 & 4557$\pm$118 & 10495$\pm$208 & 44.23$\pm$0.02 & 41.73$\pm$3.87 & 45.50$\pm$0.01\\
S82X1905 & 6200--6900 & 1.0 & 920$\pm$58 & -- & 5219$\pm$96 & 43.14$\pm$0.03 & -- & 44.11$\pm$0.01\\
S82X1940 & 6200--6900 & 0.8 & 426$\pm$117 & 3001$\pm$713 & 6430$\pm$1138 & 42.78$\pm$1.45 & 43.15$\pm$4.49 & 43.69$\pm$0.28\\
S82X2058 & 6200-6900 & 0.9 & 481$\pm$75 & 3243$\pm$192 & 8809$\pm$455 & 42.43$\pm$0.11 & 40.88$\pm$3.82 & 43.95$\pm$0.03\\
\hline
\end{tabular}
\end{table*}

\begin{table*}
\centering
\caption{H$\alpha$ line fitting parameters for the M1 model. The emission line modelling using the M1 method for J1441+0454 and J1549+1245 remained unconstrained. Only those targets are shown for which the line fitting results were different from the ones in Table \ref{table:intspec_results_kakkad_model} i.e. for those targets where only H$\alpha$ components were sufficient to reproduce the H$\alpha$ complex.}
\label{table:intspec_results_carniani_model}
\begin{tabular}{ccccccccc}
\hline
Target & $\lambda_{\rm range}$ & Aperture & \multicolumn{3}{c}{FWHM} & \multicolumn{3}{c}{$L_{\rm H\alpha}$}\\
& & & $v_{1}$ & $v_{2}$ & $v_{\rm BLR}$ & $L_{1}$ & $L_{2}$ & $L_{\rm BLR}$ \\
 & \AA & arcsec & km/s & km/s & km/s & erg/s & erg/s & erg/s\\
\hline
X\_N\_160\_22 & 6200--6900 & 0.9 & 296$\pm$44 & 2368$\pm$60 & 6742$\pm$188 & 42.68$\pm$0.07 & 43.95$\pm$0.02 & 44.28$\pm$0.01\\
X\_N\_81\_44 & 6200--6900 & 0.9 & 1034$\pm$65 & 3168$\pm$104 & 8336$\pm$321 & 43.31$\pm$0.05 & 44.17$\pm$0.02 & 44.15$\pm$0.02\\
X\_N\_53\_3 & 6200--6900 & 0.8 & 1527$\pm$623 & -- & 4718$\pm$266 & 42.71$\pm$0.52 & -- & 43.66$\pm$0.02\\
X\_N\_66\_23 & 6200--6900 & 0.9 & 150$\pm$150 & -- & 5638$\pm$167 & 41.44$\pm$0.33 & -- & 43.72$\pm$0.01\\
X\_N\_115\_23 & 6200--6900 & 0.9 & 527$\pm$27 & 3306$\pm$114 & 9339$\pm$344 & 43.10$\pm$0.02 & 43.82$\pm$0.03 & 44.10$\pm$0.02\\
cid\_1605 & 6200--6900 & 0.3 & 1200$\pm$409 & -- & 3887$\pm$217 & 42.26$\pm$0.35 & -- & 43.27$\pm$0.02\\
J1333+1649 & 6350--7000 & 0.9 & 2354$\pm$657 & 4444$\pm$909 & 9058$\pm$492 & 44.47$\pm$0.33 & 45.38$\pm$3.01 & 45.30$\pm$0.05\\
\hline
\end{tabular}
\end{table*}


\begin{figure*}
\centering
\includegraphics[scale=0.28]{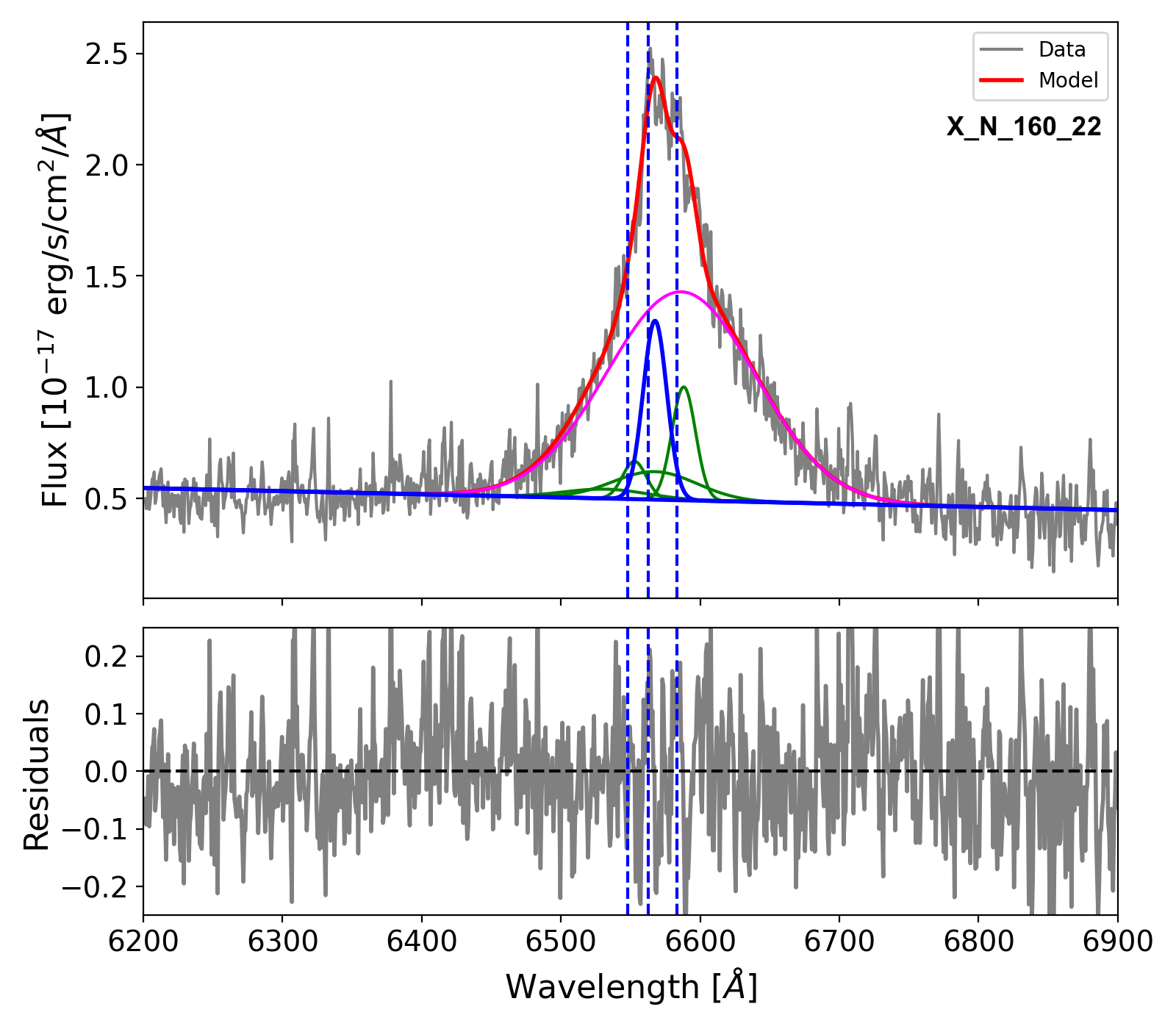}
\includegraphics[scale=0.28]{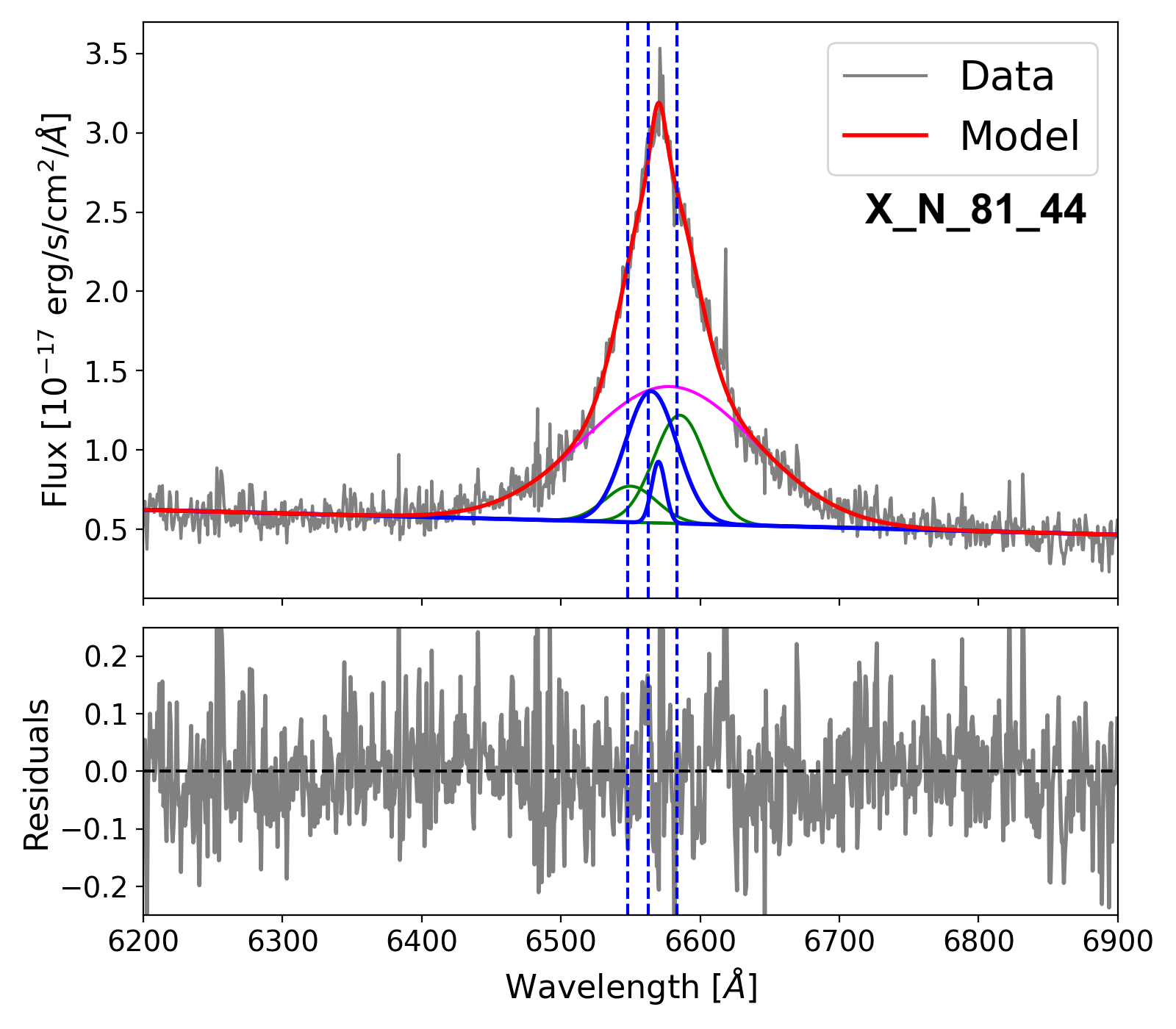}
\includegraphics[scale=0.28]{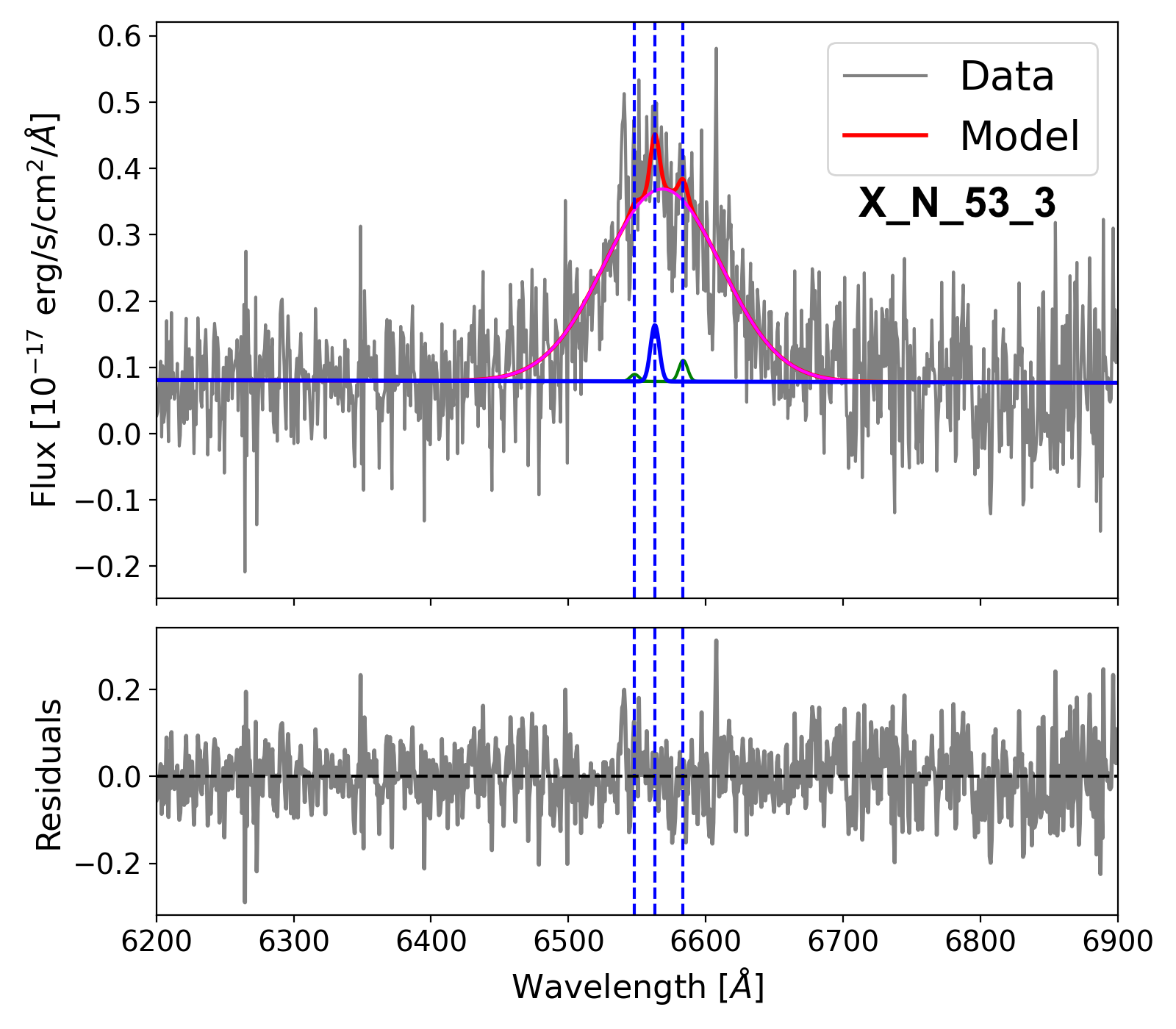}
\includegraphics[scale=0.28]{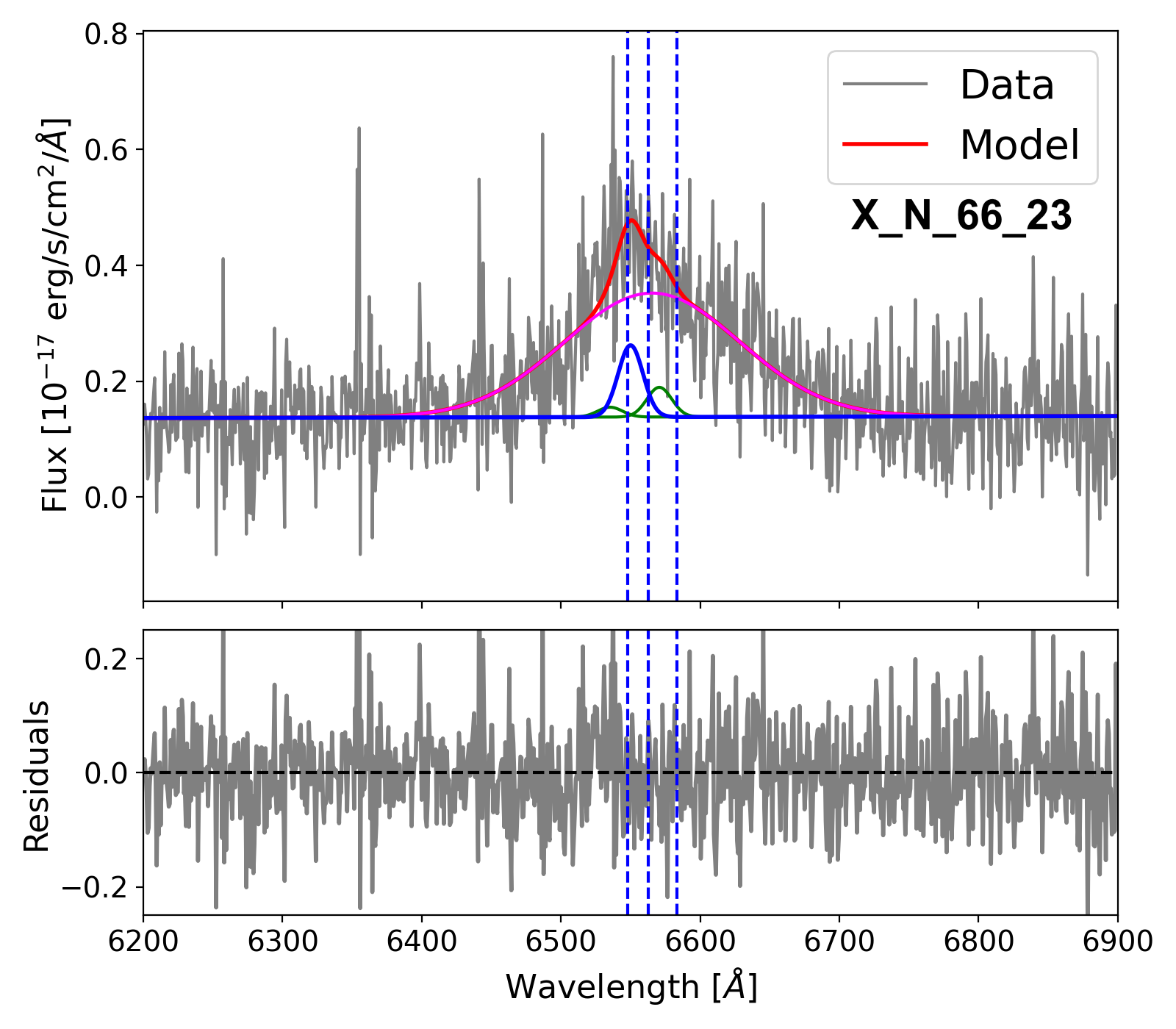}
\includegraphics[scale=0.28]{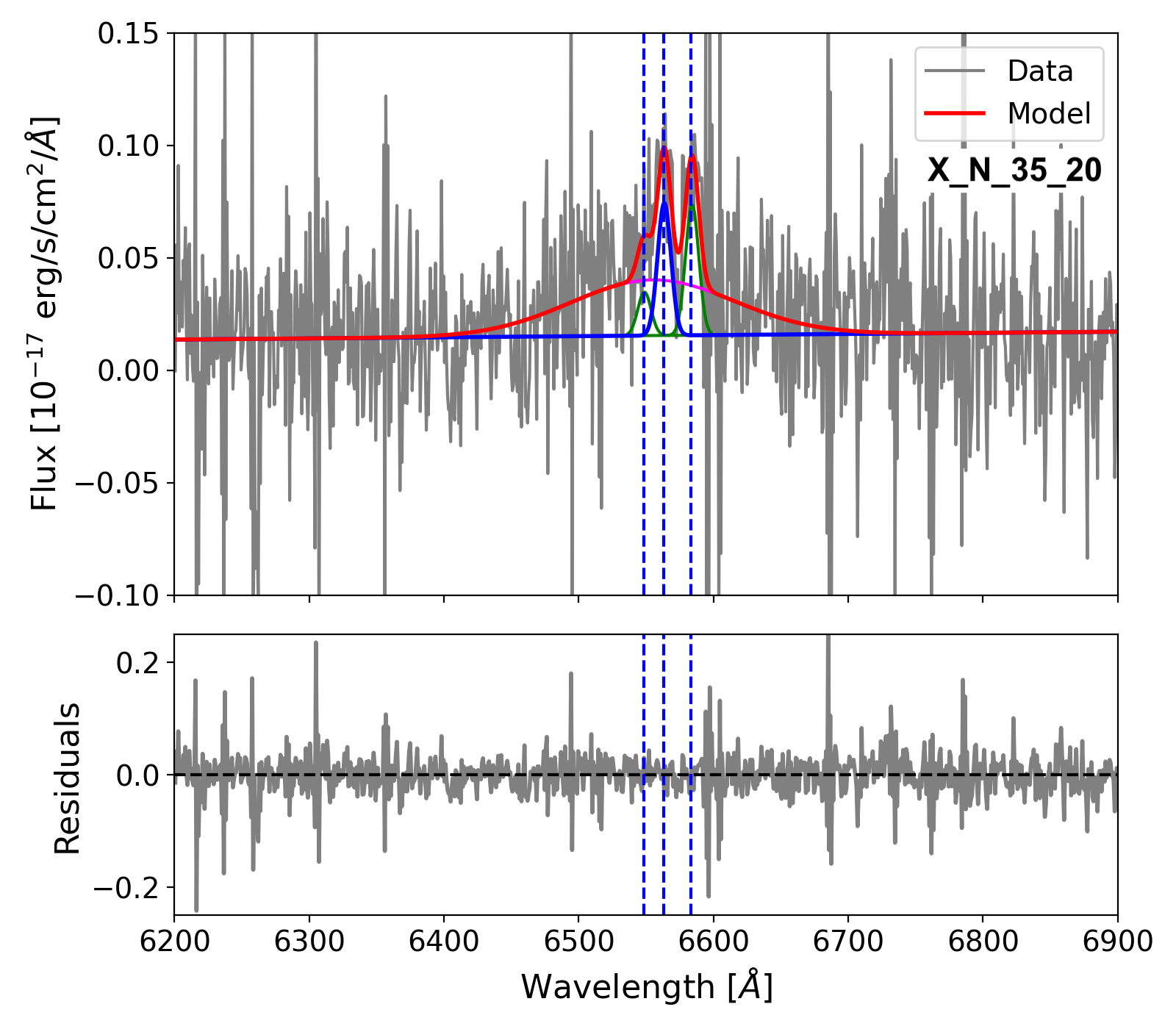}
\includegraphics[scale=0.28]{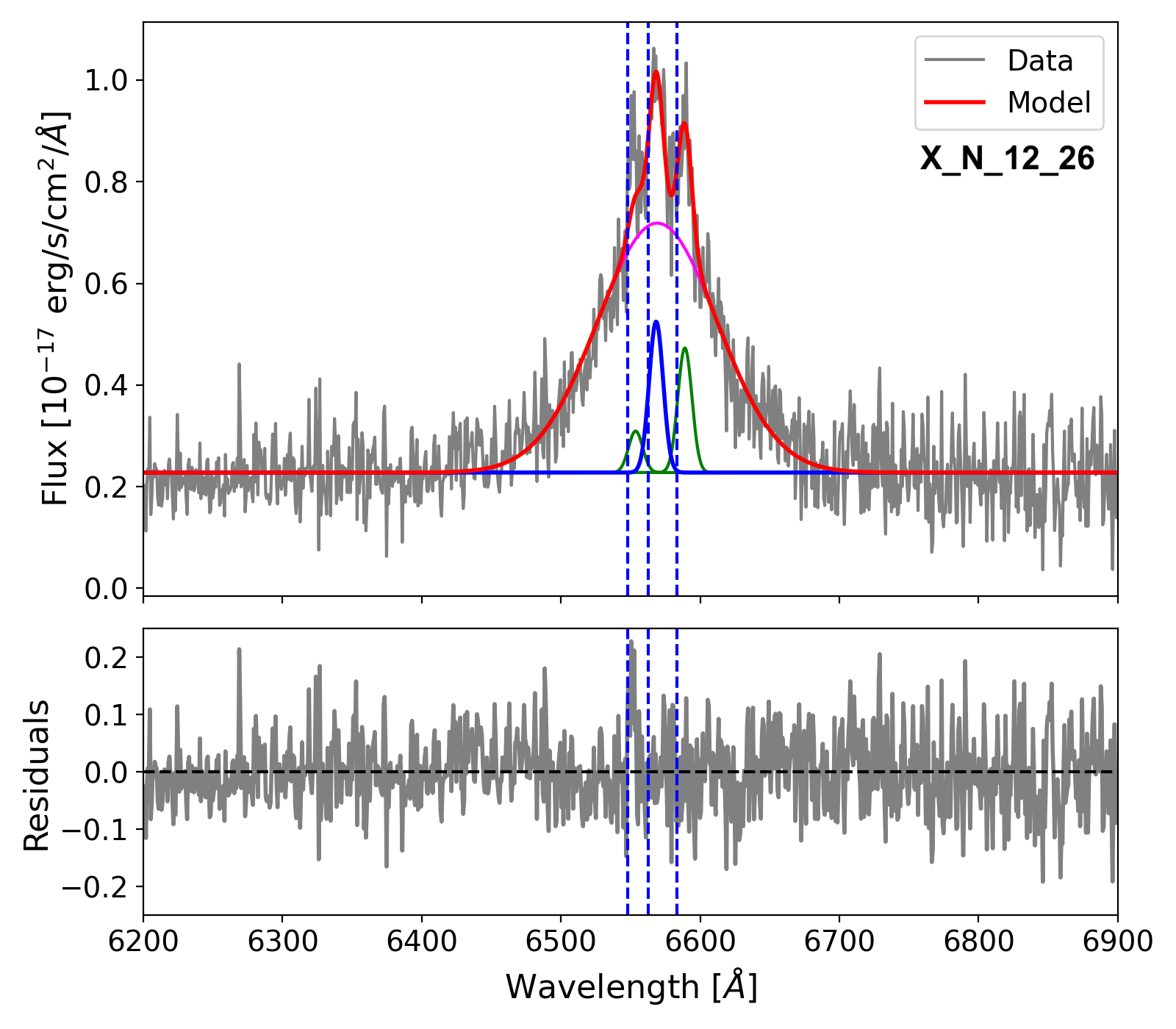}
\includegraphics[scale=0.28]{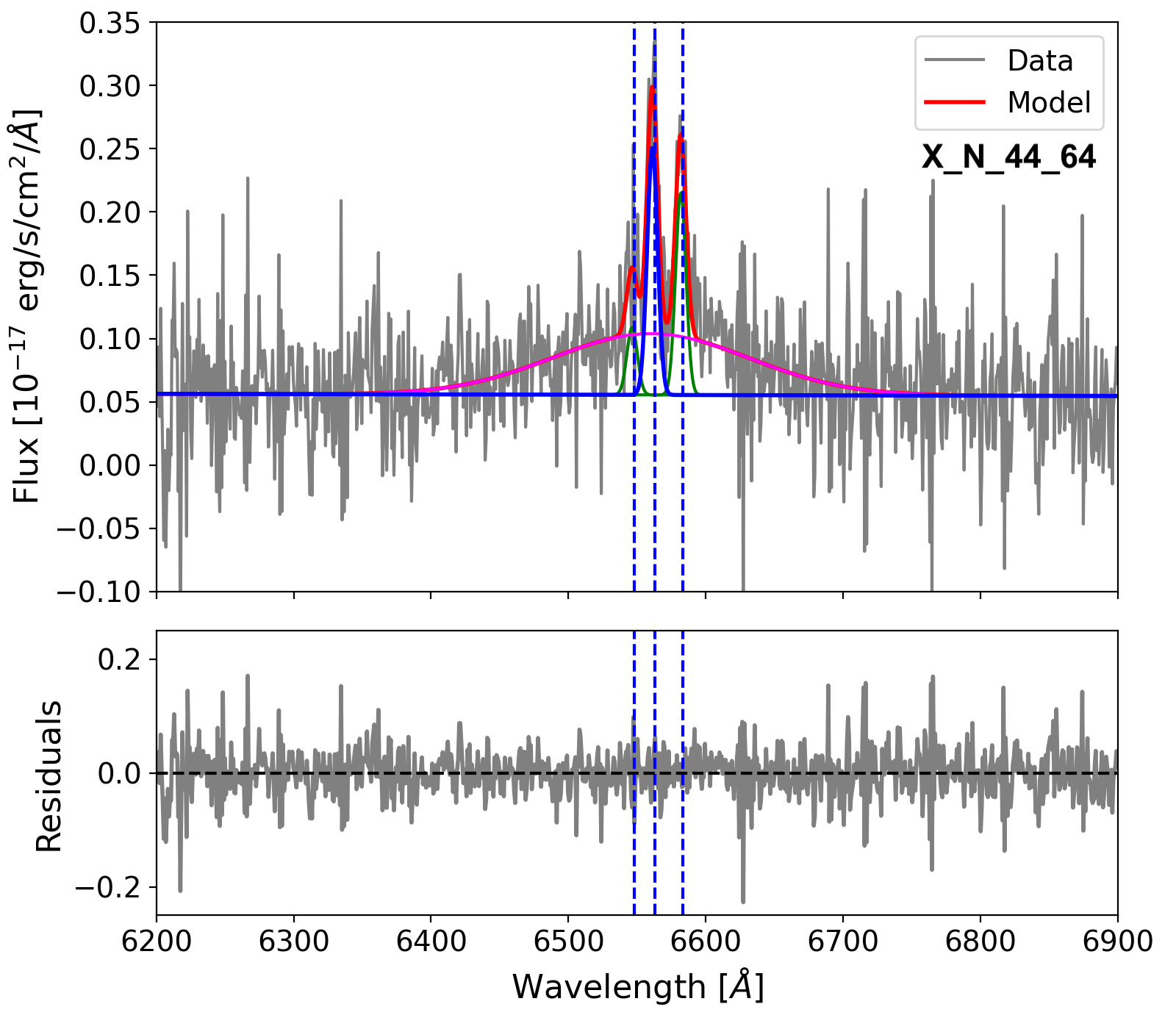}
\includegraphics[scale=0.28]{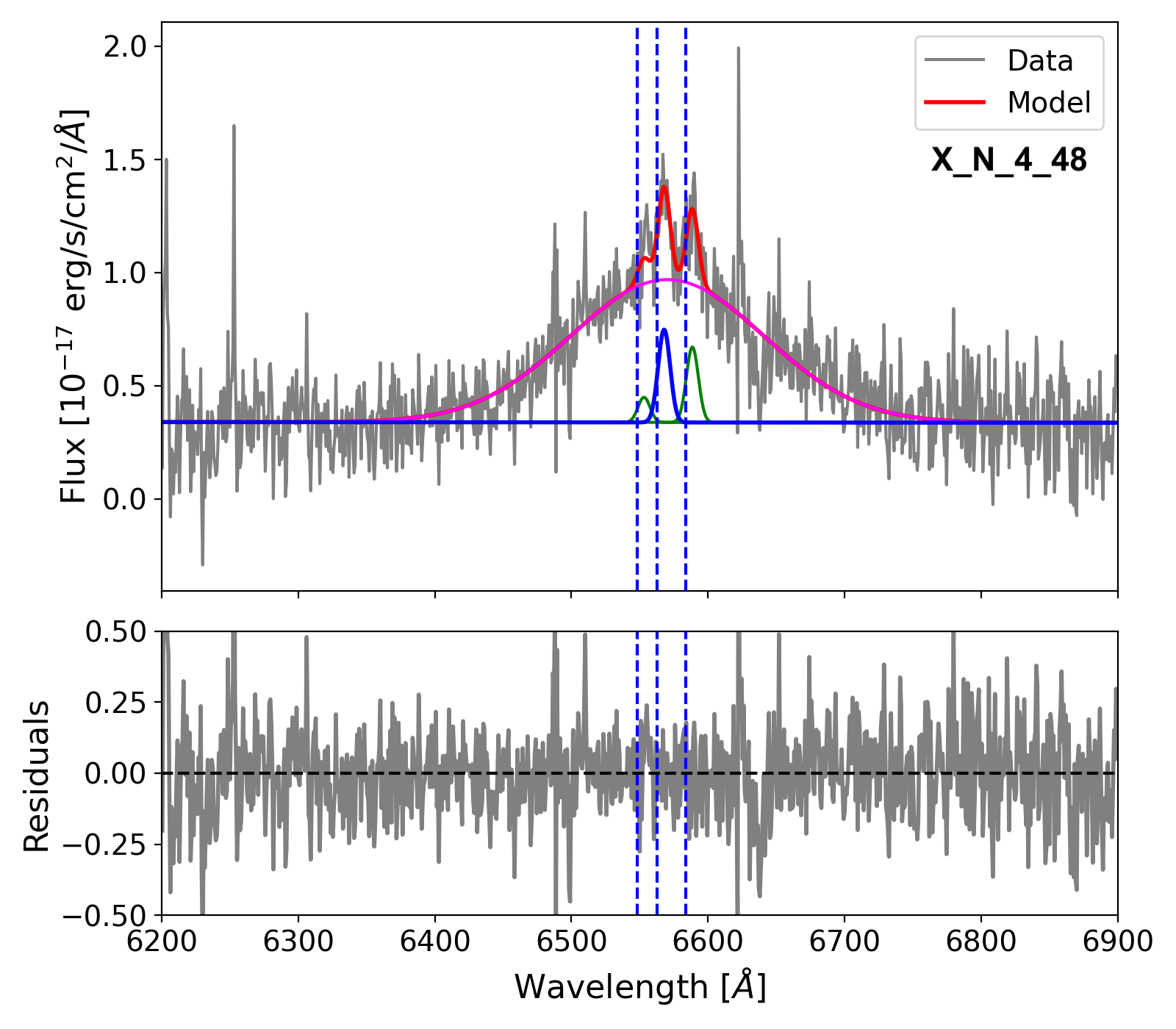}
\includegraphics[scale=0.28]{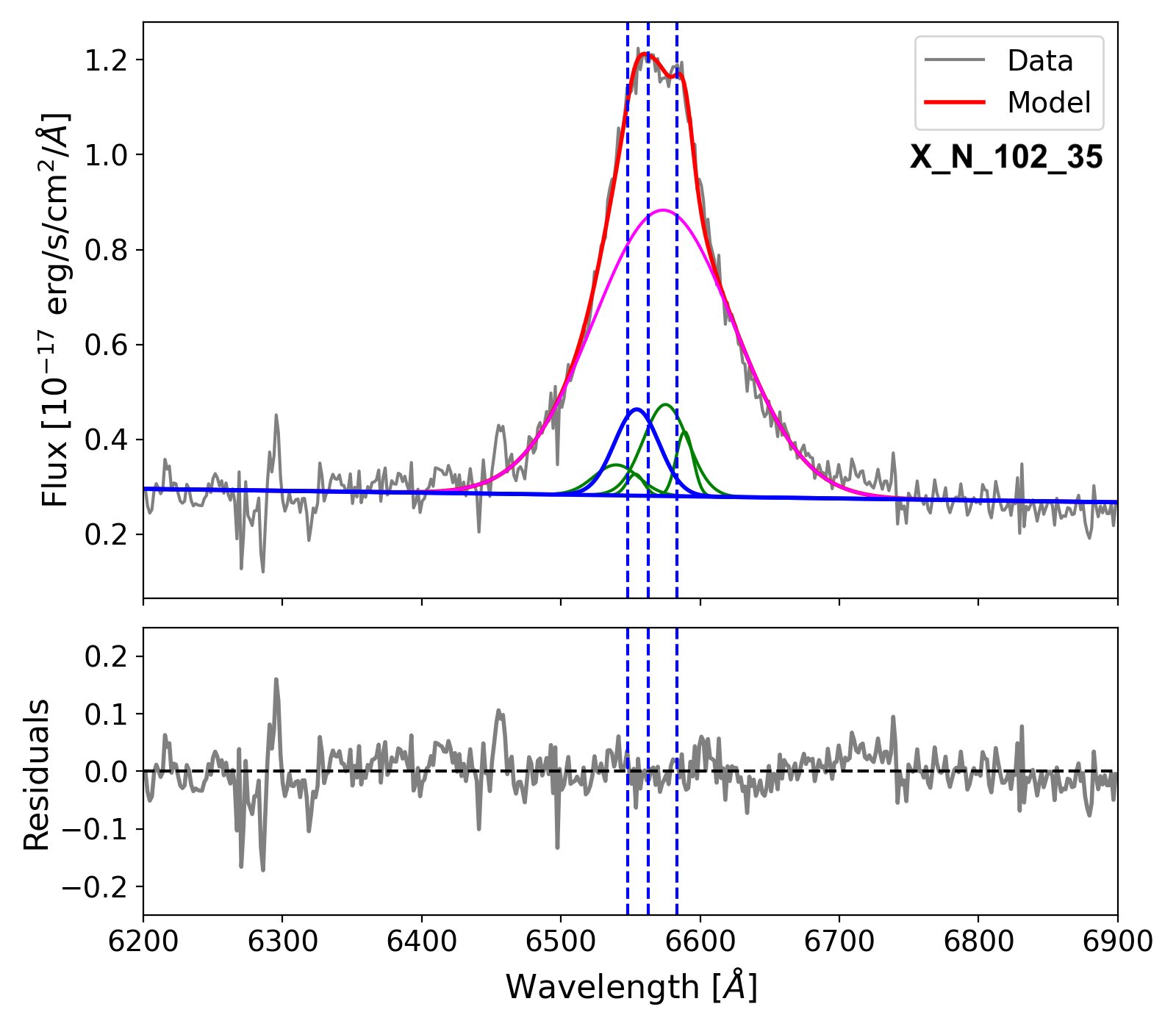}
\includegraphics[scale=0.28]{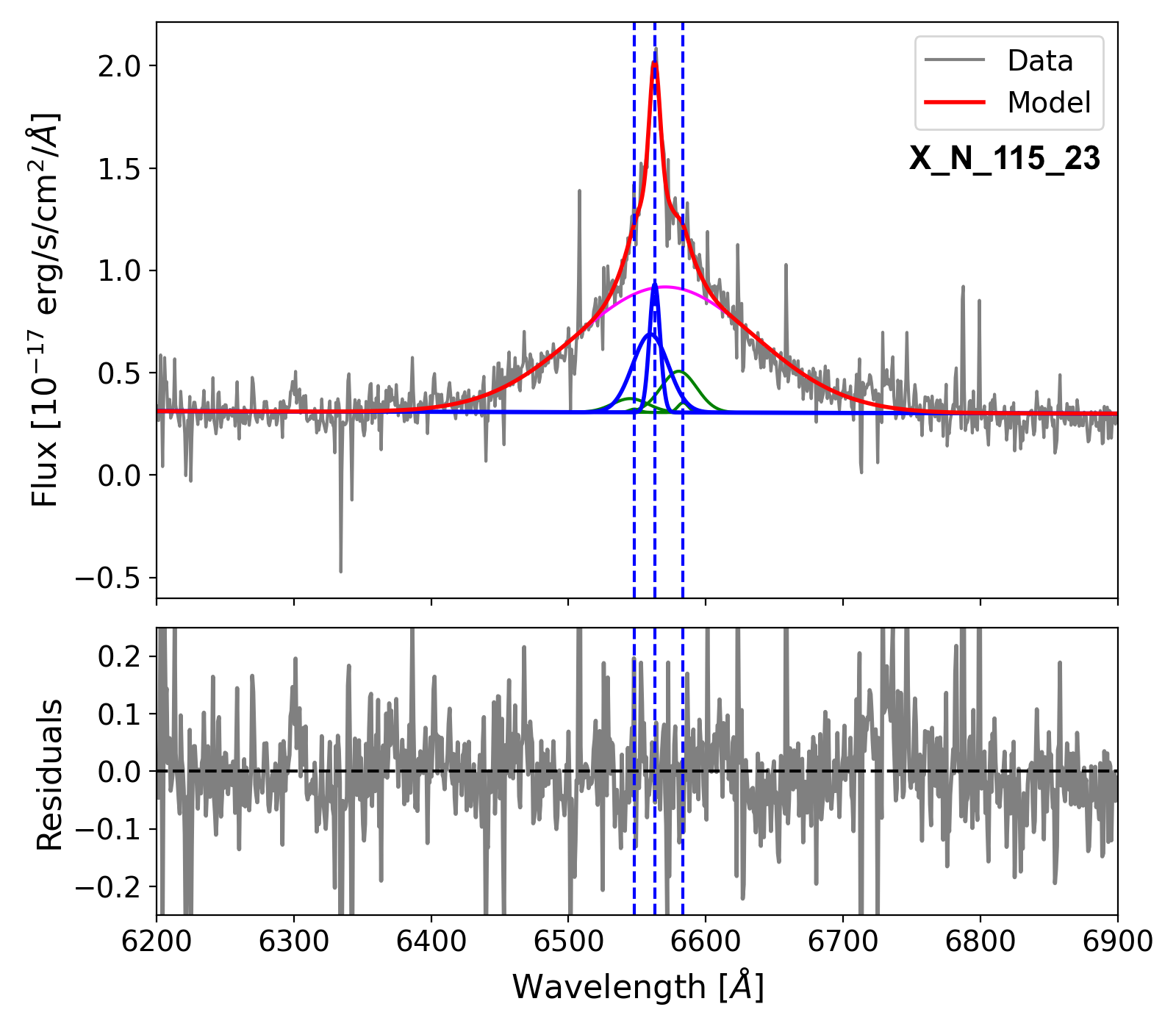}
\includegraphics[scale=0.28]{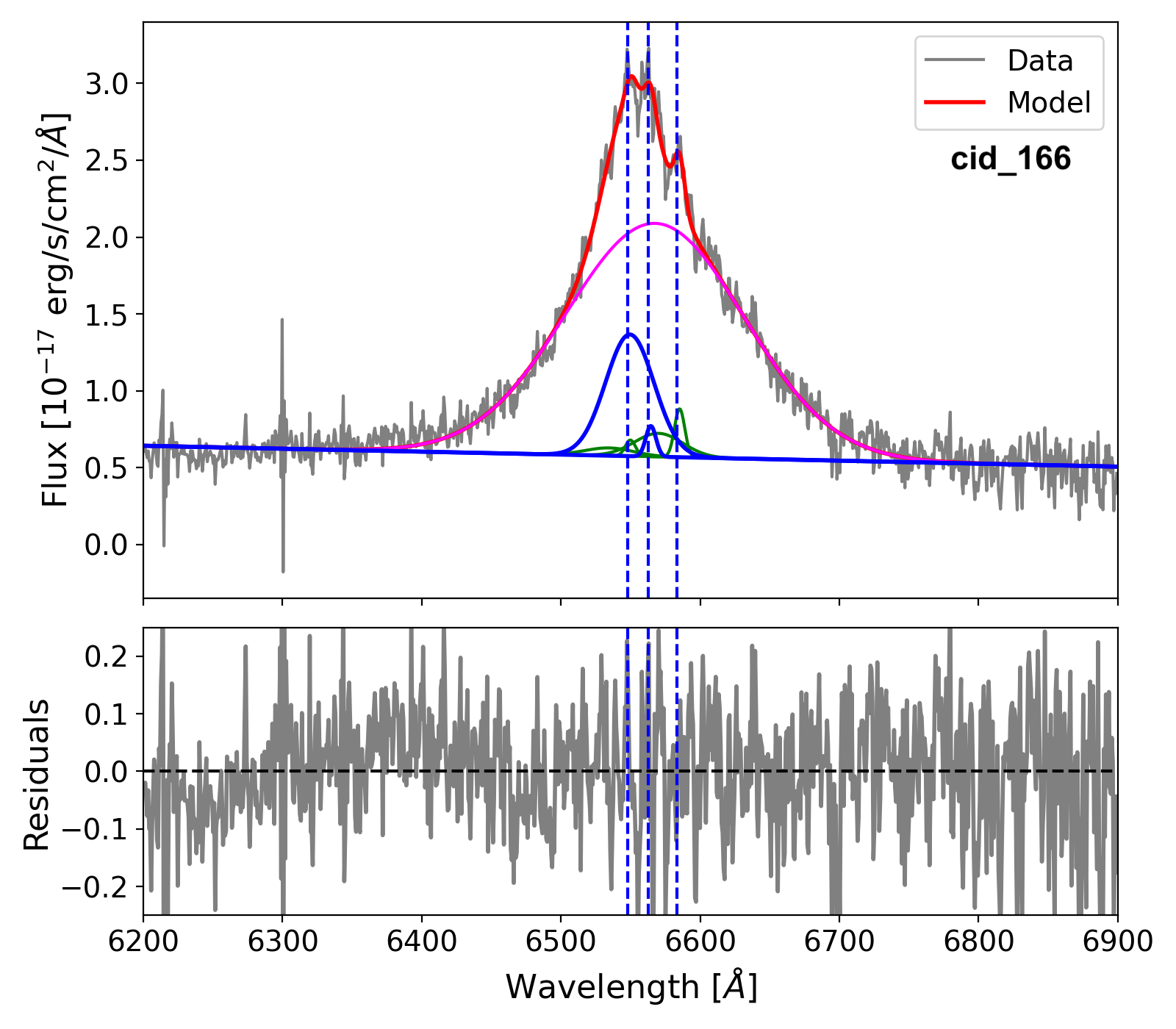}
\includegraphics[scale=0.28]{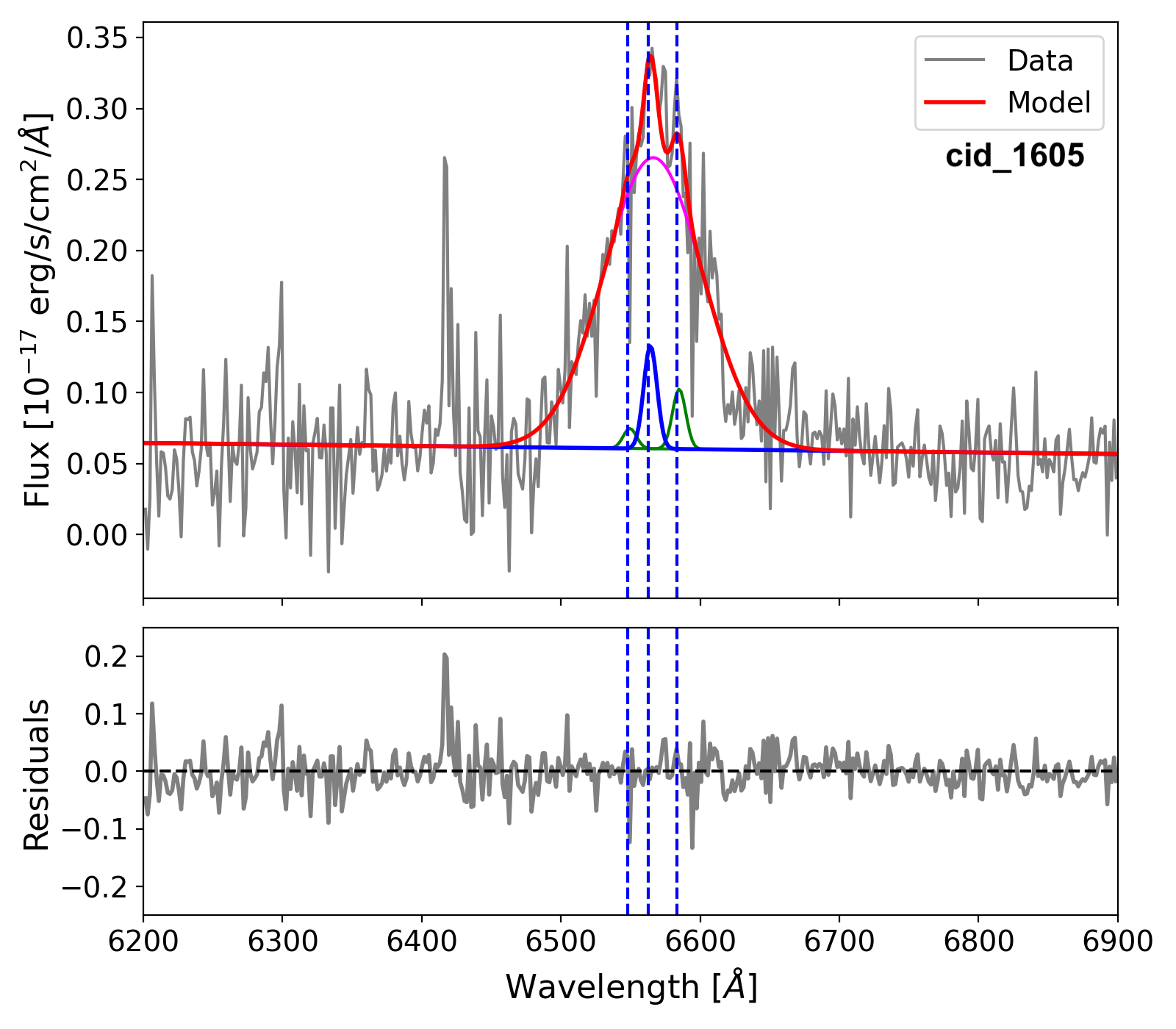}
\caption{Integrated K-band spectra of all targets for the fitting model used in this paper.}
\label{fig:intspec_alltargets1}
\end{figure*}

\begin{figure*}
\centering
\includegraphics[scale=0.28]{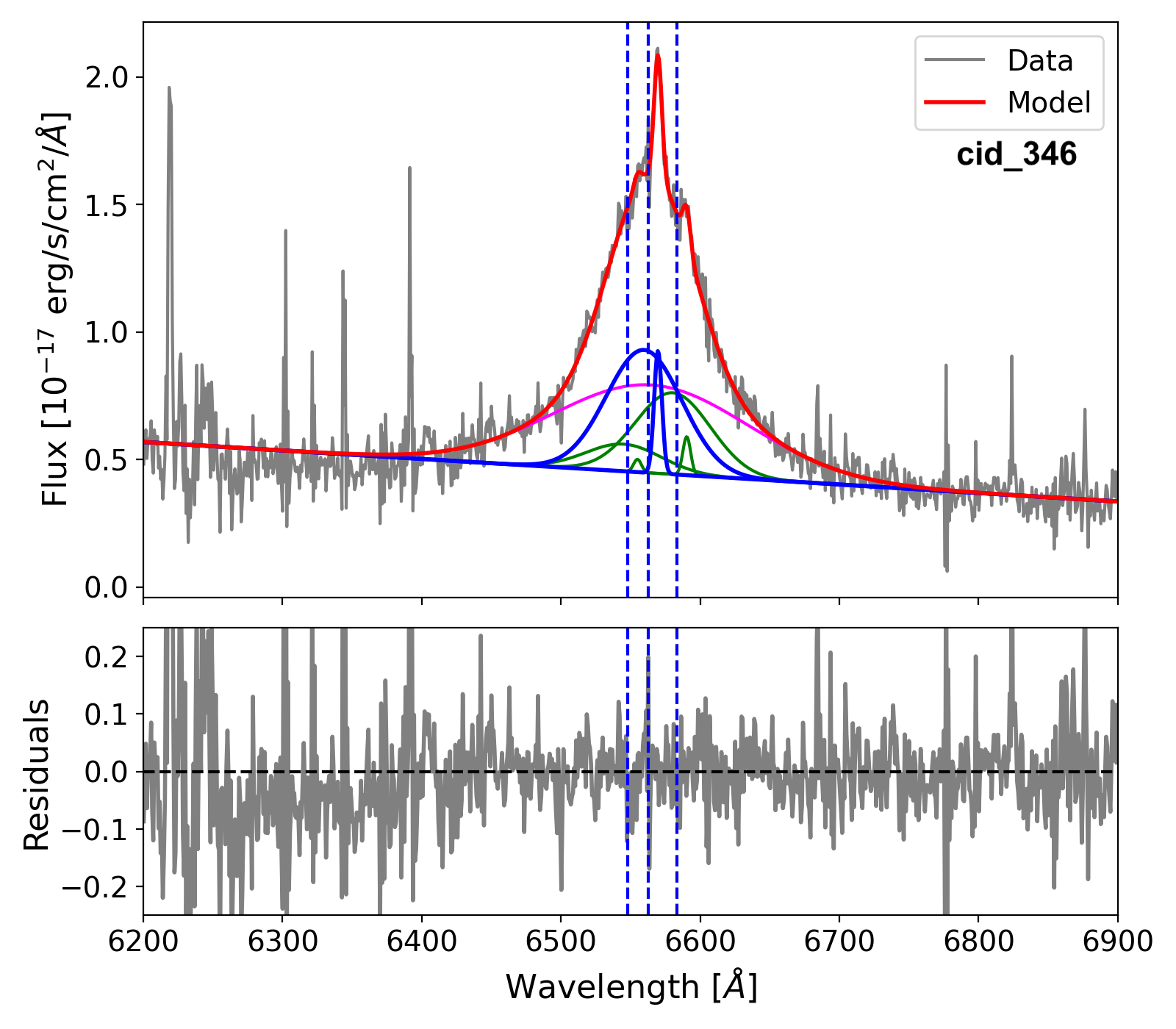}
\includegraphics[scale=0.28]{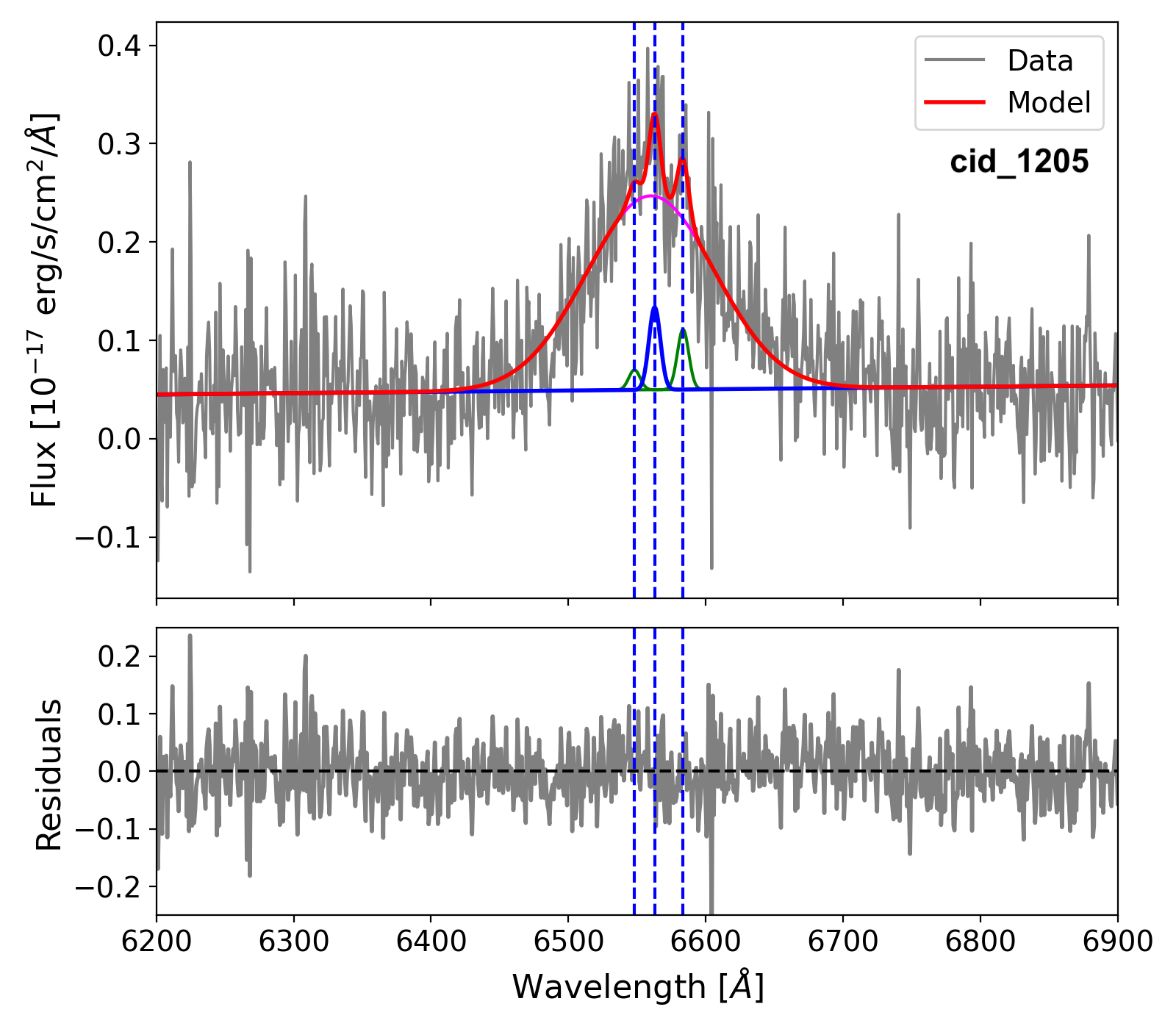}
\includegraphics[scale=0.28]{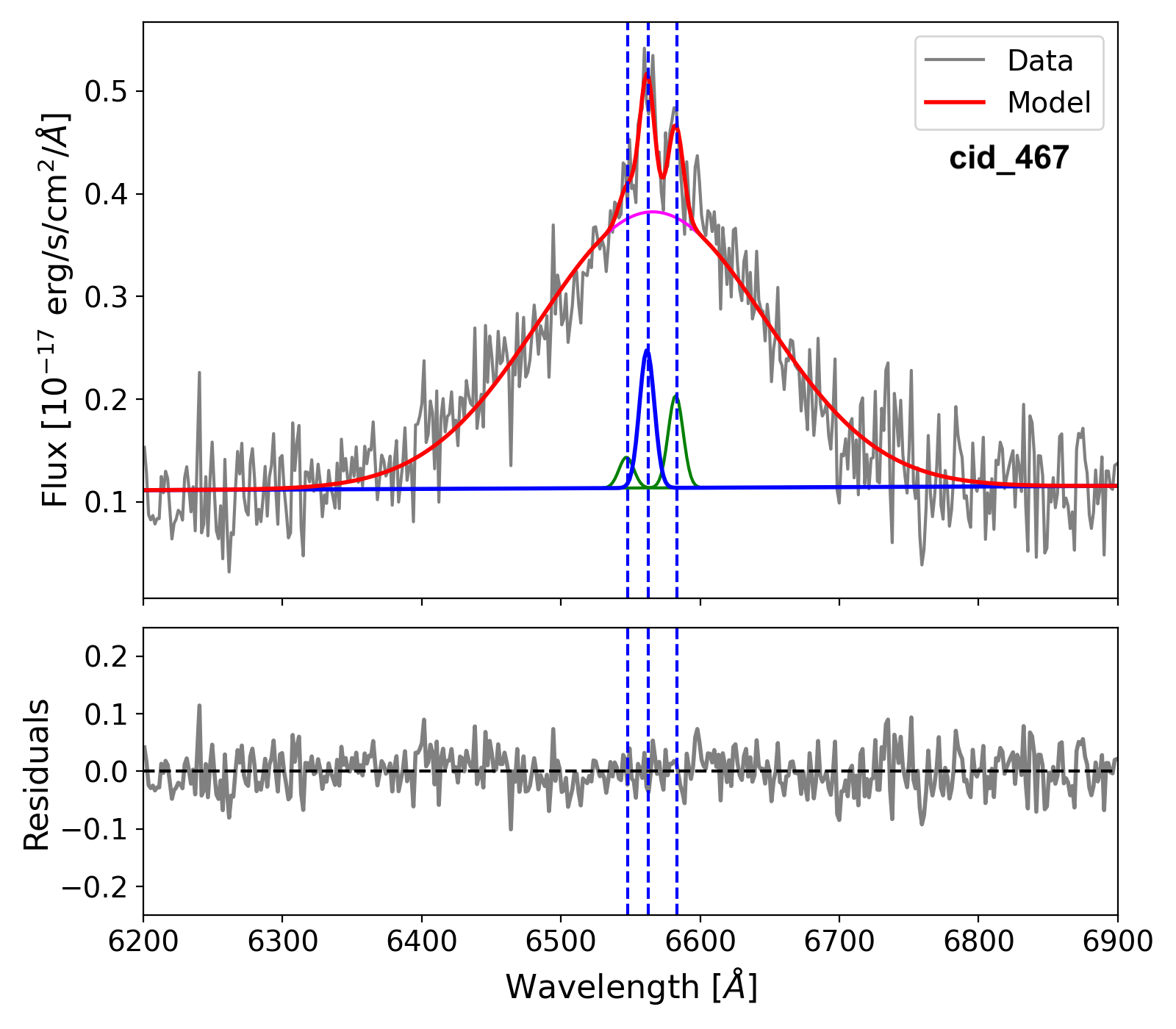}
\includegraphics[scale=0.28]{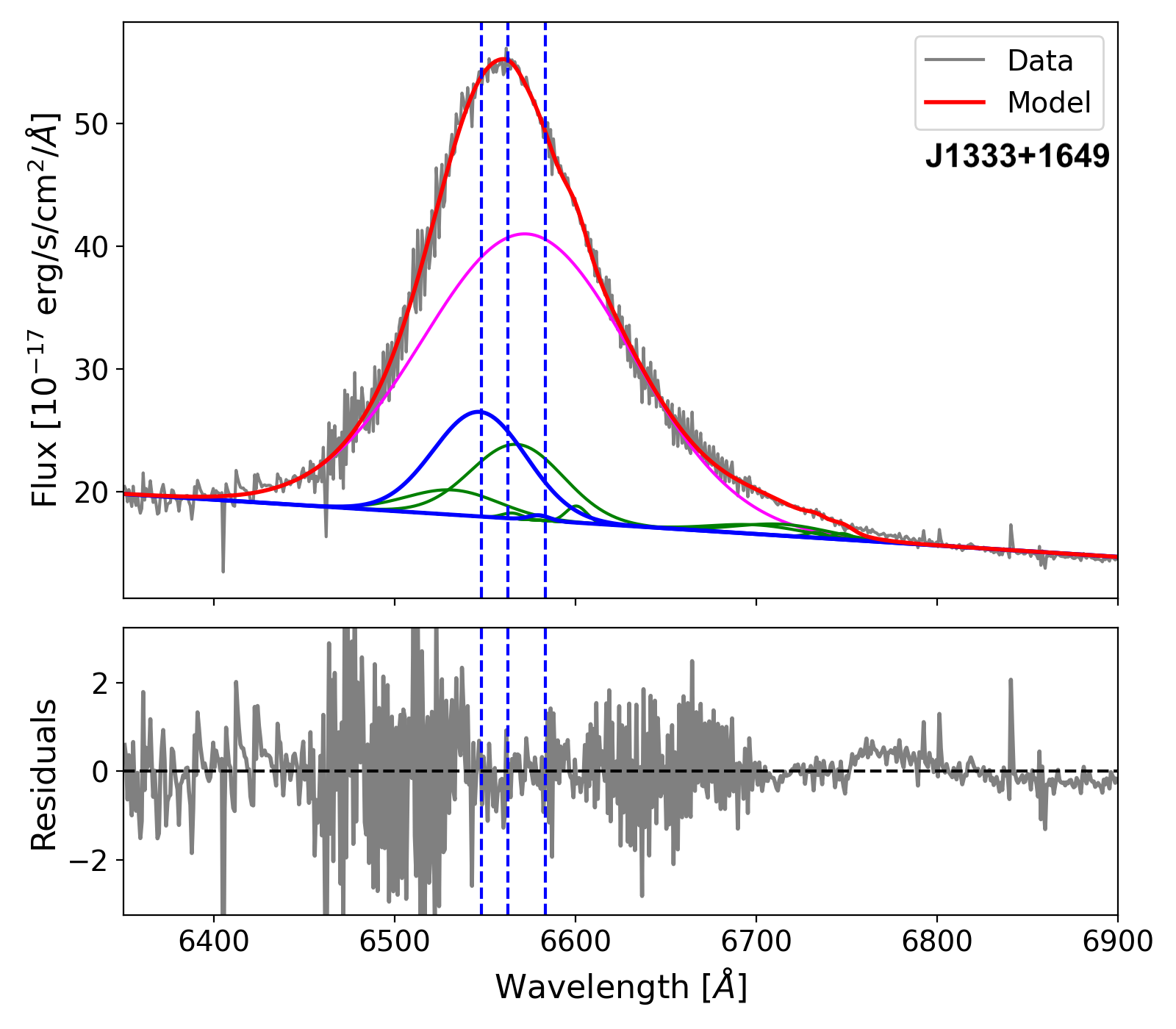}
\includegraphics[scale=0.28]{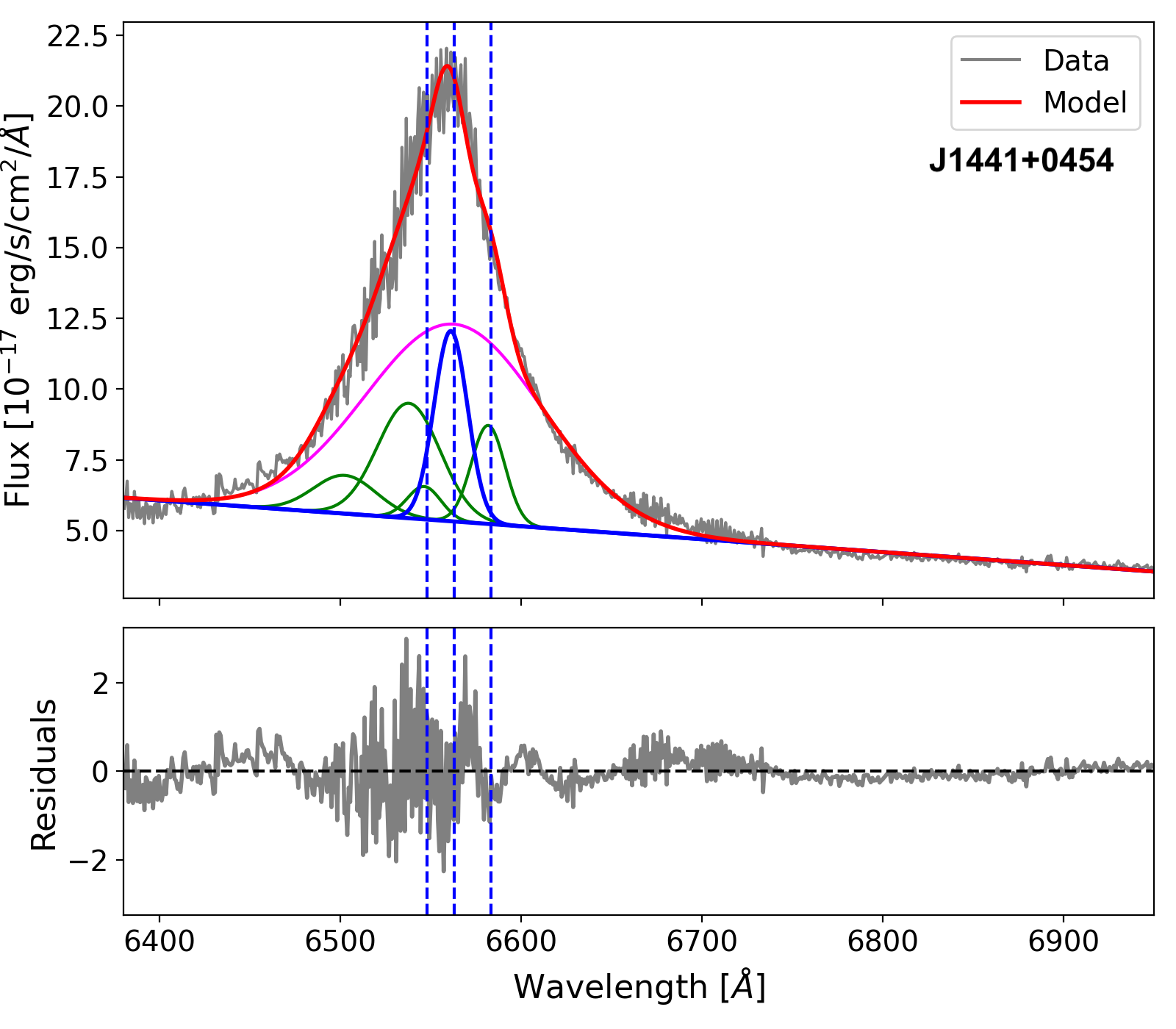}
\includegraphics[scale=0.28]{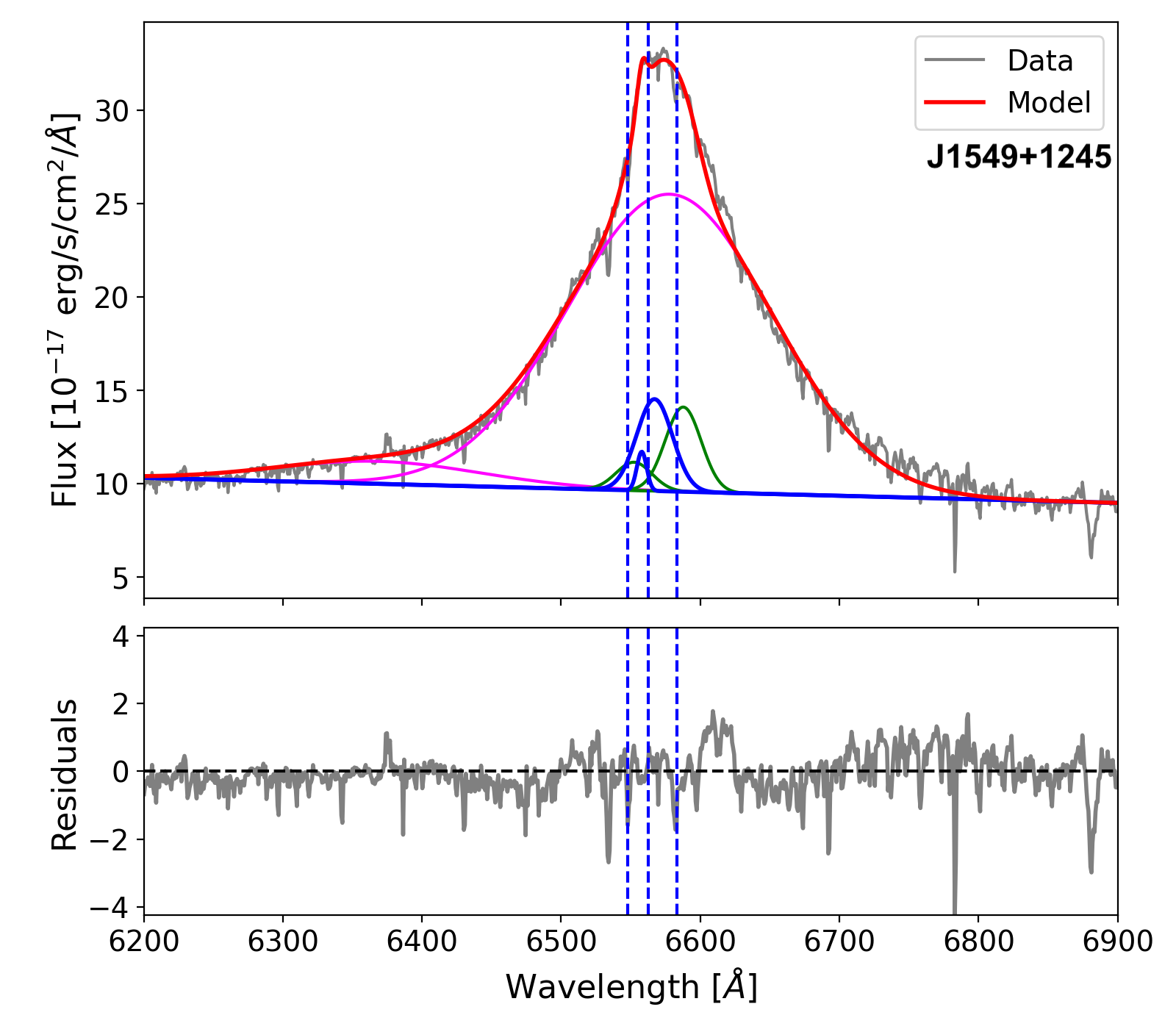}
\includegraphics[scale=0.28]{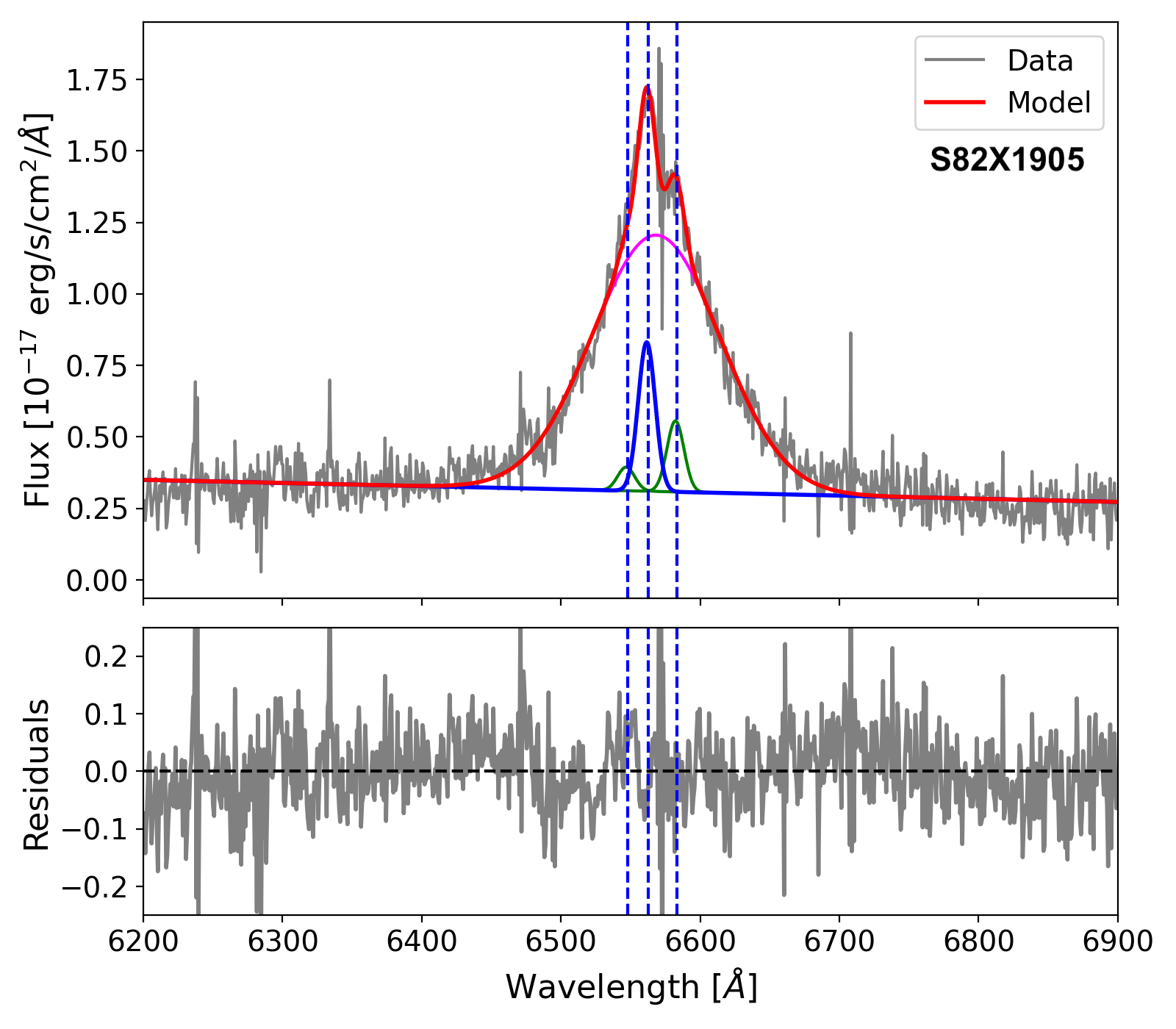}
\includegraphics[scale=0.28]{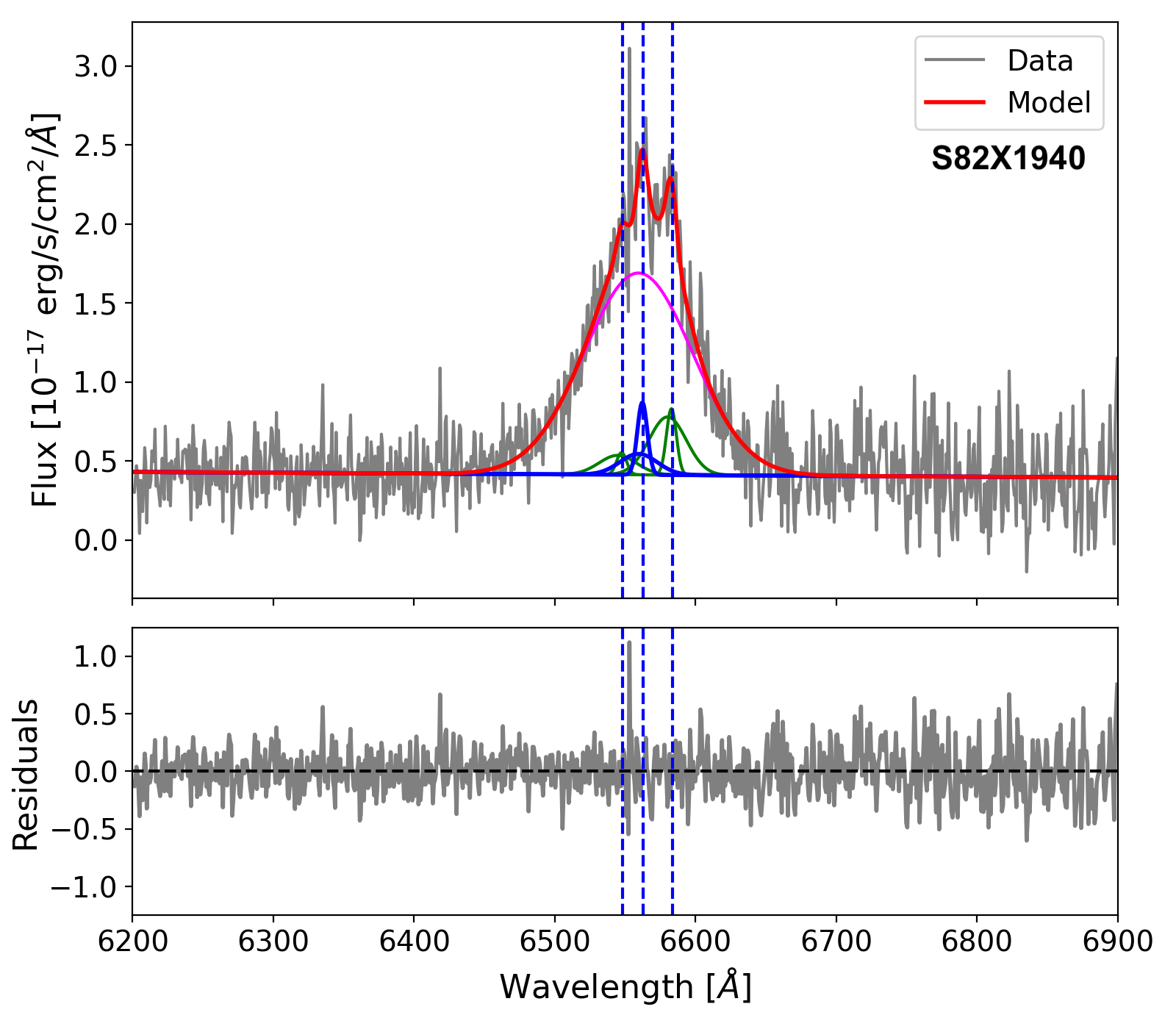}
\includegraphics[scale=0.28]{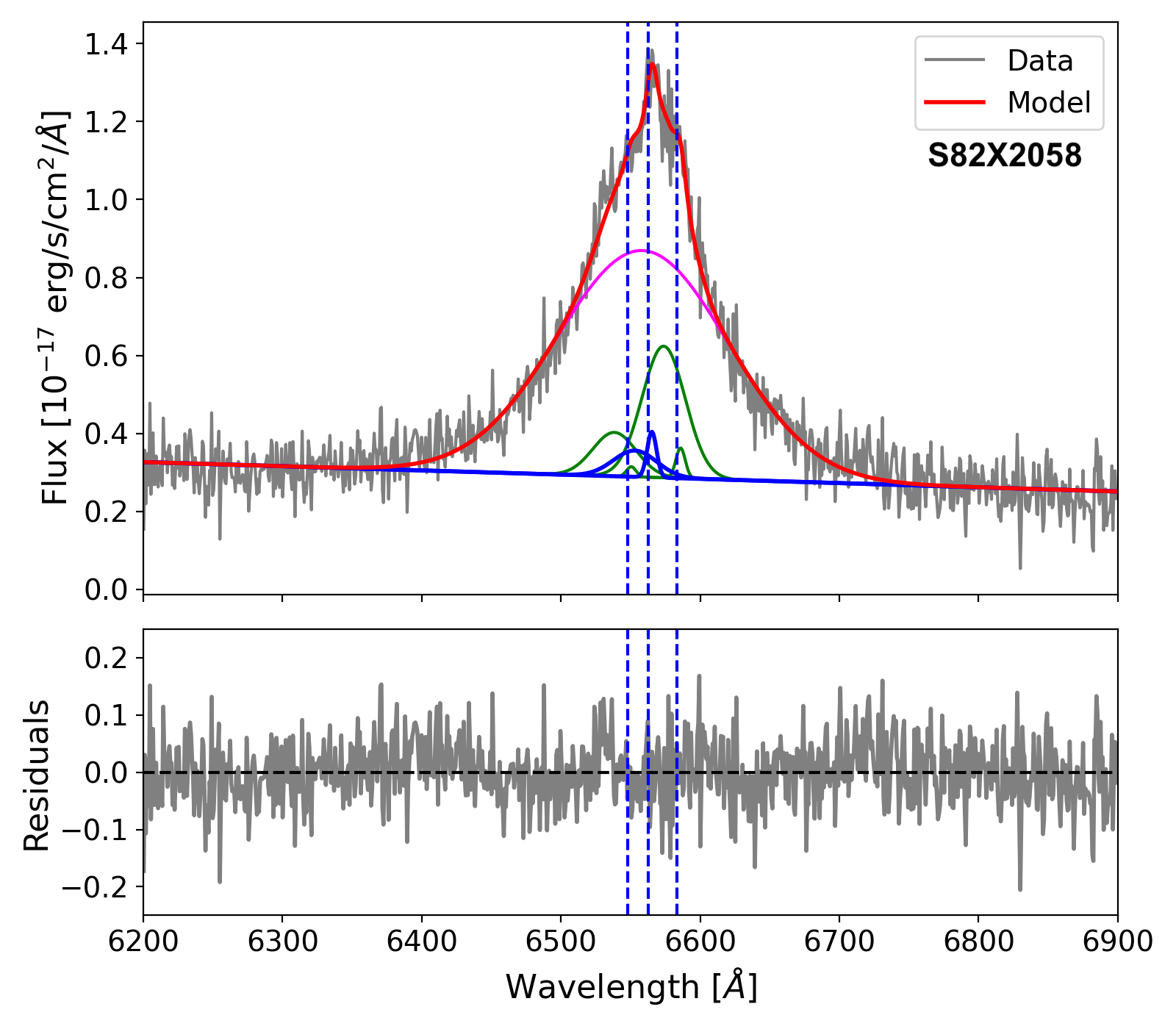}
\caption{Integrated K-band spectra of all targets for the fitting model used in this paper (continued from Figure \ref{fig:intspec_alltargets1}).}
\label{fig:intspec_alltargets2}
\end{figure*}

\begin{figure*}
\centering
\includegraphics[scale=0.38]{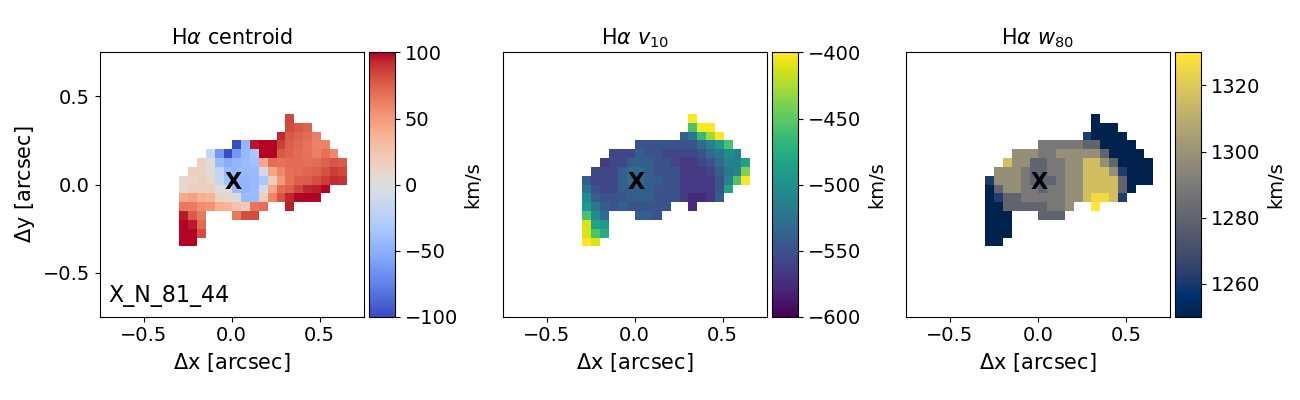}
\includegraphics[scale=0.38]{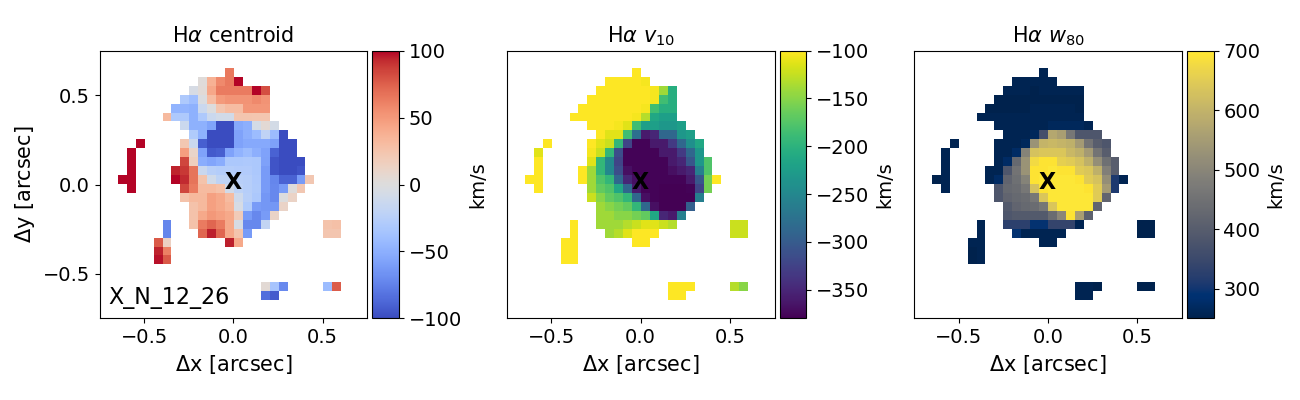}
\includegraphics[scale=0.38]{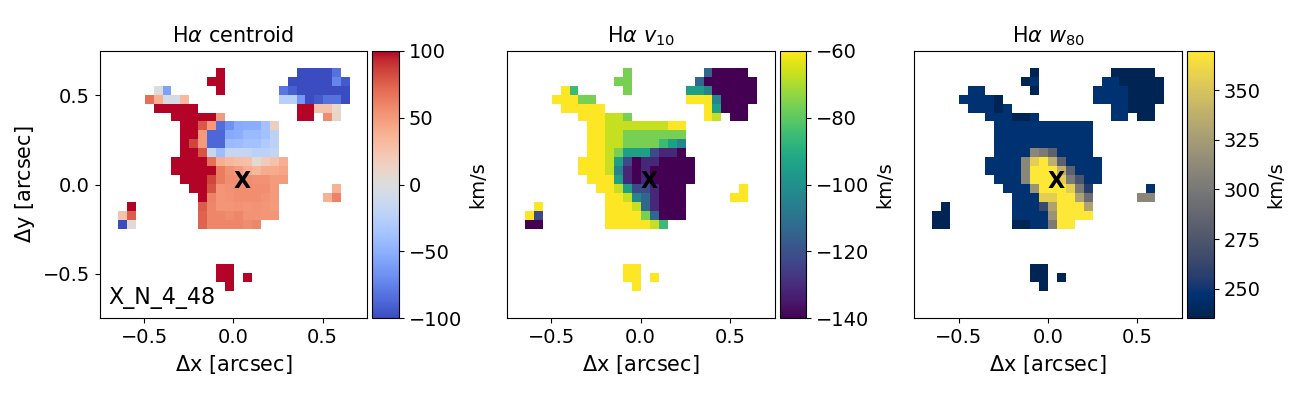}
\includegraphics[scale=0.38]{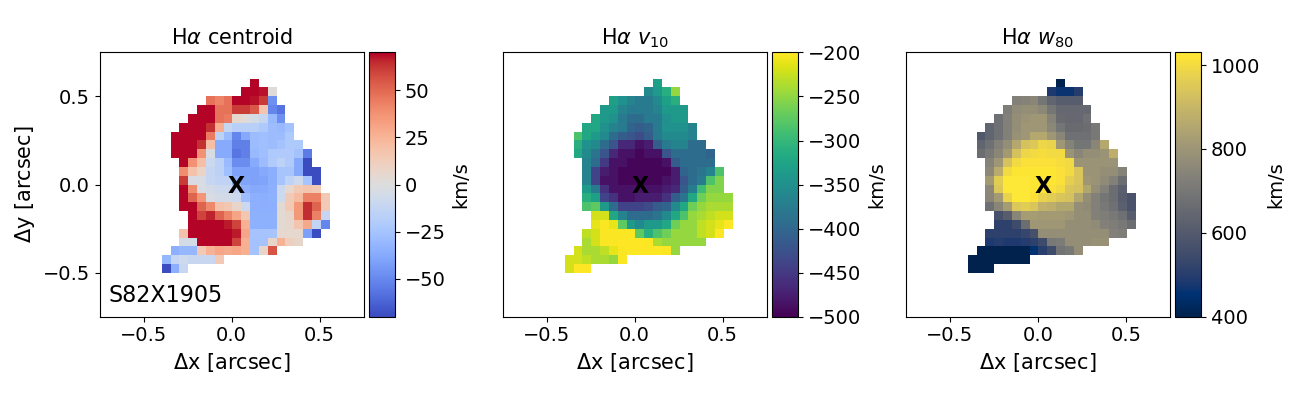}
\includegraphics[scale=0.38]{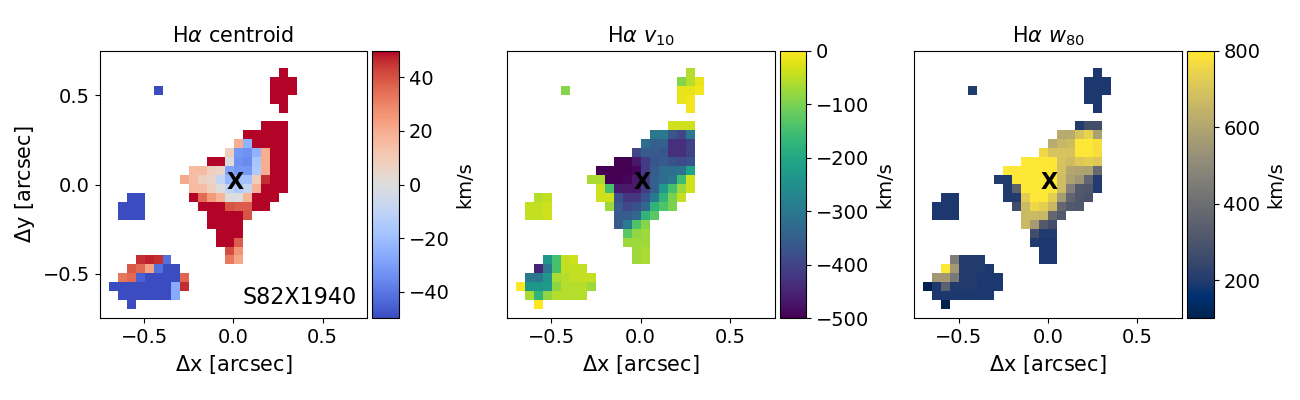}
\caption{Same as Fig. \ref{fig:velocity_maps}} 
\label{fig:velocity_maps_2}
\end{figure*}

\bsp	
\label{lastpage}
\end{document}